\documentclass[fleqn,usenatbib]{mnras}

\usepackage[T1]{fontenc}

\DeclareRobustCommand{\VAN}[3]{#2}
\let\VANthebibliography\thebibliography
\def\thebibliography{\DeclareRobustCommand{\VAN}[3]{##3}\VANthebibliography}

\usepackage{graphicx}	
\usepackage{amsmath}	
\usepackage{amssymb}	

\usepackage{newtxtext,newtxmath}

\PassOptionsToPackage{hyphens}{url}
\usepackage{xcolor}
\definecolor{Brown}{rgb}{0.647,0.165,0.165}
\definecolor{NavyBlue}{rgb}{0.0,0,0.5}
\definecolor{Burgundy}{rgb}{0.5,0.0,0.125}
\definecolor{lime}{rgb}{0.651,0.808,0.224}
\hypersetup{
    breaklinks=true,
    colorlinks=true,
    citecolor={Burgundy},
    linkcolor={NavyBlue},
    urlcolor={Brown}
}
\usepackage{multirow}

\usepackage{tikz,hyperref}

\DeclareRobustCommand{\orcidicon}{%
    \begin{tikzpicture}
    \draw[lime, fill=lime] (0,0) 
    circle [radius=0.16] 
    node[white] {{\fontfamily{qag}\selectfont \tiny ID}};
    \draw[white, fill=white] (-0.0625,0.095) 
    circle [radius=0.007];
    \end{tikzpicture}
    \hspace{-2mm}
}

\def\apj{Astrophys.\ J.}

\def\mnras{Mon.\ Not.\ R.\ Astron.\ Soc.}
\def\aap{Astron.\ Astrophys.}
\def\apjl{Astrophys.\ J.\ Lett.}

\def\physrep{Phys.\ Rep.}
\def\pre{Phys.\ Rev.\ E}

\def\apjs{ApJS}

\def\pasa{PASA}

\def\nat{Nature}
\def\pasp{PASP}

\def\araa{Ann.\ Rev.\ Astron.\ Astrophys.}

\newcommand{\Mach}{\mathcal{M}}      
 
\renewcommand{\vec}[1]{\mathbf{#1}}	
\newcommand{\dd}{\mathrm{d}}        

\newcommand{\cm}{{\rm cm}}    
\newcommand{\m}{{\rm m}}      
\newcommand{\pc}{{\rm pc}}     
\newcommand{\kpc}{{\rm kpc}}  

\newcommand{\muG}{\mu{\rm G}} 

\newcommand{\K}{{\rm K}}      

\newcommand{\rad}{{\rm rad}}

\newcommand{\brms}{b_{\rm rms}}

\renewcommand{\ne}{n_{\rm e}}

\newcommand{\RM}{\text{RM}}

\newcommand{\IHa}{{\rm I_{H\alpha}}}

\newcommand{\ku}{\mathcal{K}}
\newcommand{\pk}{\mathcal{P}_{k}}
\newcommand{\plasmabeta}{\beta_{\rm plasma}}

\newcommand{\sftwo}{{\rm SF}^{\rm(2\,pts)}_2({\rm RM})}
\newcommand{\sfthree}{{\rm SF}^{\rm(3\,pts)}_2({\rm RM})}
\newcommand{\sffour}{{\rm SF}^{\rm(4\,pts)}_2({\rm RM})}
\newcommand{\sffive}{{\rm SF}^{\rm(5\,pts)}_2({\rm RM})}

\newcommand\Eq[1]{Eq.\,\ref{#1}}
\newcommand\Fig[1]{Fig.~\ref{#1}}
\newcommand\Sec[1]{Sec.~\ref{#1}}
\newcommand\Tab[1]{Table~\ref{#1}}
\newcommand\App[1]{Appendix~\ref{#1}}

\newcommand\rev[1]{#1}
\newcommand\revb[1]{#1}

\graphicspath{{./}{figs/}}

\title[RM structure function with higher-order stencils]{Rotation measure structure functions with higher-order stencils as a probe of small-scale magnetic fluctuations and its application to the Small and Large Magellanic Clouds}

\author[Seta et al.]{
Amit Seta\href{https://orcid.org/0000-0001-9708-0286}{\orcidicon}, 
\thanks{E-mail: \href{mailto:amit.seta@anu.edu.au}{amit.seta@anu.edu.au}}
Christoph Federrath\href{https://orcid.org/0000-0002-0706-2306}{\orcidicon}, 
Jack D. Livingston\href{https://orcid.org/0000-0002-4090-8000}{\orcidicon}, and
N. M. McClure-Griffiths\href{https://orcid.org/0000-0003-2730-957X}{\orcidicon}
\\
Research School of Astronomy and Astrophysics, 
Australian National University, Canberra, ACT 2611, Australia\\
}

\date{Accepted XXX. Received YYY; in original form ZZZ}

\pubyear{2021}

\begin{document}
\label{firstpage}
\pagerange{\pageref{firstpage}--\pageref{lastpage}}
\maketitle

\begin{abstract}
Magnetic fields and turbulence are important components of the interstellar medium (ISM) of star-forming galaxies. It is challenging to measure the properties of the small-scale ISM magnetic fields (magnetic fields at scales smaller than the turbulence driving scale). Using numerical simulations, we demonstrate how the second-order rotation measure (RM, which depends on thermal electron density, $n_{\rm e}$, and magnetic field, $b$) structure function can probe the properties of small-scale $b$. We then apply our results to observations of the Small and Large Magellanic Clouds (SMC and LMC). First, using Gaussian random $b$, we show that the characteristic scale where the RM structure function flattens is approximately equal to the correlation length of $b$. We also show that computing the RM structure function with a higher-order stencil (more than the commonly-used two-point stencil) is necessary to accurately estimate the slope of the structure function. Then, using Gaussian random $b$ and lognormal $n_{\rm e}$ with known power spectra, we derive an empirical relationship between the slope of the power spectrum of $b$, $n_{\rm e}$, and RM. We apply these results to the SMC and LMC and estimate the following properties of small-scale $b$: correlation length ($160~\pm 21~{\rm pc}$ for the SMC and $87~\pm~17~{\rm pc}$ for the LMC), strength ($14~\pm 2~\mu{\rm G}$ for the SMC and $15~\pm 3~\mu{\rm G}$ for the LMC), and slope of the magnetic power spectrum ($-1.3~\pm~0.4$ for the SMC and $-1.6~\pm~0.1$ for the LMC). We also find that $n_{\rm e}$ is practically constant over the estimated $b$ correlation scales.
\end{abstract}




\section{Introduction} \label{sec:intro}
Magnetic fields play an important dynamical role in the interstellar medium (ISM) of star-forming galaxies. Locally, the magnetic field provides pressure support against gravity \citep{BoularesC1990}, reduces the star formation rate \citep{MestelS1956,Federrath2015,KrumholzF2019}, controls the propagation of cosmic rays \citep{Cesarsky1980,ShukurovEA2017}, heats up the ISM via magnetic reconnection \citep{Raymond1992}, alters the gas flow morphology \citep{ShettyO2016}, and also suppresses galactic outflows \citep{EvirgenEA17}. Globally, the role of magnetic fields in the formation and evolution of galaxies is still unknown but recent cosmological simulations suggest that they might be important \citep{VoortEA2021}. Thus, it is crucial to study the strength and structure of galactic magnetic fields, especially in different types of galaxies (such as spiral, elliptical, and irregular) and also in their different evolutionary stages.

Magnetic fields in galaxies can be probed by a variety of observational methods such as optical polarisation, dust polarisation, synchrotron emission, the Zeeman effect, and the Faraday rotation of the polarised light \citep[see Chapter 3 of][]{KleinFletcher2015}. Except for the Zeeman effect, which is accessible only in the low-temperature and high-density regions of the ISM \citep{HeilesT2004}, usually one requires additional assumptions or other independent observations to extract magnetic field properties from these observations. For example, to infer magnetic field properties from the total synchrotron intensity (which is due to cosmic ray electrons and the magnetic field component perpendicular to the line of sight), one needs information about the cosmic ray electron number density. In the absence of such information, energy equipartition between cosmic rays and magnetic fields is usually assumed to extract the magnetic field strength from synchrotron observations. However, the energy equipartition assumption is not always applicable and might give erroneous results \citep{SetaB2019}. Similarly, to extract magnetic field information from Faraday rotation (which is due to the thermal electrons and the magnetic field component parallel to the line of sight), one requires an estimate of the thermal electron density. The thermal electron density can be inferred from H$\alpha$ observations or dispersion measures in the case of pulsars, but in the absence of such observations, additional assumptions (e.g.~a constant thermal electron density) are required. Such assumptions might give incorrect magnetic field information \citep{BeckEA2003}. In this paper, we study the impact of such assumptions on the estimated small-scale magnetic field properties, primarily using the structure function of rotation measure ($\RM$) calibrated with numerical simulations. We then apply the structure function method to Faraday rotation observations of the Small and Large Magellanic Clouds (SMC and LMC).

Based on radio observations, magnetic fields in a typical spiral galaxy can be divided into \rev{fluctuating (or small-scale) and mean (or large-scale) components.} The small-scale fields are correlated at scales less than the driving scale of turbulence ($\sim 100~\pc$ in spirals) and the large-scale fields are correlated over several $\kpc$s. The linearly polarised and total synchrotron intensity traces the ordered field and total field components perpendicular to the line of sight, respectively. Information about small-scale fluctuating fields can be obtained from depolarisation values \rev{\citep[which refers to a decrease in the polarisation fraction due to random magnetic fields along the path length, see][for further details]{SokoloffEA1998}}. The magnetic field component parallel to the line of sight can be studied using Faraday rotation measurements, where the mean and standard deviation of $\RM$ roughly probes the large- and small-scale components, respectively. Using these observations for the Milky Way and nearby spiral galaxies, the total field strength is estimated to be of the order of $5~\muG$ \citep{Haverkorn2015, Beck2016}. The small-scale field strength is found to be comparable to (or even slightly stronger than) the large-scale field strength. The large-scale field roughly follows the spiral arms \citep{FletcherEA2011, Beck2016} and the small-scale random field has both Gaussian and non-Gaussian components \rev{\citep[see Appendix A in][for further details]{SetaEA2018}}. Physically, the strength and structure of magnetic fields in galaxies can be explained by the turbulent dynamo, which is a mechanism for the conversion of turbulent kinetic energy to magnetic energy \citep{BrandenburgS2005, Federrath2016, Rincon2019}. Motivated by the synchrotron and Faraday rotation observations, the dynamo theory can also be divided into a small-scale (fluctuation) and large-scale (mean-field) dynamo. The small-scale dynamo requires turbulence \citep[see][for further details]{SetaEA2020, SetaF2021b}, whereas the large-scale dynamo, besides turbulence, also requires large-scale galaxy properties such as differential rotation and density stratification \citep[see][for a detailed description]{ShukurovS2008}. Moreover, the small-scale random field can also be generated by the tangling of the large-scale field \citep[Section 4.1 in][]{SetaF2020}. 

Irregular galaxies, such as the SMC and LMC, also show synchrotron polarisation (albeit weaker than that in spiral galaxies) and Faraday rotation \citep{Gaenslar2005, MaoEA2008, MaoEA2012, LivingstonEA2021b}. This implies the presence of a small- and (a weaker) large-scale magnetic field. Also, for the SMC and LMC, a `Pan-Magellanic' magnetic field, i.e., a large-scale field connecting the SMC and LMC is discussed \citep{KaczmarekEA2017, LivingstonEA2021b}. \rev{Also, see \citet{ChyzyEA2016} for a detailed investigation of magnetic fields in an irregular galaxy, IC~10.}

In most of these different types of galaxies, except for the strength of the small-scale magnetic field (which can be obtained from depolarisation observation, \rev{polarised} synchrotron fluctuations, or $\RM$ variations), it is very difficult to estimate the other properties (such as the length scales and structure) of the small-scale random magnetic field. In external galaxies, this is because of limited resolution, with telescope beams often corresponding to hundreds of parsecs, which is larger than the expected correlation length of small-scale magnetic fields \citep[for example, see][]{KierdorfEA2020}. While resolution effects are less problematic for observations in the Milky Way, the main difficulty is confusion due to the observer being positioned inside the Galaxy and the influence of the Local Bubble \citep{AlvesEA2018, PelgrimsEA2020, WestEA2021}. \rev{Synchrotron polarisation gradient techniques can also be used to study properties of small-scale magnetic fields \citep{Gaenslar2011, Burkhart2012, IacobelliEA2014, HerronEA2017, LazarianY2018}. However, especially for external galaxies, these techniques are difficult to use due to a lack of resolution, \revb{missing pixels} in the data \revb{due to an insufficient signal-to-noise ratio}, and noise (these are common problems when taking a derivative of sparsely sampled data in observations).}

Here, we start by using simple numerical experiments and magnetohydrodynamic (MHD) simulations to show that the second-order structure function of $\RM$ from point radio sources in the background of a magneto-ionic medium can be used to obtain the strength and important length scales of the small-scale fluctuating magnetic field. We also discuss the role of the thermal electron density in the analysis, which \rev{was previously either completely} ignored \rev{or considered in a very simplified manner (for example, assuming a constant thermal electron density)}. Finally, we apply our methods to observations of $\RM$ from background polarised sources located behind the SMC and LMC to probe the small-scale magnetic fields in those systems. 

We discuss our methods to compute and characterise the structure function of $\RM$ in \Sec{sec:met}. In \Sec{sec:sim}, we demonstrate how the properties of the $\RM$ structure function can be translated to the properties of small-scale magnetic fields and thermal electron densities. For this, we use simple numerical experiments with a fixed distribution of random magnetic fields (\Sec{sec:simgrf}) and thermal electron density (\Sec{sec:simlnrf}). Then, in \Sec{sec:obs}, we use the results from \Sec{sec:sim} to probe the small-scale magnetic fields in the SMC and LMC. Finally, we summarise our results and conclusions in \Sec{sec:con}. In \Tab{tab:notdef}, we provide the notations used throughout the text with their descriptions.

\begin{table*} 
\caption{Description of the notations used throughout the text. }
\label{tab:notdef}
\begin{tabular}{ll} 
\hline 
Notation & Description \\ 
\hline
$\ne$ & thermal electron density \\ 
$\langle \ne \rangle$ & mean $\ne$ \\
$\vec{b}$ & small-scale fluctuating magnetic field \\ 
$\brms$ & root mean square strength of $\vec{b}$ \\
$\RM$ & rotation measure \\
$\sigma_{\RM}$ & standard deviation of $\RM$ \\
$\pk$(variable) & power spectrum of the variable \\ 
$\beta$ & slope of the $\RM$ power spectrum, $\pk(\RM) \sim k^{-\beta}, \beta > 0$ \\ 
$\alpha$ & slope of the $\vec{b}$ power spectrum, $\pk(\vec{b}) \sim k^{-\alpha}, \alpha > 0$ \\ 
$\gamma$ & slope of the thermal electron density power spectrum, $\pk(\ne) \sim k^{-\gamma}, \gamma > 0$ \\
${{\rm SF}^{\rm({\rm np}\,pts)}_2({\text{variable}})}$ & second-order structure function of the variable computed using $\rm np$-point stencil \\
$\ell$(variable) & correlation length of the variable \\
$L$ & size of the numerical domain \\
$\rm PDF$ & probability distribution function \\
\hline
\end{tabular}
\end{table*}

\section{Methods: $\RM$ structure function computation} \label{sec:met}
For a given thermal electron density, $\ne$, and magnetic field, $\vec{b}$, of the magneto-ionic medium, the rotation measure is
\begin{align} \label{eq:rm}
\frac{\RM}{\rad~\m^{-2}} = 0.812 \int_{L_{\rm pl}/\pc} \frac{\ne}{\cm^{-3}} \,  \frac{b_{\parallel}}{\muG} \,  \left(\frac{\dd l}{\pc}\right),
\end{align}
where $b_{\parallel}$ is the component of the magnetic field parallel to the line of sight and $L_{\rm pl}$ is the path length \rev{(from source to observer)}. Observationally, $\RM$ can be determined from the slope of the synchrotron polarisation angle vs.~the square of wavelength of light or more accurately using the $\RM$ synthesis technique \citep[see][for further details]{BrentjensBruyn05}. 
\rev{Faraday rotation of radiation can be due to the medium in front of a background polarised source (usually referred to as point source/background polarisation) or that emitted by the medium itself (usually referred to as the diffuse polarisation). Most of this work deals with $\RM$ due to background polarised sources, but the ideas and results presented here can also be extended to diffuse polarised emission.}

\rev{Throughout the text, we consider $\RM$ sources as point sources, \revb{i.e., the projected size of these sources is much smaller than the typical ISM scales,} because the angular size of these sources is of the order of $1''$ -- $10''$ \citep{OortEA1987,WindhorstEA1993,RudnickO2014} and we probe scales much larger than these scales. Moreover, we assume that the intrinsic $\RM$ fluctuations of sources are negligible in comparison to the $\RM$ due to the intervening medium \citep[see][and \Sec{sec:obsstr} for further details]{Schnitzeler2010,ShahS2021}.}

\rev{Some of the properties of} a small-scale random magnetic field can be studied with power spectrum analysis, which can be further probed by the power spectrum of $\RM$ (knowing some properties of the thermal electron density distribution). However, it is usually very difficult to compute the power spectrum of $\RM$ from observations because the polarised sources are located at random positions and even for the diffuse polarisation there can be gaps in the observational data. In such a situation, it is easier to compute the structure function of $\RM$ which in principle contains the same information as the $\RM$ power spectrum \citep[e.g.,][]{StutzkiEA1998}.

The second-order structure function of $\RM$ (with a two-point stencil, where stencil refers to the arrangement of points around the point of interest for the numerical calculation) as a function of the scale, $r=|\vec{r}|$, $\sftwo$, can be computed as,
\begin{align} \label{eq:sftwo}
\sftwo = \langle | \RM(\vec{x} + \vec{r}) - \RM(\vec{x}) |^{2} \rangle,
\end{align}
where $\vec{x}$ is the (two-dimensional) positional vector, over which the average, $\langle \rangle$, is computed. This has been used to study magnetic fields in a variety of observations \citep{MinterSpangler1996, HaverkornEA2008, StilEA2011, AndersonEA2015, LivingstonEA2021, RaychevaEA2022} and simulations \citep{HollinsEA2017}. 

If the power spectrum of $\RM$, $\pk(\RM)$, follows a power law with slope $-\beta$ ($\beta$ being positive), i.e., $\pk(\RM) \sim k^{-\beta}$ ($k$ being the wavenumber), then ideally \citep{StutzkiEA1998},
\begin{align}  \label{eq:sftwocases}
\sftwo \propto  \begin{cases}
r^{\beta - 1}, \quad & \text{if} \, \beta < 3 \\
r^{2}, \quad & \text{if} \, \beta \ge 3.
\end{cases}
\end{align}
Furthermore, from the $\RM$ spectra, the correlation length of $\RM$, $\ell(\RM)$, can be computed as
\begin{align} \label{eq:lbrm}
\ell(\RM) = \frac{\int~k^{-1} ~ \pk(\RM)~ \dd k}{\int~ \pk(\RM)~ \dd k}.
\end{align}
If $\ell(\RM)$ is smaller than the size of the \rev{medium} probed, at larger scales, the second-order structure function (with two points) would approach a constant value of $2~\sigma_{\RM}^{2}$, where $\sigma_{\RM}$ is the standard deviation of the $\RM$ distribution, i.e., 
\begin{align} \label{eq:sftwosat}
\sftwo_{r \rightarrow \infty} = 2~\sigma_{\RM}^{2}. 
\end{align}
The scale around which $\sftwo$ flattens, is approximately equal to $\ell(\RM)$. \rev{ We note that $r \rightarrow \infty$ refers to scales much larger than the correlation scale, not actually infinitely large scales.}

To study the slope of $\pk(\RM)$, $\sftwo$ is only helpful when $\beta < 3$ (see \Eq{eq:sftwocases}). However, the slope of the $\RM$ power spectrum can well be steeper than $3$, and in that case, to accurately capture the fluctuations at smaller scales, multiple points (or stencils) are required to compute the second-order structure function \citep{LazarianPogosyan2008}. The second-order $\RM$ structure function computed using three ($\sfthree$), four ($\sffour$), and five ($\sffive$) points, and the corresponding expectation for $\pk(\RM) \sim k^{-\beta}$ and at high values of $r$ (assuming $\ell(\RM)$ is smaller than the size of the region) is given below:

\begin{itemize}

\item The second-order $\RM$ structure function with a three-point stencil:
\begin{align} \label{eq:sfthree}
\sfthree = \langle | \RM(\vec{x} + \vec{r}) - 2~\RM(\vec{x}) + \RM(\vec{x} - \vec{r}) |^{2} \rangle,
\end{align}
\begin{align} \label{eq:sfthreecases}
\sfthree \propto  \begin{cases}
r^{\beta - 1}, \quad & \text{if} \, \beta < 5 \\
r^{4}, \quad & \text{if} \, \beta \ge 5,
\end{cases}
\end{align}
and 
\begin{align} \label{eq:sfthreesat}
\sfthree_{r \rightarrow \infty} = 6~\sigma_{\RM}^{2}.
\end{align}

\item The second-order $\RM$ structure function with a four-point stencil:
\begin{align} \label{eq:sffour}
\sffour = \langle | \RM(\vec{x} + 3~\vec{r}) - 3~\RM(\vec{x} + \vec{r}) ~ + \\ \nonumber  \quad 3~\RM(\vec{x} - \vec{r}) -  \RM(\vec{x} - 3~\vec{r}) |^{2} \rangle,
\end{align}
\begin{align} \label{eq:sffourcases}
\sffour \propto  \begin{cases}
r^{\beta - 1}, \quad & \text{if} \, \beta < 7 \\
r^{6}, \quad & \text{if} \, \beta \ge 7,
\end{cases}
\end{align}
and 
\begin{align} \label{eq:sffoursat}
\sffour_{r \rightarrow \infty} = 20~\sigma_{\RM}^{2}.
\end{align}

\item The second-order $\RM$ structure function with a five-point stencil:
\begin{align} \label{eq:sffive}
\sffive = \langle | \RM(\vec{x} + 2~\vec{r}) - 4~\RM(\vec{x} + \vec{r}) + 6~\RM(\vec{x}) ~ - \\ \nonumber 4~\RM(\vec{x} - \vec{r}) +  \RM(\vec{x} - 2~\vec{r}) |^{2} \rangle,
\end{align}
\begin{align} \label{eq:sffivecases}
\sffive \propto  \begin{cases}
r^{\beta - 1}, \quad & \text{if} \, \beta < 9 \\
r^{8}, \quad & \text{if} \, \beta \ge 9,
\end{cases}
\end{align}
and 
\begin{align} \label{eq:sffivesat}
\sffive_{r \rightarrow \infty} = 70~\sigma_{\RM}^{2}.
\end{align}
\end{itemize}

Second-order structure functions must be computed with these higher-order stencils in case of very steep spectral slopes of the underlying fields. The computed $\RM$ structure function with a different number of points per stencil is also generally useful in order to study the convergence of the slope of the $\RM$ structure function. 

In the next section, we apply these concepts to $\RM$ computed from magnetic field and thermal electron density distributions with known spectral properties, i.e., with a range of controlled spectral slopes, in order to determine how the properties of thermal electron density and magnetic fields are translated to the $\RM$ structure function. \rev{Throughout the next (numerical) section, we express quantities of interest in terms of normalised units, i.e., magnetic fields in terms of their root mean square (rms) value, thermal electron densities in terms of their mean value, and lengths in terms of the path length. This keeps the analysis and results more generic and can be applied to a wide range of systems. For reference, the rms strength of small-scale random magnetic fields in the Milky Way and nearby galaxies is $\approx 10~\muG$ \citep{Haverkorn2015, Beck2016}, the mean thermal electron density is $\approx 0.02~\cm^{-3}$ \citep{BerkhuijsenM2008,GaenslerEA2008,YaoEA2017}, and the path length depends on the region probed (see \Sec{sec:obs} for specific cases of the SMC and LMC).}

\section{Rotation measure structure function from known magnetic fields and thermal electron densities} \label{sec:sim}

\subsection{Gaussian random magnetic fields and constant thermal electron density} \label{sec:simgrf}
\begin{figure*}
    \includegraphics[width=\columnwidth]{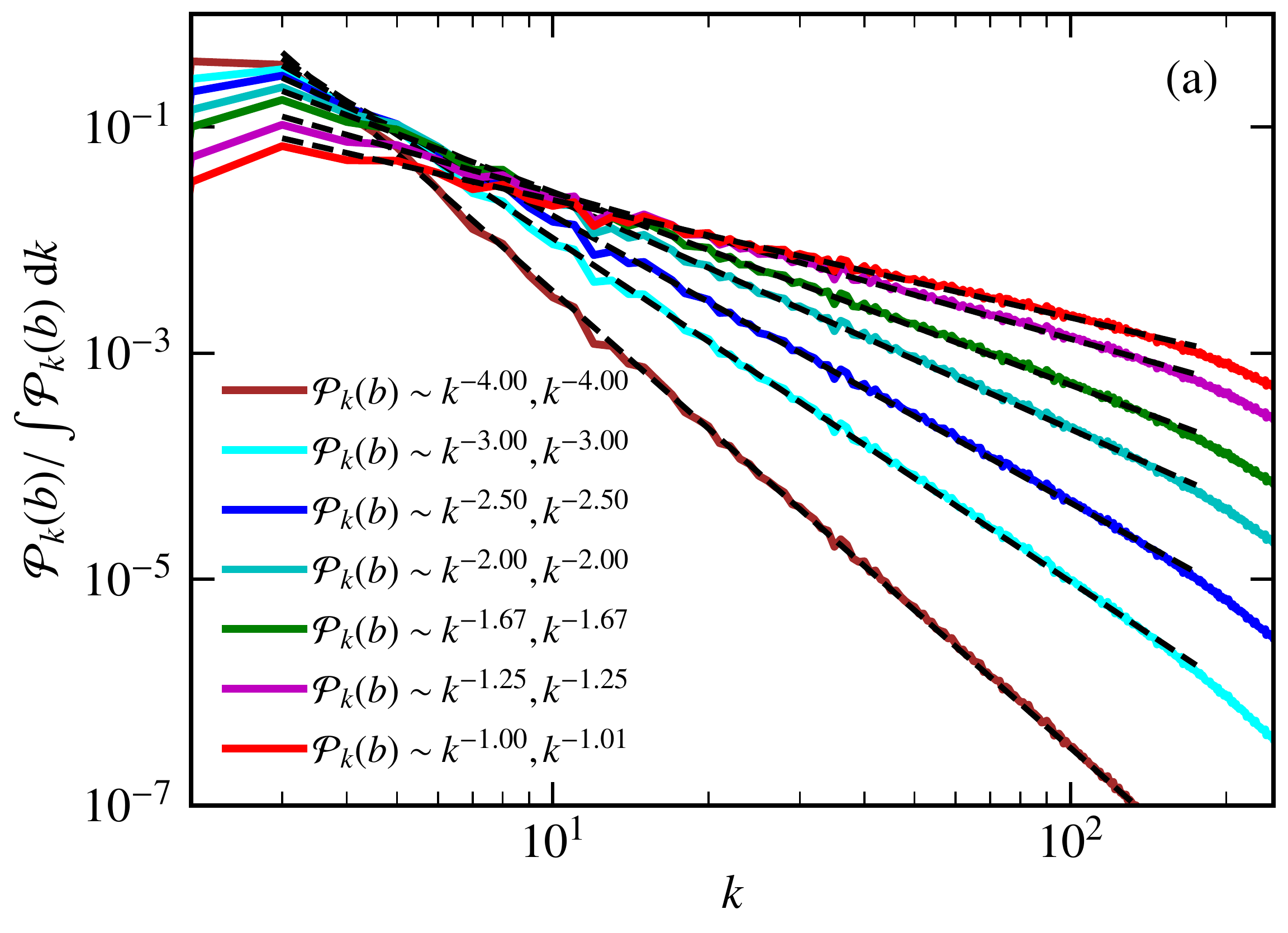} \hspace{0.5cm}
    \includegraphics[width=\columnwidth]{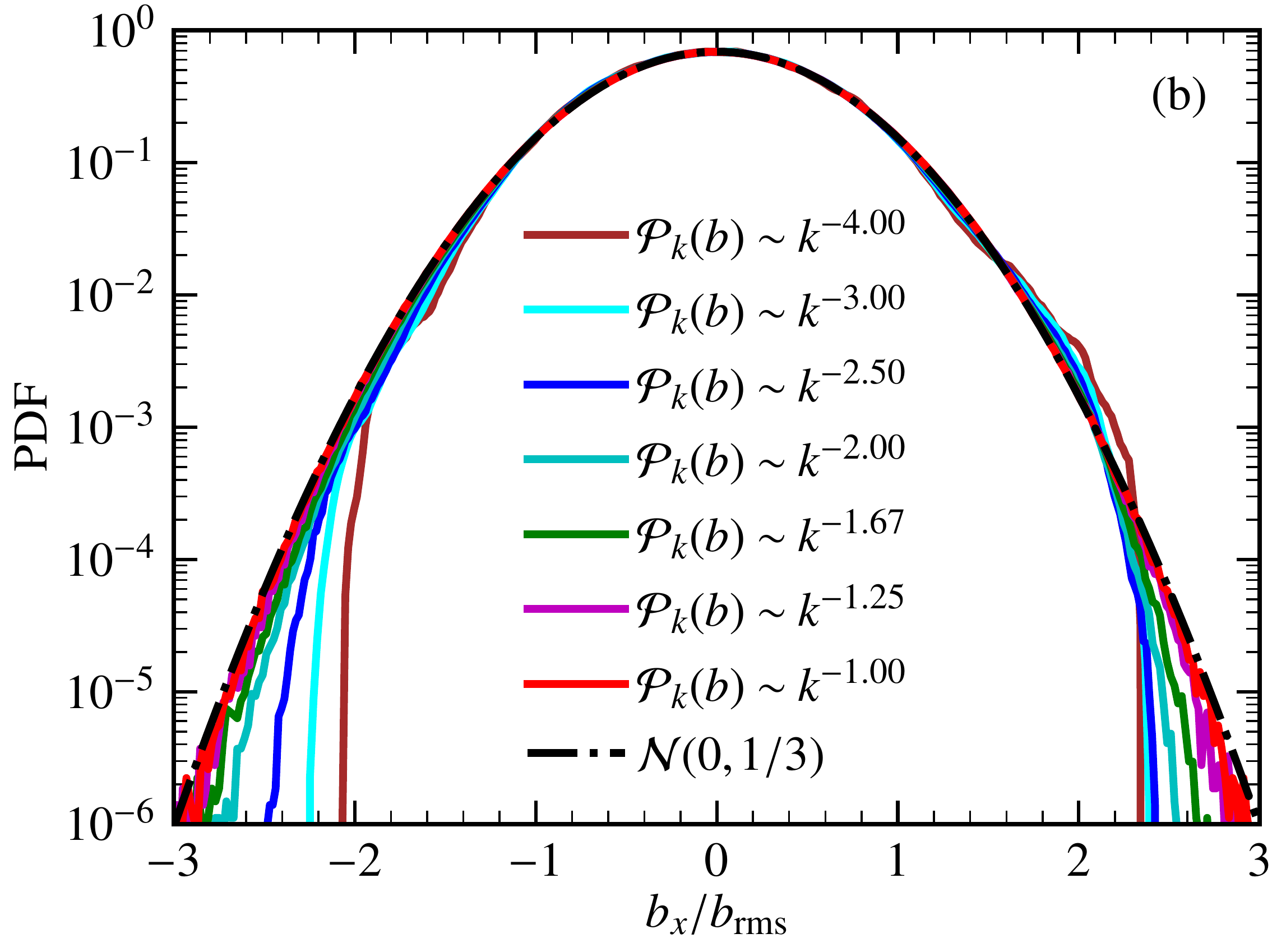}
    \caption{One-dimensional shell averaged power spectrum of three-dimensional magnetic fields ($\vec{b}$), $\pk(b)$, normalised by its total power (a) and probability distribution functions (PDFs) of a single component, $b_{x}$, normalised to the root mean square (rms) value, $\brms$ (b), for various slopes of the magnetic power spectrum ($\alpha$). The fitted slope matches the input slope well and the PDFs for all slopes approximately follow a Gaussian distribution with mean zero and one-third standard deviation, $\mathcal{N}(0, 1/3)$.}
    \label{fig:grfbspecpdf}
\end{figure*}

\begin{figure*}
    \includegraphics[width=2\columnwidth]{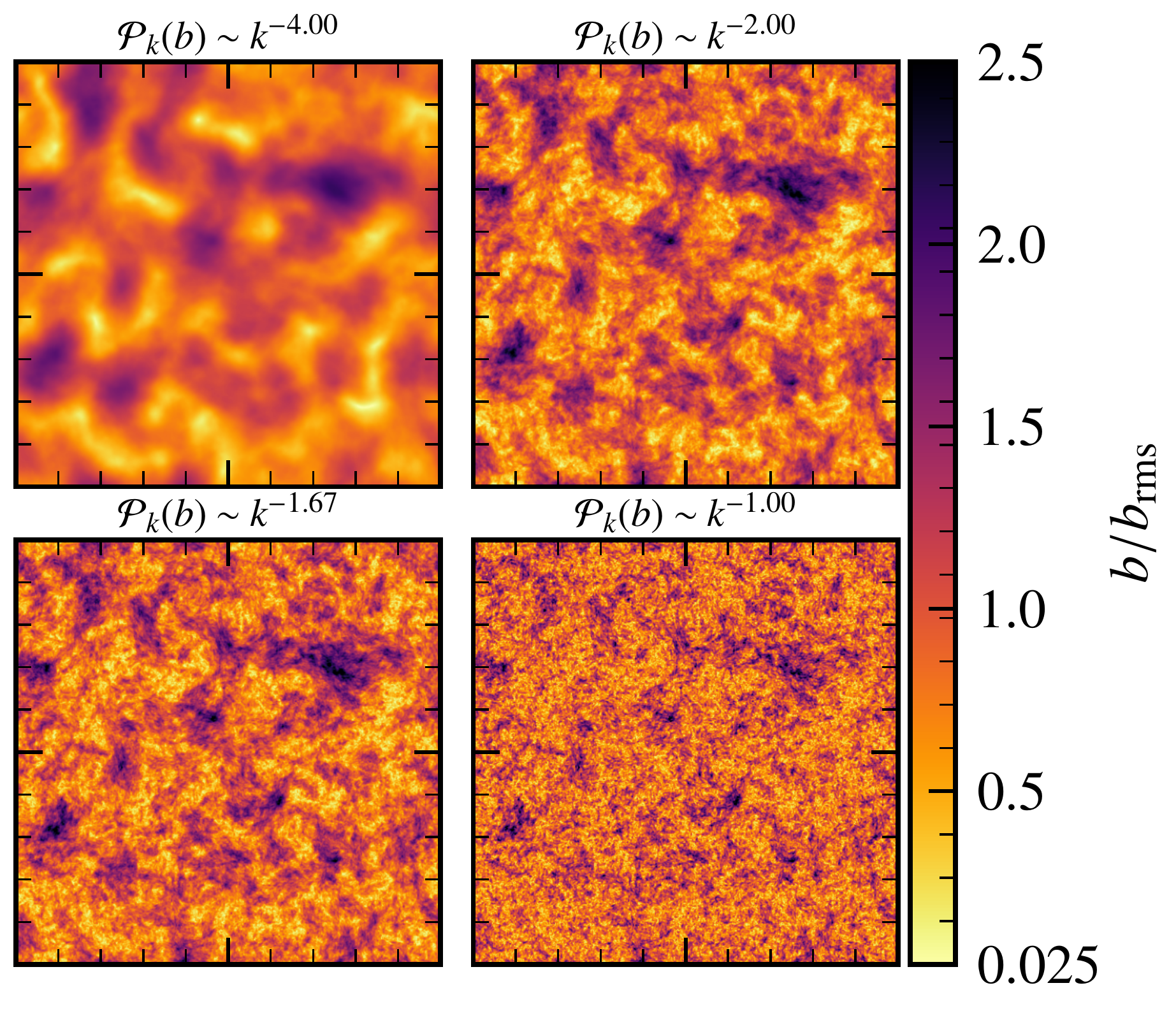}
    \caption{Two-dimensional slices of the Gaussian random magnetic field strength, normalised to the corresponding rms values, for various slopes of the magnetic power spectrum. As the power spectrum becomes shallower, the magnetic field contains more power in small-scale structures.}
    \label{fig:grfb2d}
\end{figure*}

First, we consider the simplest possible case of constant thermal electron density and Gaussian random magnetic fields ($\vec{b}$) with mean zero and a power-law spectrum with a controlled slope, $\alpha$, i.e., $\pk(b) \sim k^{-\alpha}$ (where $\alpha$ is positive) \footnote{While constructing the Gaussian random magnetic fields, we ensure that the divergence-free nature of magnetic fields is preserved. \rev{This is done via the Helmholtz decomposition, where we only keep the modes transverse to the wave vector and then normalise the field to have the desired $\brms$.}}. The field is constructed on a three-dimensional cubic uniform triply-periodic grid with side length $L$ and $512^{3}$ grid points. \Fig{fig:grfbspecpdf} shows the computed magnetic power spectrum and probability distribution function (PDF) from the constructed Gaussian random magnetic fields with $\alpha$ in the range $[1,4]$ and maximum power at $k=2$ (see \App{sec:kmin} for a discussion on the effects of varying this scale) for all values of $\alpha$. As expected, the slope of the fitted power spectrum in \Fig{fig:grfbspecpdf}a agrees well with the prescribed $\alpha$, and the corresponding PDF for all $\alpha$ in \Fig{fig:grfbspecpdf}b is very close to a Gaussian distribution. In \Fig{fig:grfb2d}, we show the magnetic strength, normalised to its corresponding root mean square (rms) computed from the entire three-dimensional grid, in \rev{a slice centered on the middle of the numerical domain} for $\alpha=4, 2, 5/3,$ and $1$. We see that the \rev{relative} size of magnetic structures decreases as the slope of the magnetic power spectrum becomes shallower.

\begin{figure*}
    \includegraphics[width=2\columnwidth]{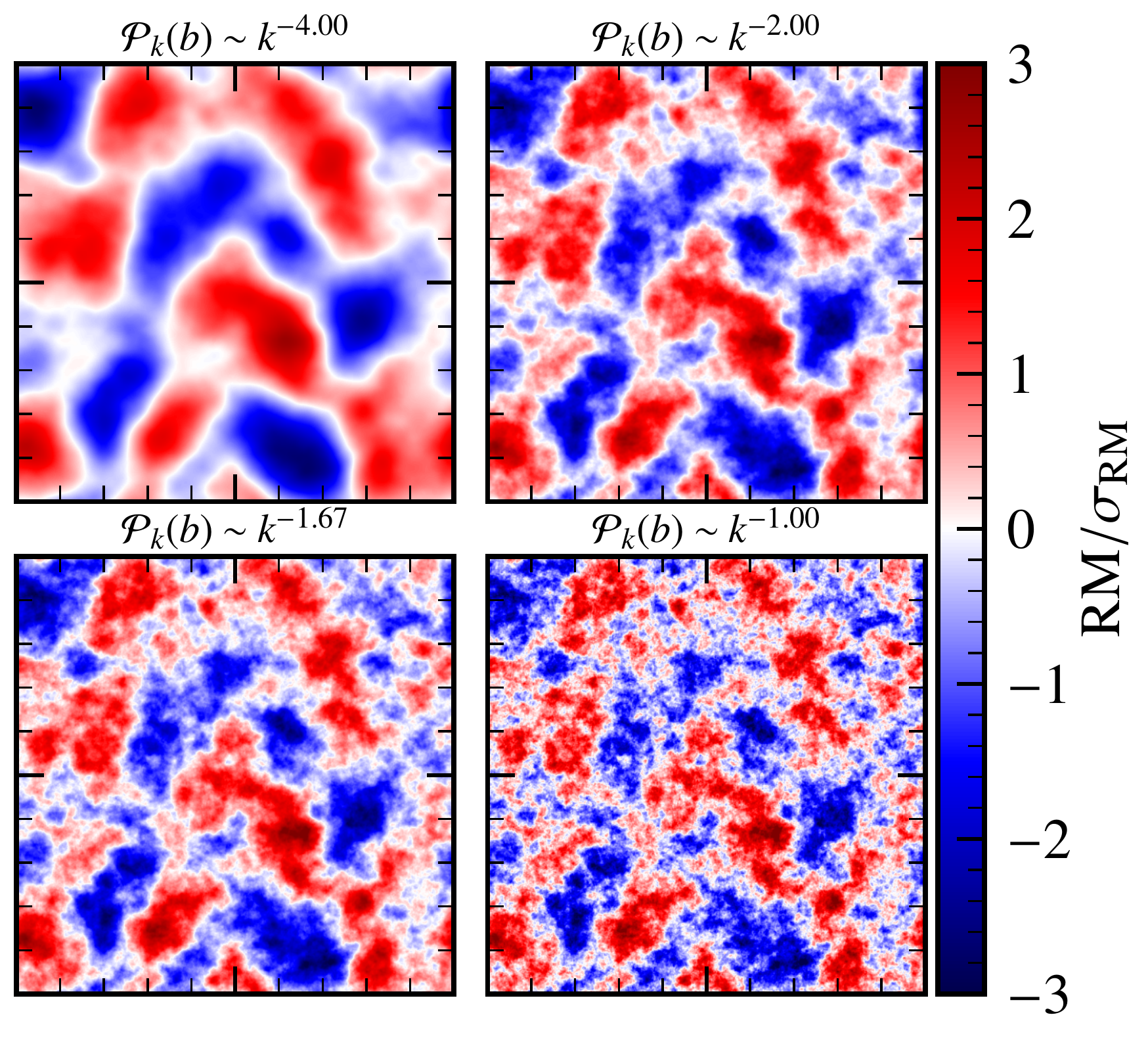}
    \caption{$\RM$ maps (computed by assuming $\ne=\mathrm{const}$) for four different slopes of the magnetic field power spectrum, $\pk(b) \sim k^{-\alpha}, \, \alpha=4, 2, 1.67, 1$. As the slope becomes shallower, the $\RM$ maps exhibit more small-scale structures, but the large-scale features remain roughly the same.}
    \label{fig:grfrm}
\end{figure*}

\begin{figure*}
    \includegraphics[width=\columnwidth]{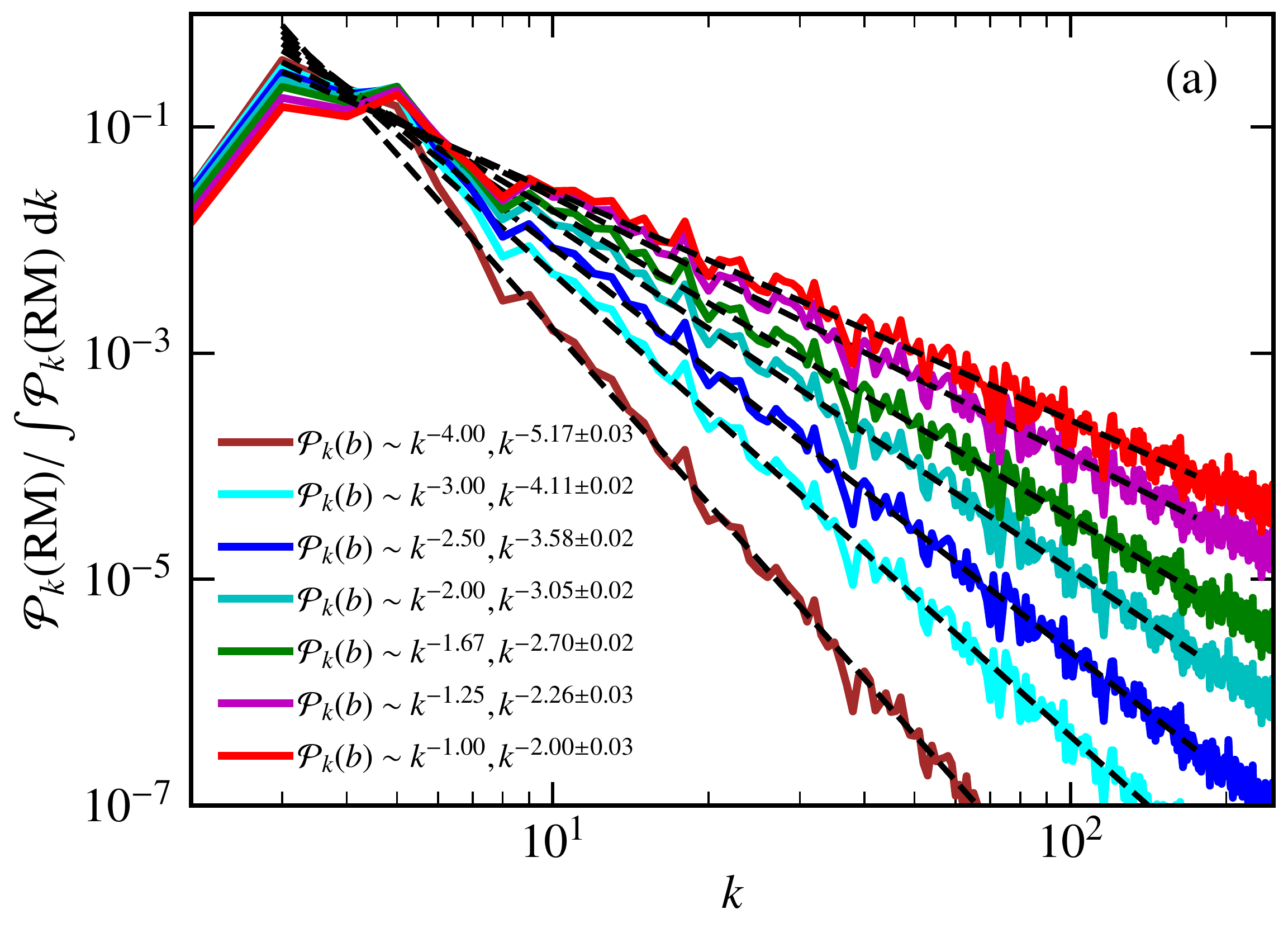} \hspace{0.5cm}
    \includegraphics[width=\columnwidth]{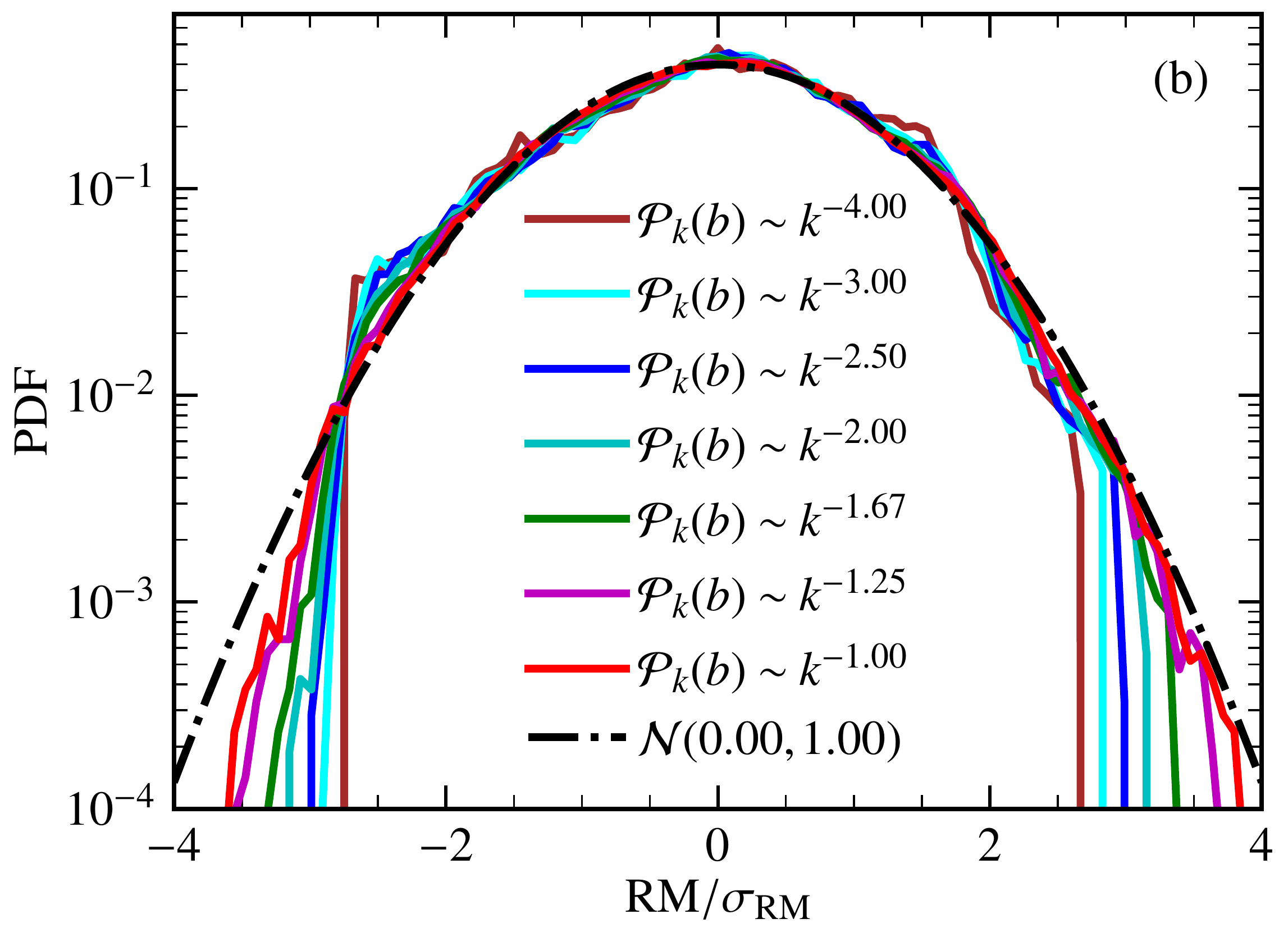}
    \caption{One-dimensional shell averaged power spectrum of the two-dimensional $\RM$ maps from see \Fig{fig:grfrm} (assuming $\ne$ is constant), $\pk(\RM)$, normalised by total power (a) and PDFs of $\RM$ normalised by its standard deviation, $\sigma_{\RM}$ (b), for different magnetic field power spectrum slopes, $\pk(b) \sim k^{-\alpha}, \alpha $ in $[1,4]$. We see that $\pk(\RM)$ approximately follows a power law with slope $-\alpha-1$. The $\RM$ PDF is roughly Gaussian with mean zero and standard deviation one, $\mathcal{N}(0, 1)$, for all $\alpha$.}
    \label{fig:grfrmspecpdf}
\end{figure*}

Assuming a uniform density of background polarised sources (the path length is the entire simulation domain, $L$), we compute $\RM$ due to these Gaussian random magnetic fields ($\ne$ is constant) using \Eq{eq:rm}. \Fig{fig:grfrm} shows $\RM$ maps, normalised by the standard deviation in $\RM$, $\sigma_{\RM}$, for four different magnetic field power spectrum slopes ($\alpha= 4, 2, 1.67,$ and $1$). Even though the $\RM$ structures at larger scales look similar between various cases, structures with smaller sizes are seen for shallower slopes. \Fig{fig:grfrmspecpdf} shows the power spectrum of $\RM$, $\pk(\RM)$, and the PDF of $\RM / \sigma_{\RM}$ for $\alpha=[1, 4]$. We see that the power spectrum of $\RM$ is also a power law and for $\pk(b) \sim k^{-\alpha}$, $\pk(\RM) \sim k^{-\alpha-1}$. This holds true for all $\alpha$ tested here. Thus, knowing the slope of the $\RM$ power spectrum, one can compute the slope of the magnetic field power spectrum (provided the electron density is constant). The PDF of $\RM$ roughly follows a Gaussian distribution with mean zero and standard deviation one. Since $\pk(\RM) \sim k^{-\alpha - 1}$, the slope of the second-order structure function (when it captures the fluctuations accurately, see \Eq{eq:sftwocases},  \Eq{eq:sfthreecases}, \Eq{eq:sffourcases}, and \Eq{eq:sffivecases}),  would be $-(-\alpha-1)-1 = \alpha$. Thus, for Gaussian random magnetic fields, $\pk(b) \sim k^{-\alpha}$ implies ${\rm SF}_{2} (\RM) \sim r^{\alpha}$.
 
\begin{figure*}
    \includegraphics[width=\columnwidth]{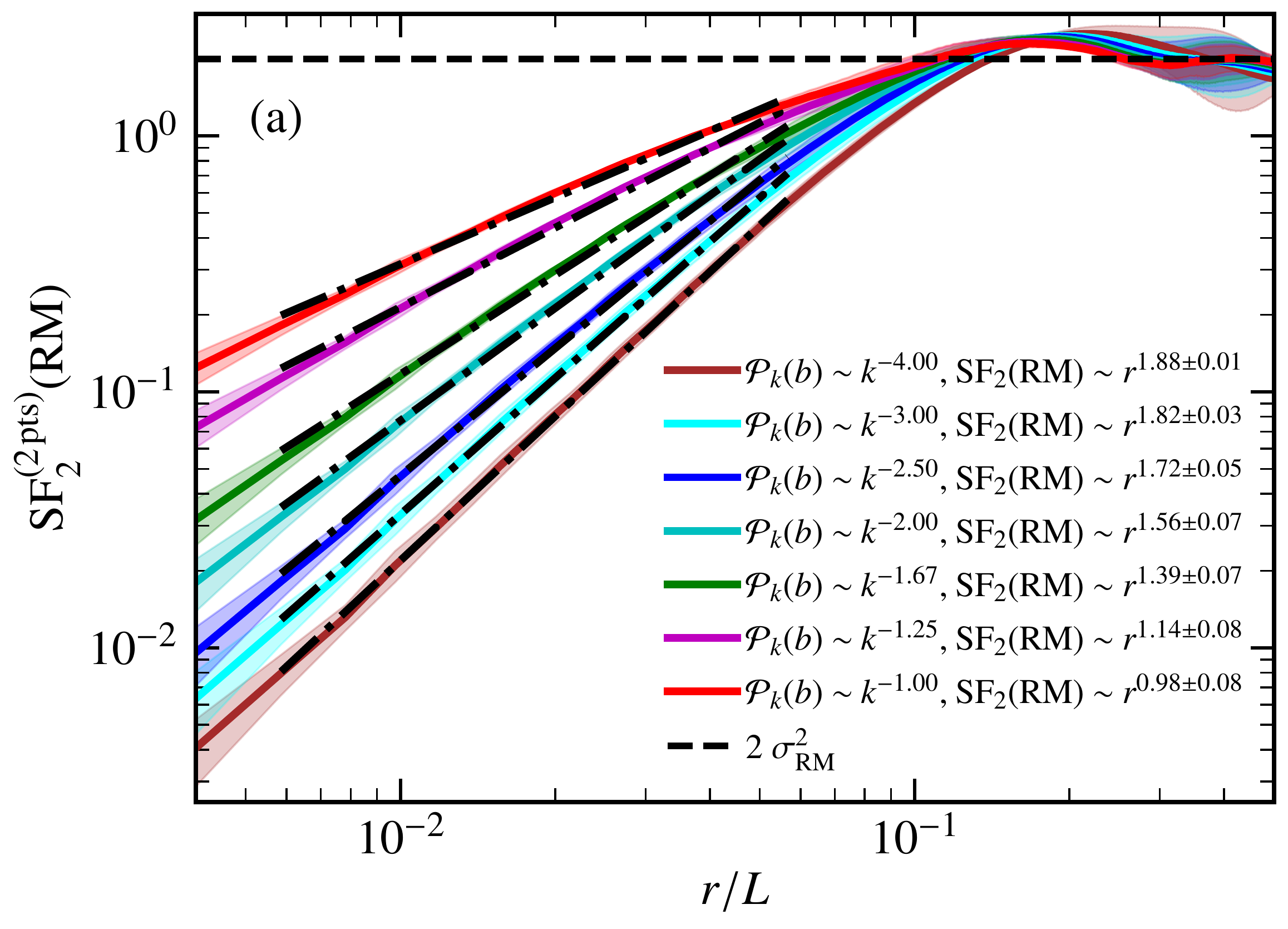} \hspace{0.5cm}
    \includegraphics[width=\columnwidth]{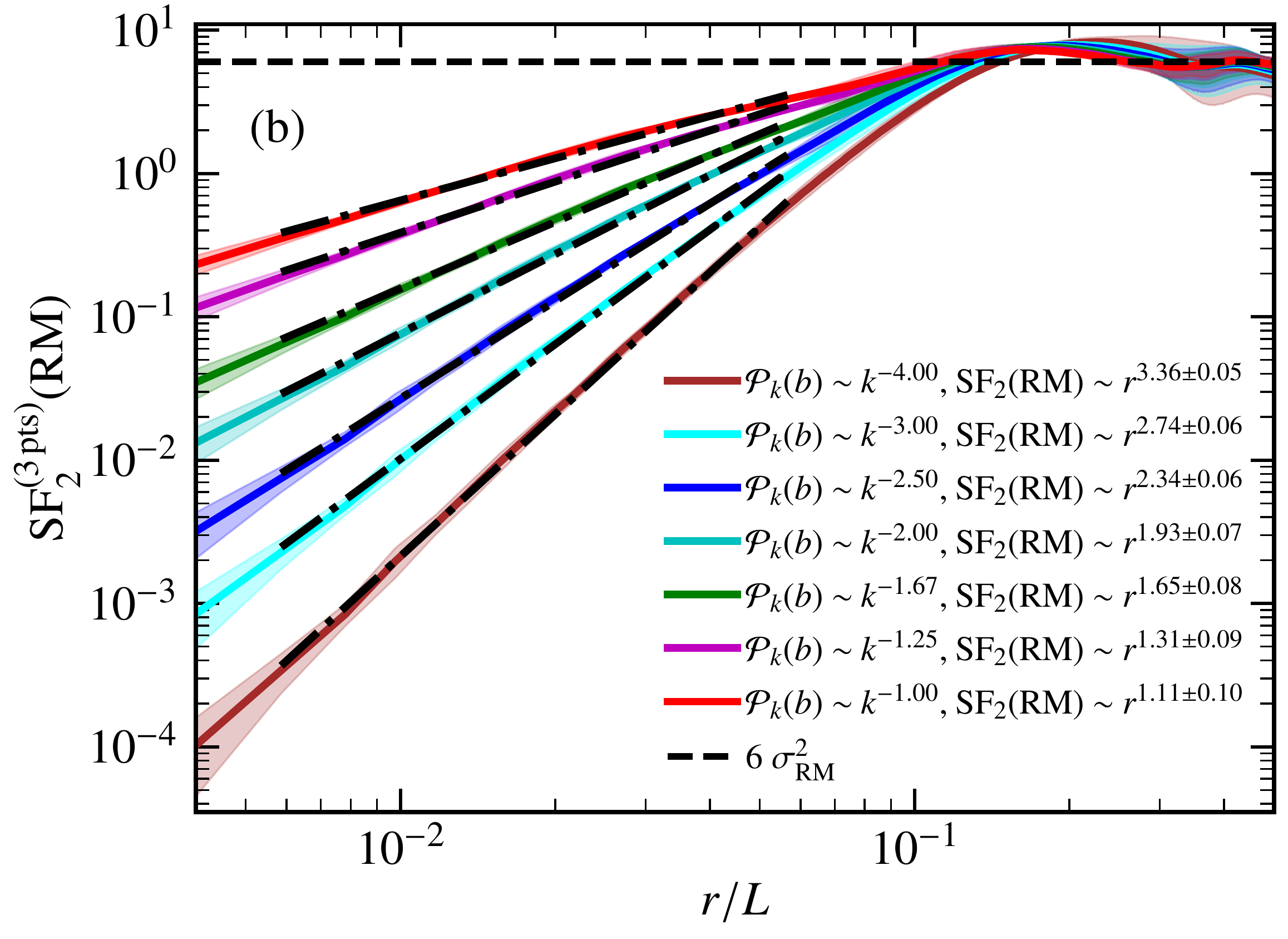} 
    \includegraphics[width=\columnwidth]{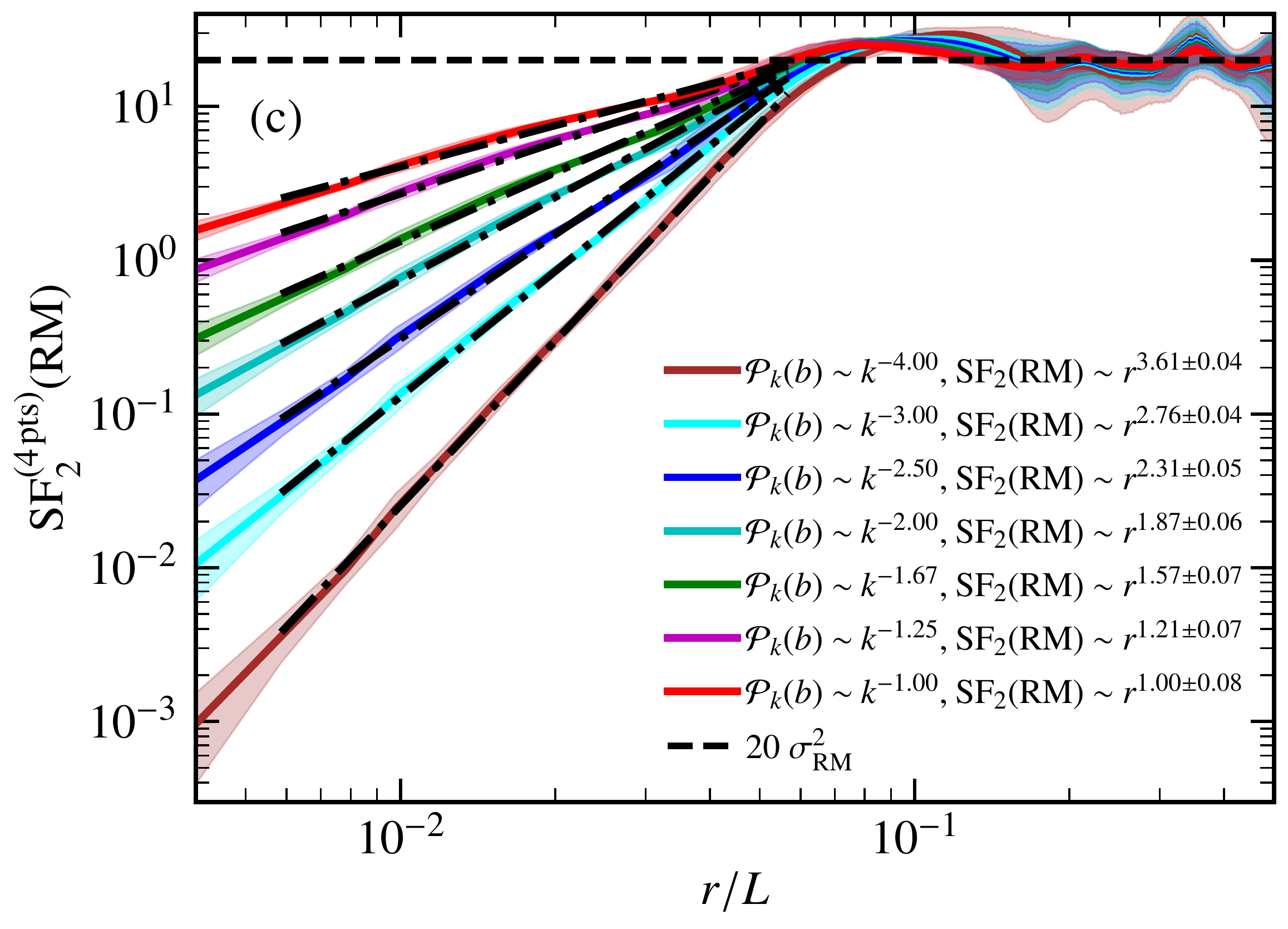} \hspace{0.5cm}
    \includegraphics[width=\columnwidth]{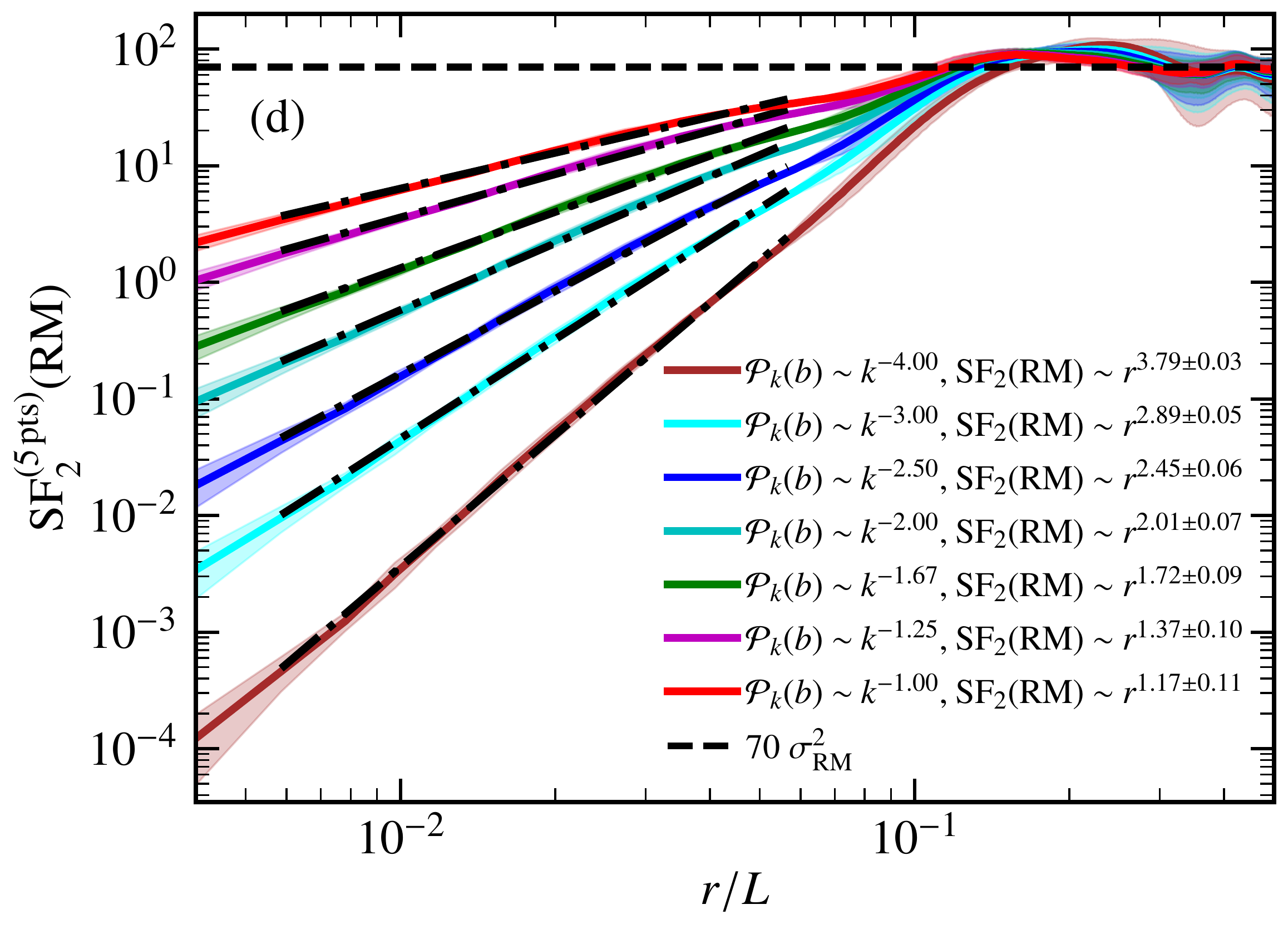} 
    \caption{Second-order $\RM$ structure function computed using stencils with a different number of points: two (a), three (b), four (c), and five (d), for Gaussian random magnetic fields with different slopes of the magnetic field power spectrum ($\pk(b)$) and a constant thermal electron density. \rev{The shaded region shows the corresponding one-sigma variation in each case.} At higher values of $r$, the structure functions for all slopes tend to their expected values ($2~\sigma_{\RM}^{2}$ for the two-point stencil, $6~\sigma_{\RM}^{2}$ for the three-point stencil, $20~\sigma_{\RM}^{2}$ for the four-point stencil, and $70~\sigma_{\RM}^{2}$ for the five-point stencil). As the number of points per stencil increases, the agreement with our expectation, ${\rm SF}_{2} (\RM) \sim r^{\alpha}$ for $\pk(b) \sim k^{-\alpha}$ over the entire range of $\alpha$s, improves. Even for shallower slopes (e.g., for $\pk(b) \sim k^{-1}$), the structure function remains roughly independent of the number of points per stencil, which is a sign of convergence.}
\label{fig:grfrmsf}
\end{figure*}

\Fig{fig:grfrmsf} shows the second-order $\RM$ structure function computed using a two-point (\Eq{eq:sftwo}, \Fig{fig:grfrmsf}a), three-point (\Eq{eq:sfthree}, \Fig{fig:grfrmsf}b), four-point  (\Eq{eq:sffour}, \Fig{fig:grfrmsf}c), or five-point (\Eq{eq:sffive}, \Fig{fig:grfrmsf}d) stencil, for different magnetic field power spectrum slopes in range $\alpha=[1, 4]$. \rev{The level of fluctuation in the second-order $\RM$ structure function at each scale is computed by taking a standard deviation of values in that bin.} For all four cases, the structure function first increases and then (statistically) saturates to a value (that varies with the number of points per stencil) at higher distances ($r/L \gtrsim 0.2$). This shows that the correlation length of $\RM$ is always less than $L$. The saturated values agree well with \Eq{eq:sftwosat} (structure function with two-point stencil), \Eq{eq:sfthreesat} (structure function with three-point stencil), \Eq{eq:sffoursat} (structure function with four-point stencil), and \Eq{eq:sffivesat} (structure function with five-point stencil). For $\sftwo$, the slope of the $\RM$ structure function roughly agrees with our expectation (${\rm SF}_{2} (\RM) \sim r^{\alpha}$ for $\pk(b) \sim k^{-\alpha}$) only for shallower slopes (\rev{$\alpha = 1, 1.25,$ and $1.67$}, as also expected from \Eq{eq:sftwocases}). Using a higher number of points per stencil, the agreement with our expectations improves, with $\sffive$ providing the best agreement, where the slope of $\sffive$ is approximately equal to $\alpha$ (negative of the slope of the magnetic power spectrum). Thus, to correctly probe the magnetic power spectrum, especially with steeper slopes, one must compute the second-order $\RM$ structure function with a higher-order stencil, where the number of points per stencil must be greater than 2, depending on $\alpha$. We also find that comparing structure functions computed with different stencils is useful in checking for convergence. After exploring $\RM$ structure functions with Gaussian random magnetic fields and constant thermal electron density, we now look at the effects of including a lognormal thermal electron density distribution in the next subsection.

\subsection{Gaussian random magnetic fields and (uncorrelated) lognormal thermal electron density fields} \label{sec:simlnrf}
\begin{figure*}
    \includegraphics[width=\columnwidth]{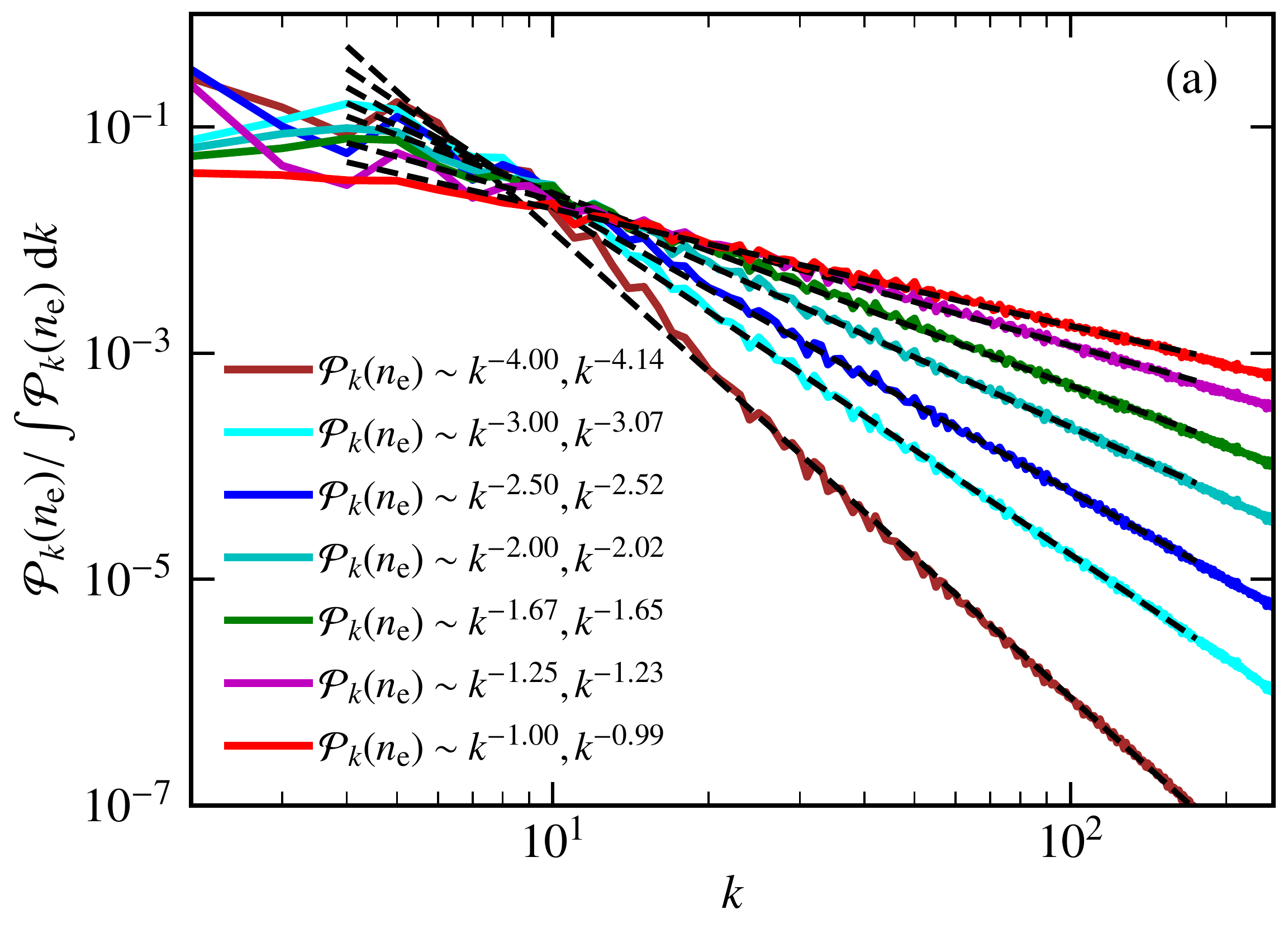} \hspace{0.5cm}
    \includegraphics[width=\columnwidth]{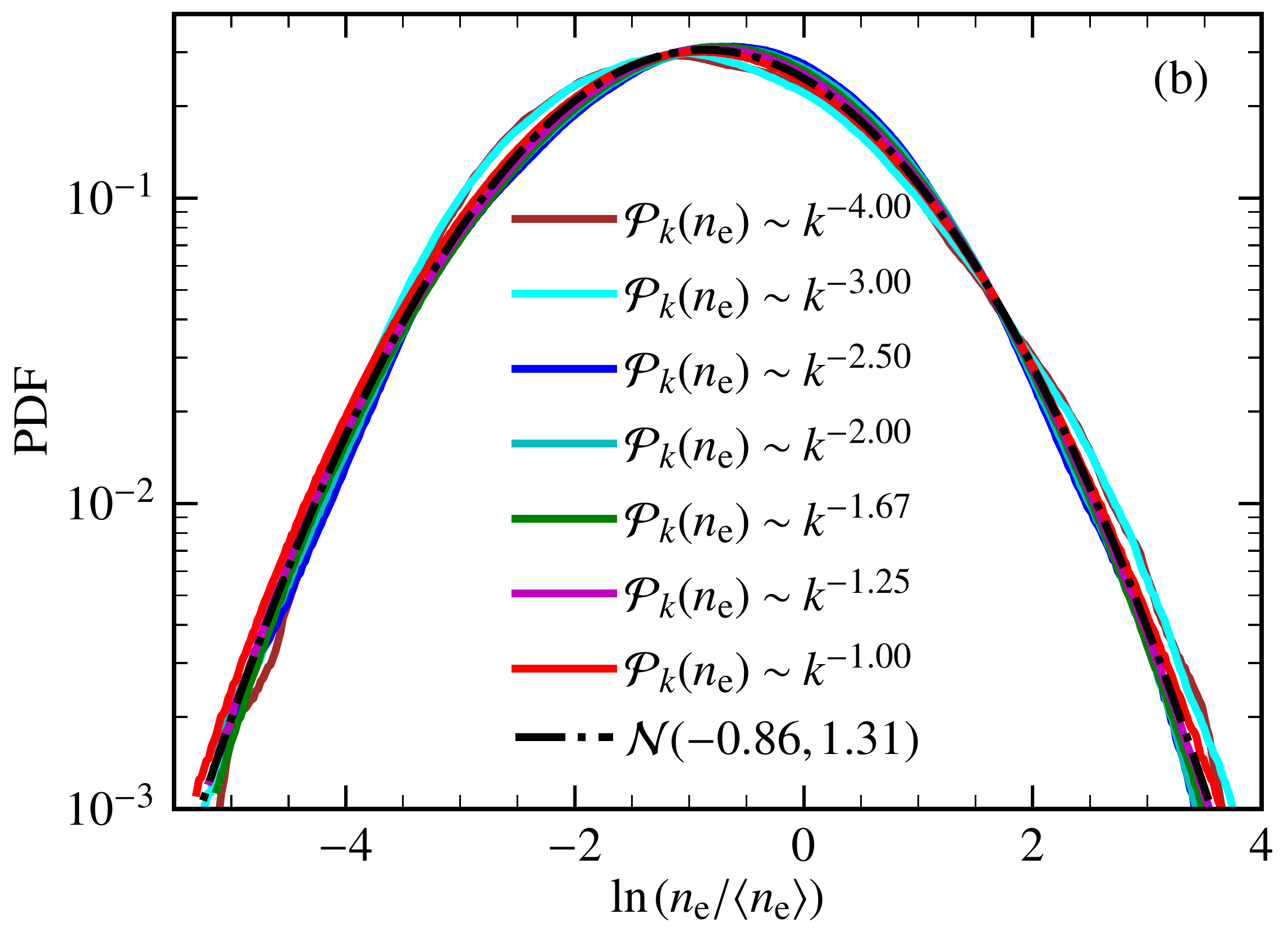}
    \caption{One-dimensional shell-averaged power spectrum of the three-dimensional lognormal thermal electron density, $\pk(\ne) \sim k^{-\gamma}$ (with $\gamma=[1, 4]$) (a), and the corresponding PDF of $\ln(\ne/\langle \ne \rangle)$ (b). The slope of the $\ne$ power spectra agrees well with the input slope and the PDF of $\ne / \langle \ne \rangle$ follows a lognormal distribution for all slopes.}
    \label{fig:lnrfnespecpdf}
\end{figure*}

\begin{figure*}
    \includegraphics[width=2\columnwidth]{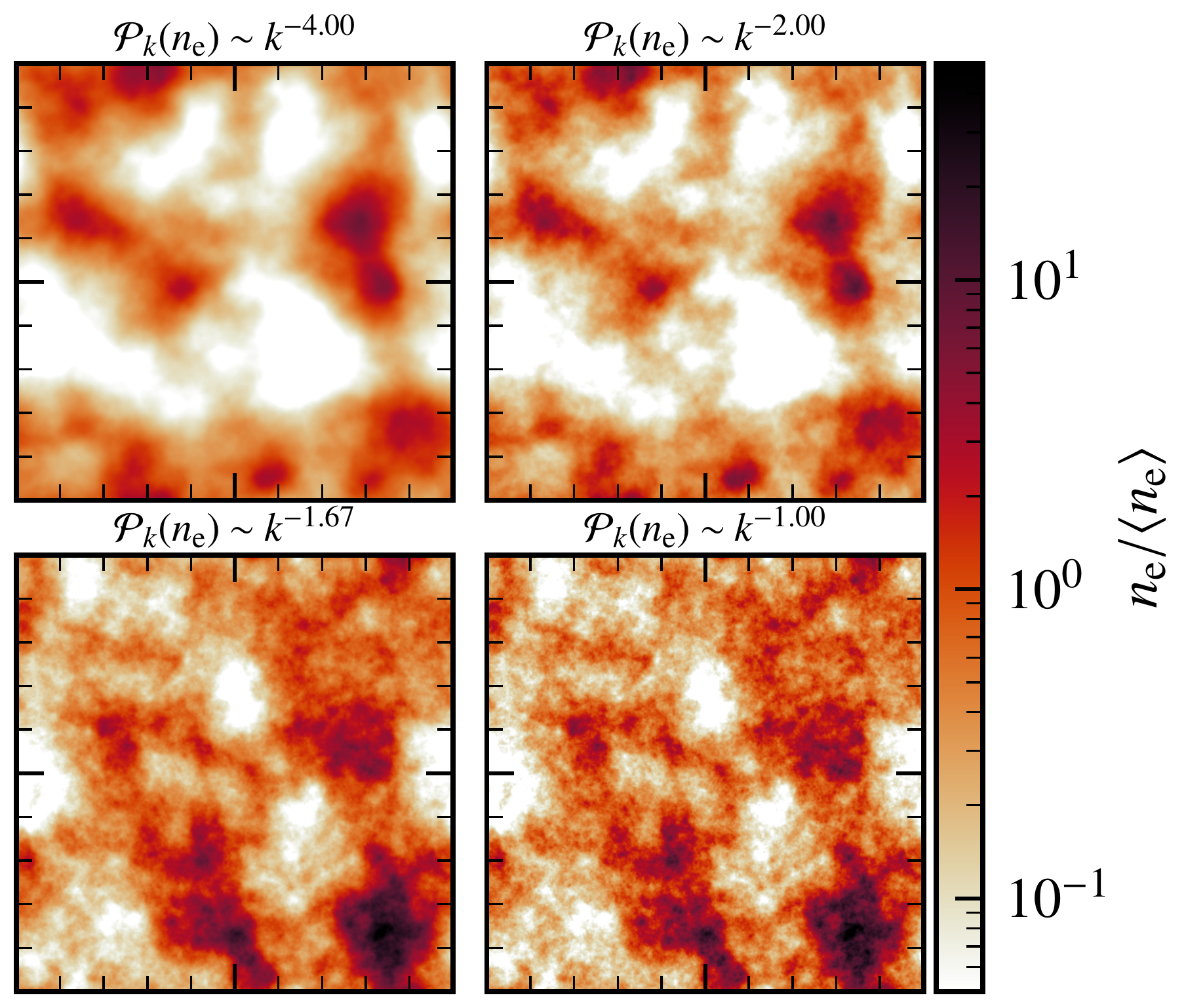}
    \caption{Two-dimensional slices of three-dimensional thermal electron density ($\ne$), normalised by the mean value, for four different slopes (indicated at the top of each panel) of the thermal electron density power spectrum. Over larger scales, the structures are similar, with smaller-scale structures emerging as the slope of $\ne$ becomes shallower.}
    \label{fig:lnrfne2d}
\end{figure*}

With Gaussian random magnetic fields, we now also consider the contribution of the thermal electron density ($\ne$) to $\RM$ and its effect on the computed $\RM$ structure function. Motivated by the studies of lognormal gas density PDFs in the ISM \citep{Vazquez1994, FederrathEA2008}, we construct lognormal PDFs of $\ne$ with a power-law spectrum and a controlled slope, $\gamma$, i.e., $\pk(\ne) \sim k^{-\gamma}$ ($\gamma>0$)\footnote{For constructing the thermal electron density with a lognormal PDF, we use the module pyFC (\href{https://bitbucket.org/pandante/pyfc}{https://bitbucket.org/pandante/pyfc}).}. Like the magnetic field, $\ne$ is also constructed on a three-dimensional cubic uniform triply-periodic grid of side length $L$ with $512^{3}$ points. The constructed thermal electron densities are uncorrelated with and independent of the Gaussian random magnetic fields. In \Fig{fig:lnrfnespecpdf}, we show the power spectrum of $\ne$ (\Fig{fig:lnrfnespecpdf}a) and the PDF of $\ln(\ne/\langle \ne \rangle)$ (\Fig{fig:lnrfnespecpdf}b) for $\gamma=[1, 4]$. The slope of the fitted power spectrum agrees with the input slope, and the PDF of $\ne/\langle \ne \rangle$ follows a lognormal distribution. In \Fig{fig:lnrfne2d}, we show the normalised thermal electron density \rev{in a slice centered on the middle of the numerical domain}. As the power spectrum of $\ne$ becomes shallower, smaller-scale structures are seen, while the large-scale structures remain roughly the same between various cases.

\begin{figure*}
    \includegraphics[width=2\columnwidth]{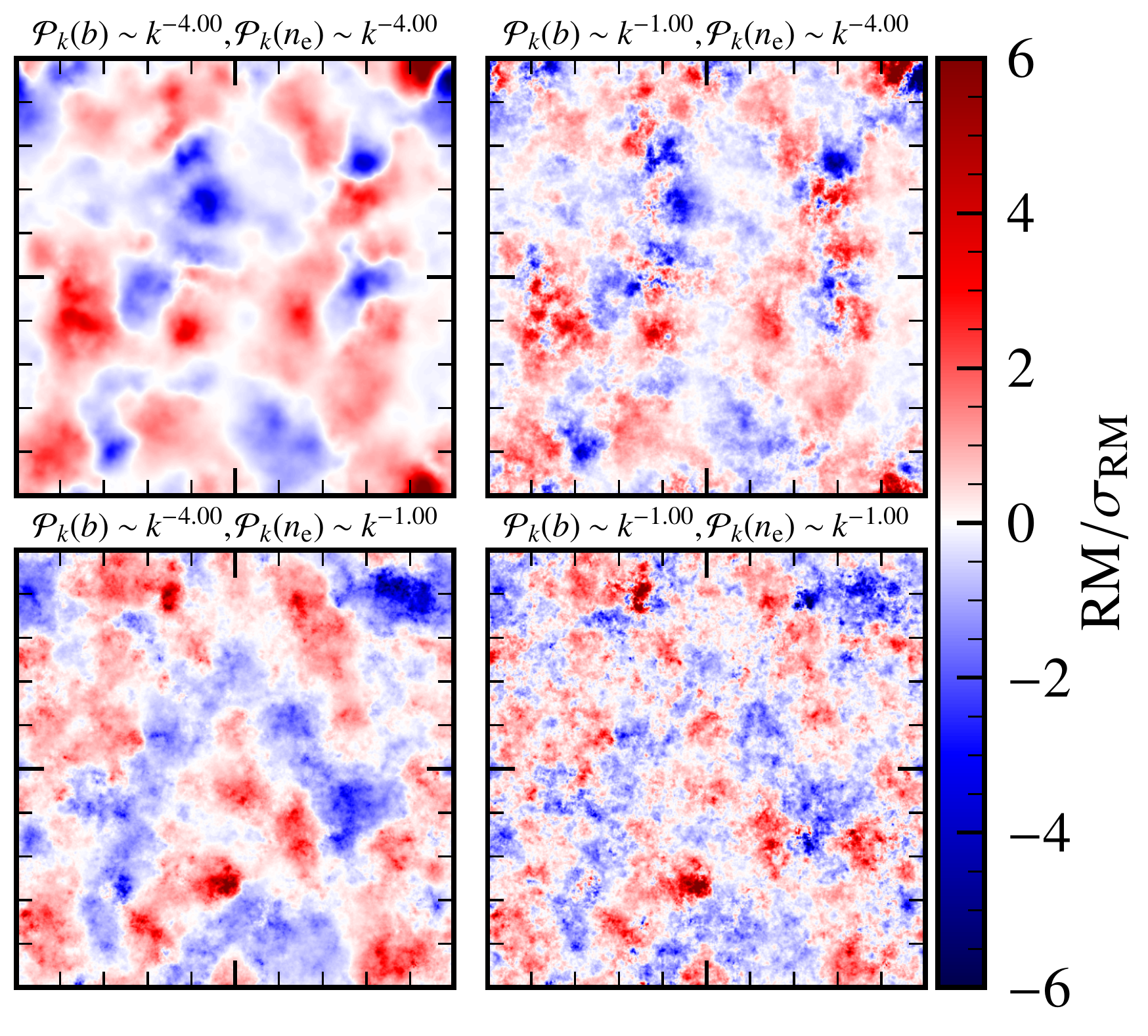}
    \caption{$\RM$ maps for different combinations of $\alpha=[1,4]$ in $\pk(b) \sim k^{-\alpha}$ and $\gamma=[1,4]$ in $\pk(\ne) \sim k^{-\gamma}$, as indicated at the top of each panel. Visually, it seems that the large-scale structures are more controlled by the steeper spectrum out of the two slopes and the smaller size structures by the shallower one (see \Fig{fig:grfb2d} and \Fig{fig:lnrfne2d} for corresponding $b$ and $\ne$, respectively).}
    \label{fig:lnrfrm}
\end{figure*}

\begin{figure*}
    \includegraphics[width=\columnwidth]{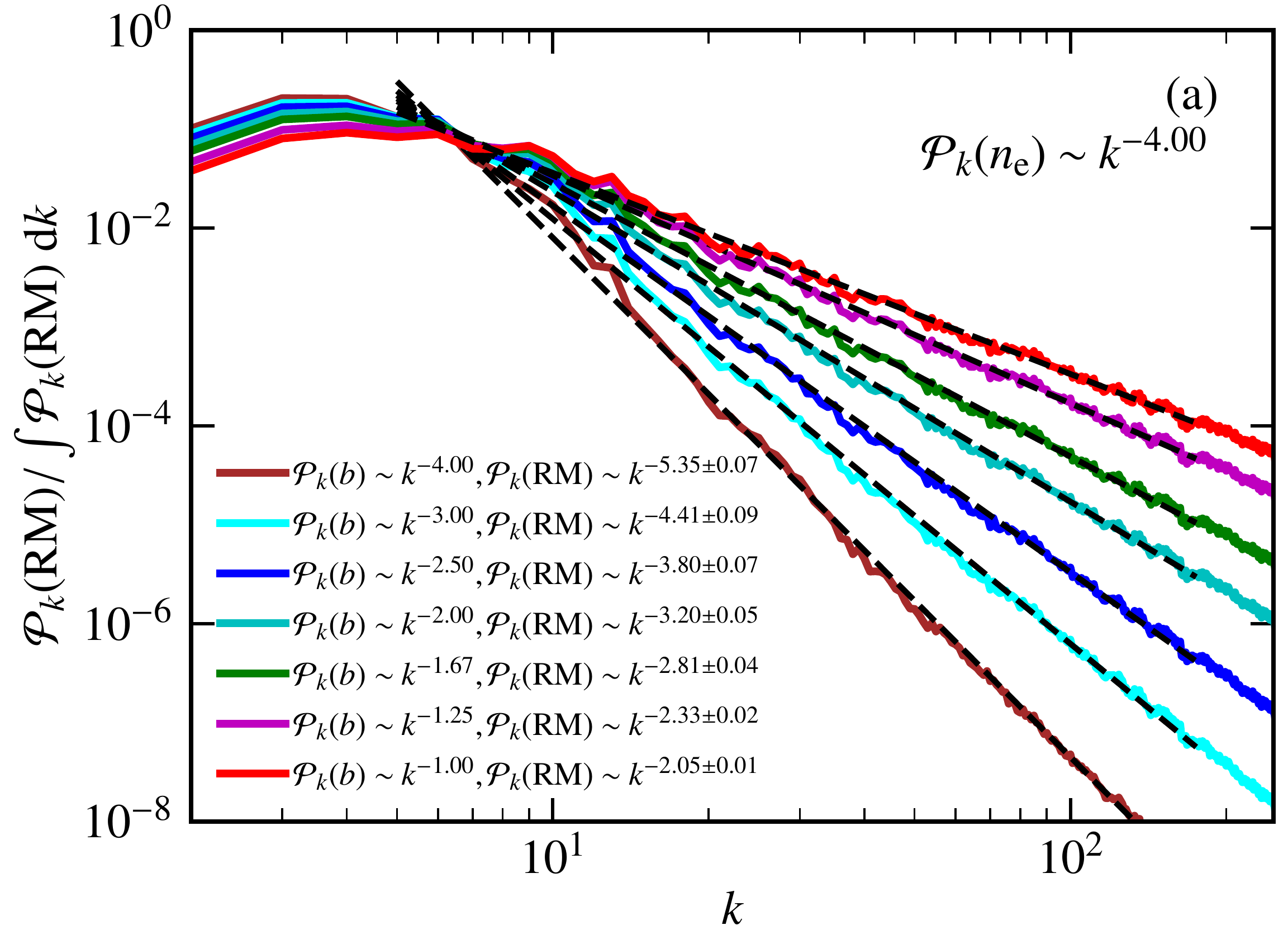} \hspace{0.5cm}
    \includegraphics[width=\columnwidth]{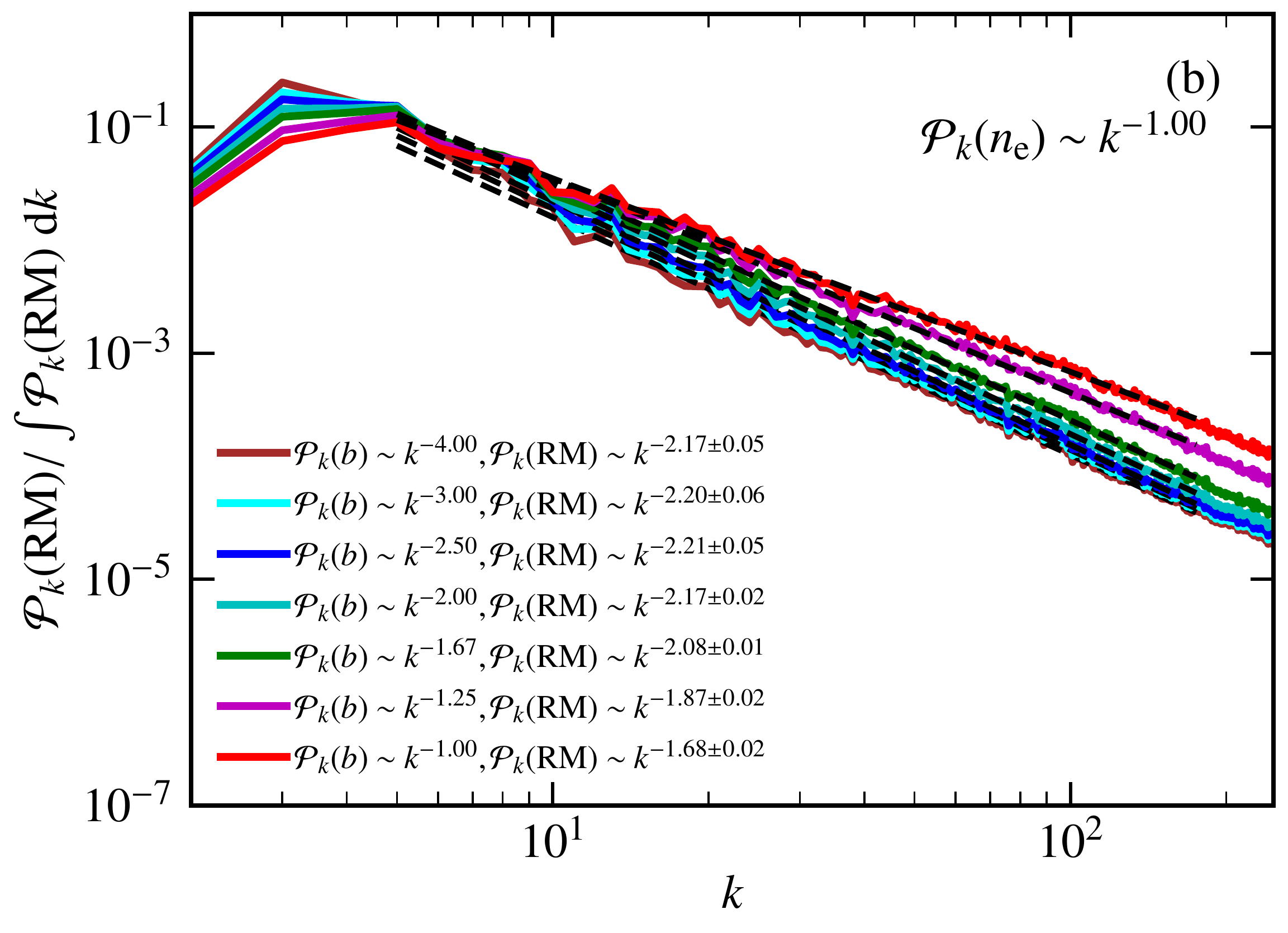}
    \caption{One-dimensional shell-averaged power spectrum of $\RM$ for $\gamma=4$ (a) and $\gamma=1$ (b) for various magnetic field power spectrum slopes, $\alpha=[1, 4]$. The slope of the $\RM$ power spectrum is more affected by the shallower of the two slopes ($-\alpha$ and $-\gamma$) and it roughly varies as $\alpha-1$ or $\gamma-1$ (whichever is shallower).}
    \label{fig:lnrfrmspec}
\end{figure*}

\begin{figure*}
    \includegraphics[width=\columnwidth]{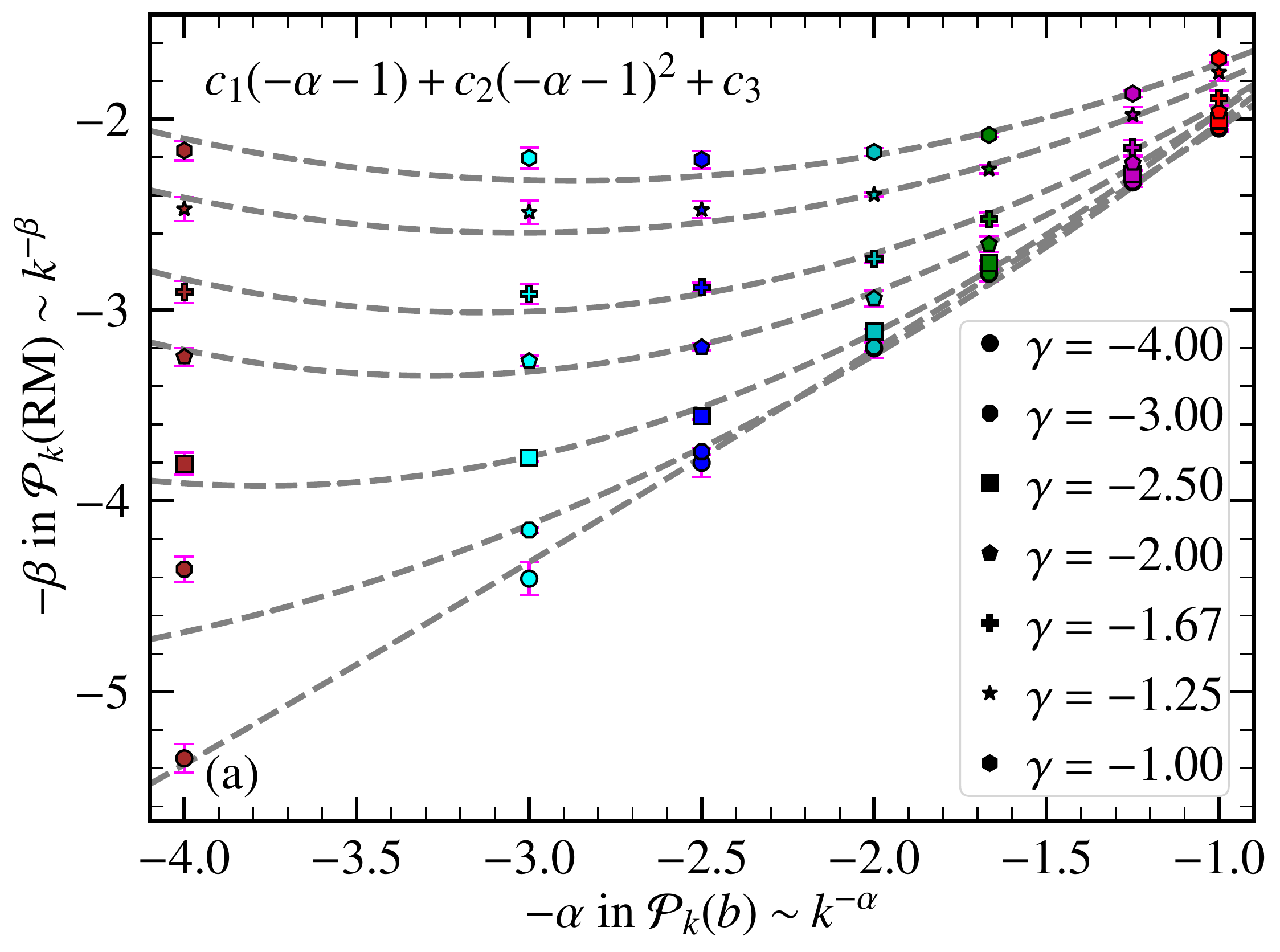} \hspace{0.5cm}
    \includegraphics[width=\columnwidth]{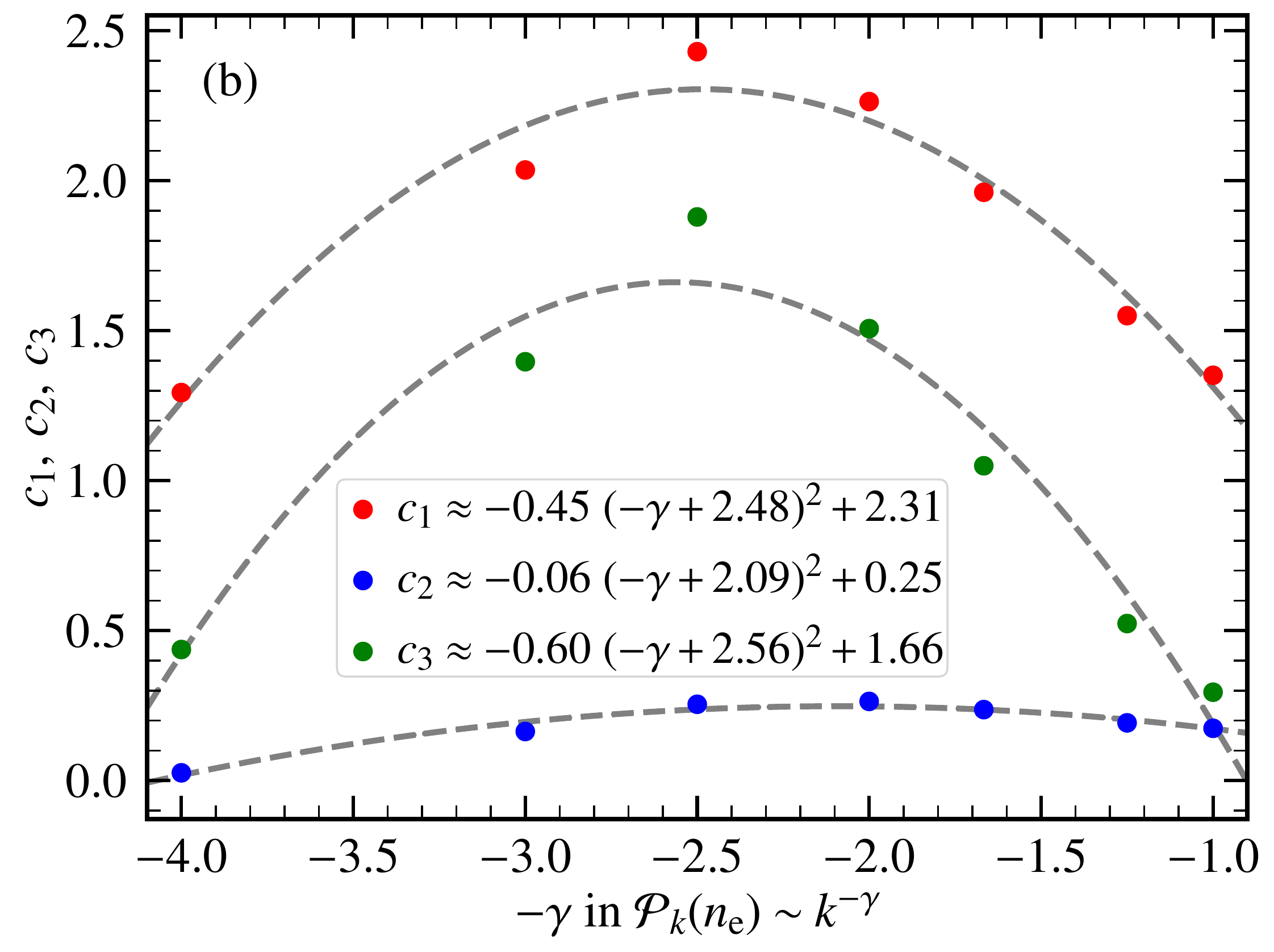}
    \caption{To estimate the slope of the $\RM$ power spectrum, $\beta$, from the slope of magnetic field power spectrum, $\alpha$, and thermal electron density power spectrum, $\gamma$, we first fit the obtained values of $\beta$ as a function of $\alpha$ (different colours) for different values of $\gamma$ (different marker shapes)  (a). We see that the function, $c_{1} (-\alpha - 1) + c_{2} (-\alpha - 1)^{2} + c_{3}$, where $c_{1}, c_{2},$ and $c_{3}$ are $\gamma$-dependent coefficients, fits the trends well for all $\alpha$. Then we fit $c_{1}, c_{2},$ and $c_{3}$ as a function of $\gamma$ (b), and obtain the dependence of those coefficients on $\gamma$. A quadratic function of $\gamma$ with different constants approximately fits $c_{1}, c_{2}$, and $c_{3}$ (see the legend in (b)). This shows that $\beta$ can be estimated for a given $\alpha$ and $\gamma$ using the derived empirical relation given by \Eq{eq:rmspecfit}.}
\label{fig:lnrfrmspecfit}
\end{figure*}

Again, we use \Eq{eq:rm} to compute $\RM$, but now we consider Gaussian random magnetic fields ($\pk(b) \sim k^{-\alpha}$ with $\alpha=[1, 4]$) combined with lognormal thermal electron densities ($\pk(\ne) \sim k^{-\gamma}$ with $\gamma=[1, 4]$). The computed $\RM$ depends now on both $\alpha$ and $\gamma$. \Fig{fig:lnrfrm} shows the $\RM$ maps for four combinations of the minimum and maximum of $\alpha=1,\,4$ and $\gamma=1,\,4$. The $\RM$ maps have smaller fine-scale structures when either the magnetic field or thermal electron density power spectrum slopes are shallower, and show the smallest-scale structures when both of them are low ($\alpha = \gamma = 1$). The structures, even at larger scales, are primarily controlled by the quantity with the shallower slope. We further investigate this using the power spectrum of $\RM$ for different values of $\alpha$ and $\gamma$.

For all combinations of $\alpha$ and $\gamma$, the $\RM$ power spectrum follows a power law with slope $\beta$, i.e., $\pk(\RM) \sim k^{-\beta}$. Here, we aim to find how $\beta$ can be estimated from a given $\alpha$ and $\gamma$. In \Fig{fig:lnrfrmspec}, we show the power spectrum of $\RM$ for a fixed $\gamma = 4$ in \Fig{fig:lnrfrmspec}a and $\gamma = 1$ in \Fig{fig:lnrfrmspec}b), with $\alpha$ in the range $[1, 4]$. In \Fig{fig:lnrfrmspec}a, for shallower magnetic field power spectra ($\alpha=1, 1.25, 1.67, 2$), we find $\beta \approx -\alpha -1$. On the other hand, for very shallow thermal electron density slope ($\gamma = 1$) in \Fig{fig:lnrfrmspec}b, for steeper magnetic field power spectrum slopes ($\alpha=4, 3, 2.5, 2$), we find $\beta \approx -\gamma-1$. Thus, if the slopes of the magnetic field power spectrum and the thermal electron density power spectrum are significantly different, the shallower of the two slopes have a larger influence on the slope of the $\RM$ power spectrum. However, when they are close, that need not be the case and both the slope of the magnetic field and thermal electron density power spectra decide the slope of the $\RM$ power spectrum.

We use our estimated values of $\beta$ (slope of the $\RM$ power spectrum) from our numerical experiments with different $\alpha$ (slope of the magnetic field power spectrum) and $\gamma$ (slope of the thermal electron density power spectrum) to construct an empirical relation between those three slopes. In \Fig{fig:lnrfrmspecfit}a, we show that the dependence of $\beta$ on $\alpha$ for all values of $\gamma$ (different marker styles) is approximately captured by a quadratic equation, $c_{1} (-\alpha - 1) + c_{2} (-\alpha - 1)^{2} + c_{3}$, where $c_{1}, c_{2}, $ and $c_{3}$ are coefficients that depend only on $\gamma$. In \Fig{fig:lnrfrmspecfit}b, we determine the functional form of the dependence of these coefficients on $\gamma$. Thus, for a Gaussian random magnetic field with power spectrum slope $\alpha$ and lognormal thermal electron density with power spectrum slope $\gamma$, the slope of the computed $\RM$ power spectrum is
\begin{align} \label{eq:rmspecfit}   
\beta (-\alpha, -\gamma) \approx \, & (-~0.45~(-\gamma + 2.48)^{2} + 2.31)~ (-\alpha - 1) ~ + \\  \nonumber
                         & (-~0.06~(-\gamma + 2.09)^{2} + 0.25)~ (-\alpha - 1)^{2} ~ - \\ \nonumber
                         & 0.60~ (-\gamma + 2.56)^{2} + 1.66. & 
\end{align}
\rev{The errors in the fit parameters of \Eq{eq:rmspecfit}, obtained from the least-squares fitting procedure, are less than $10\%$.} \revb{These numerical experiments are repeated with different seeds in the random number generator. We find that the variation in the results (estimated slopes of $\RM$, magnetic fields, and thermal electron densities, and by extension the coefficients in \Eq{eq:rmspecfit}) are within the reported fitting errors.}

\begin{figure*}
    \includegraphics[width=\columnwidth]{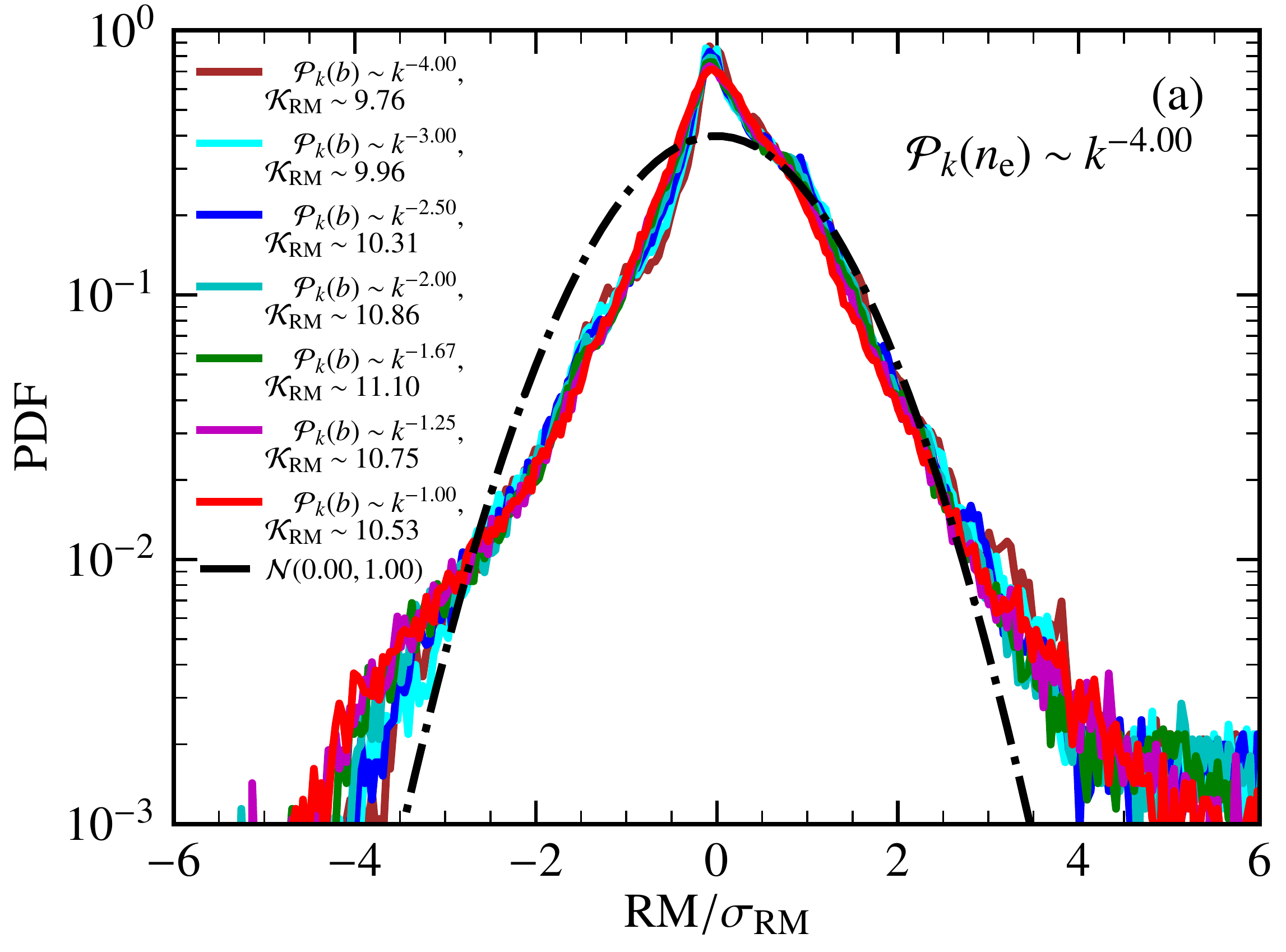} \hspace{0.5cm}
    \includegraphics[width=\columnwidth]{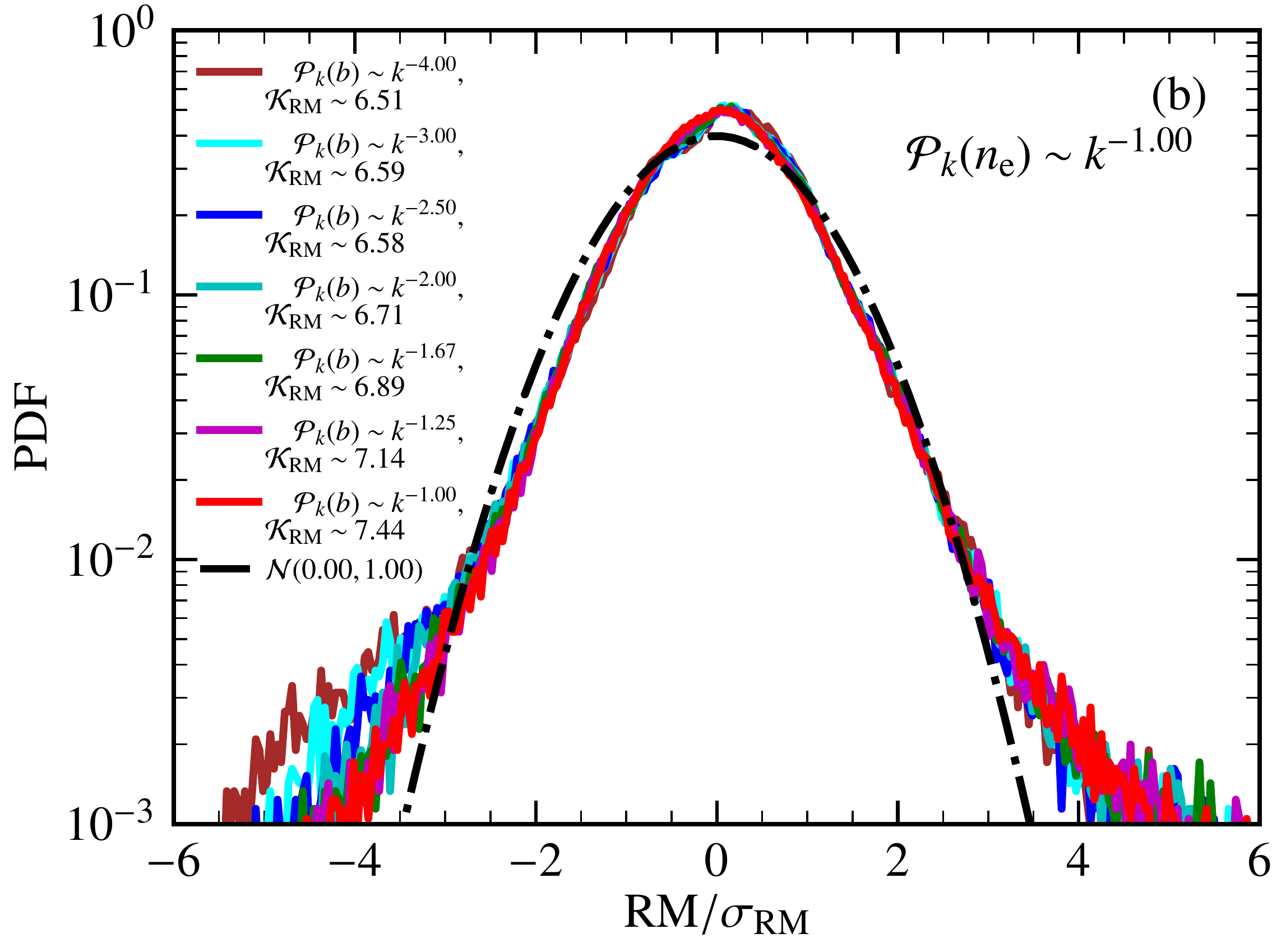}
    \caption{PDF of normalised $\RM$ for all values of $\alpha$ (magnetic field power spectrum slopes) in the range $[1, 4]$, for two extreme values of $\gamma$ (thermal electron density power spectrum slope), i.e., $\gamma=1$ and $4$. Unlike for purely Gaussian random magnetic fields in \Fig{fig:grfrmspecpdf}b, the PDF of $\RM$ is non-Gaussian and this is further confirmed by the computed kurtosis values, $\ku_{\RM}$ (given in the legend). They are always greater than three (kurtosis of a Gaussian distribution).}
    \label{fig:lnrfrmpdf}
\end{figure*}

\begin{figure*}
    \includegraphics[width=\columnwidth]{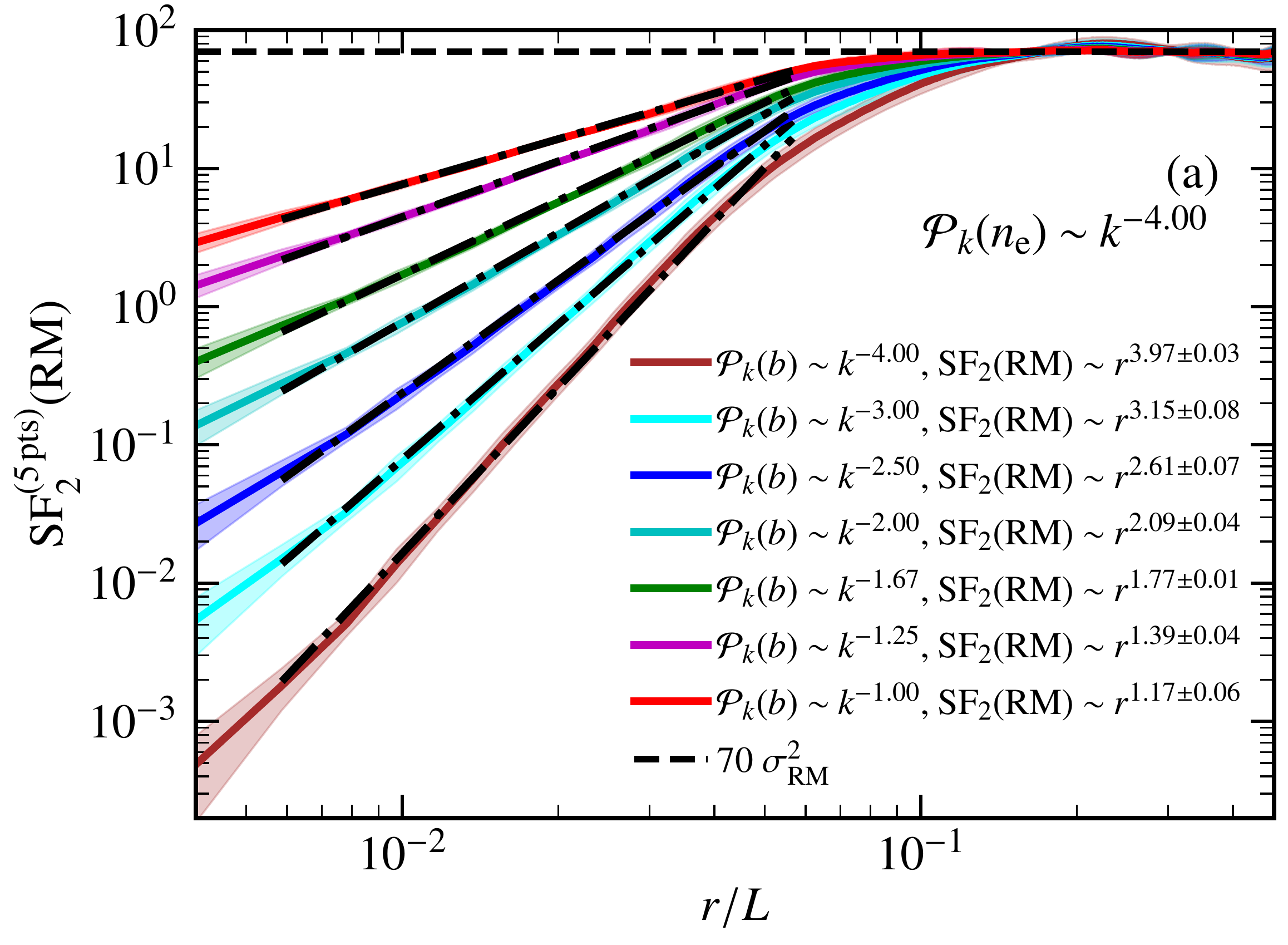} \hspace{0.5cm}
    \includegraphics[width=\columnwidth]{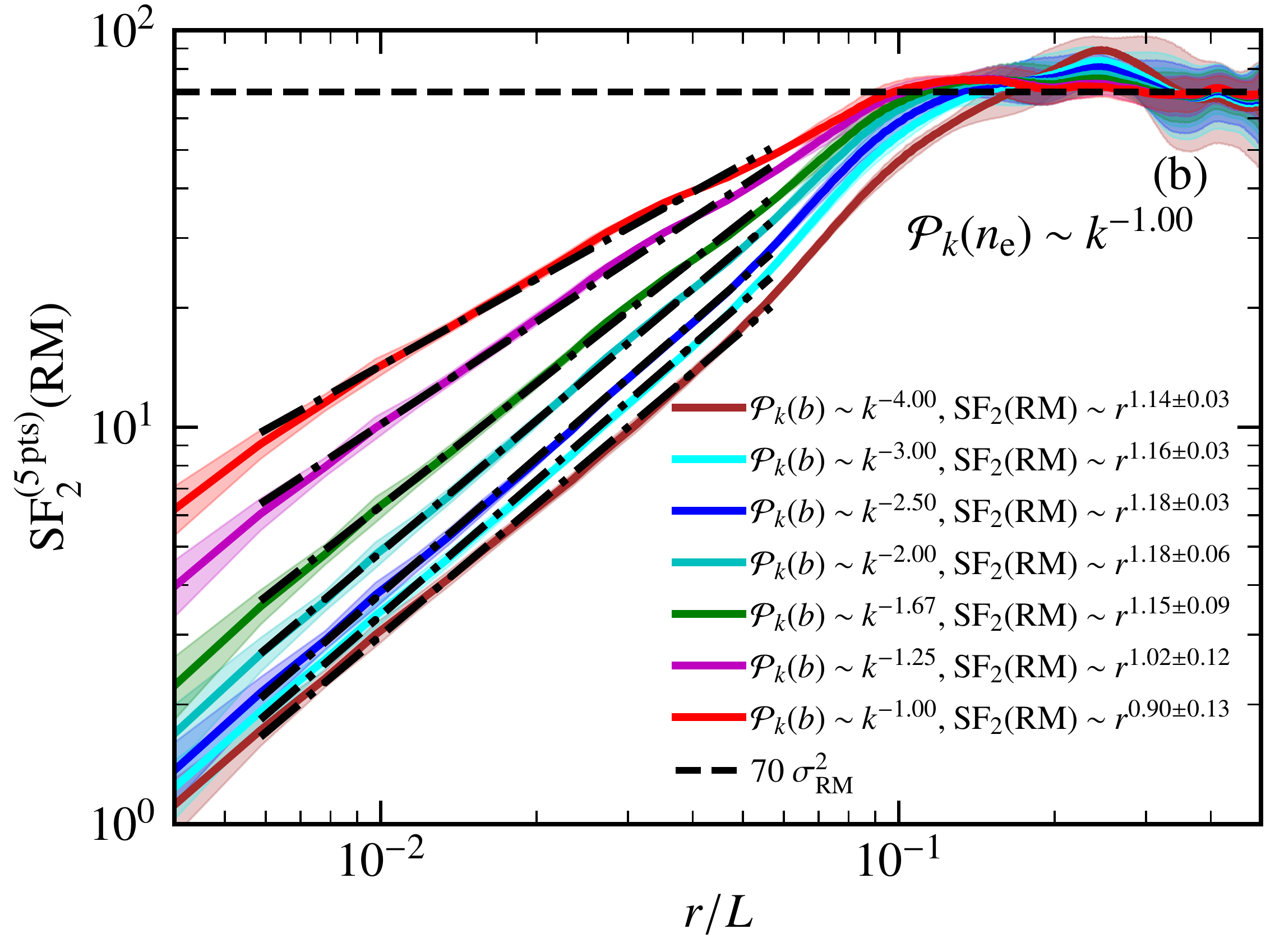}
    \caption{Second-order $\RM$ structure function computed using the five-point stencil, $\sffive$ (\Eq{eq:sffive}), for $\gamma=1$ (a) and $\gamma=4$ (b), and $\alpha=[1, 4]$. \rev{The shaded region shows the corresponding one-sigma variation in each case.} The slope of the second-order $\RM$ structure function is largely controlled by the shallower of the magnetic field and thermal electron density spectrum. Thus, for $\pk(\RM) \sim k^{-\beta}$, $\beta \approx \alpha$ (if the magnetic field spectrum is shallower) or $\gamma$ (if the electron density spectrum is shallower). More accurately, \Eq{eq:rmspecfit} should be used to calculate $\beta$ from $\alpha$ and $\gamma$, and then the slope of the second-order $\RM$ structure function is approximately $\beta - 1$.}
    \label{fig:lnrfrmsf}
\end{figure*}

\Fig{fig:lnrfrmpdf} shows the PDF of $\RM/\sigma_{\RM}$ for two extreme values of $\gamma$ ($1$ and $4$) and all $\alpha$ in the range $[1, 4]$. The $\RM$ PDFs no longer follow a Gaussian distribution, even though the magnetic field is Gaussian random (see \Fig{fig:grfrmspecpdf}b). This is the effect of the lognormal thermal electron density, and thus, including $\ne$ not only affects the spectral properties (second-order statistics), but also the PDF (which includes higher-order statistics). To further demonstrate the non-Gaussianity of these $\RM$ PDFs, we compute their kurtosis as
\begin{align} \label{eq:kurt}
\ku_{\RM} = \frac{\langle \RM \rangle^{4}}{\langle \RM^{2} \rangle^{2}},
\end{align}
where $\langle \rangle$ denotes the spatial average over the entire map. The kurtosis of a Gaussian distribution is three. The estimated kurtosis of $\RM$ for different values of $\alpha$ and $\gamma$ are given in the legend of \Fig{fig:lnrfrmpdf}. All the kurtosis values are significantly greater than three, confirming the non-Gaussian nature of the computed $\RM$.

Finally, we show the second-order structure function of $\RM$ computed with the five-point stencil ($\sffive$, \Eq{eq:sffive}) in \Fig{fig:lnrfrmsf}, for two extreme values of $\gamma$ ($1$ and $4$) and all $\alpha=[1, 4]$. The slope of the $\RM$ structure function is primarily controlled by the slope of the power spectrum of $\RM$, $\beta$. Since, $\beta$ is more affected by the shallower of the two spectra (magnetic fields, $-\alpha$, and thermal electron density, $-\gamma$), the slope of the $\RM$ structure function is roughly either $\alpha$ or $\gamma$ (depending on which one is shallower; see \Fig{fig:lnrfrmsf}). More generally, given a value of $\alpha$ and $\gamma$, $\beta$ can be estimated using \Eq{eq:rmspecfit} and then the slope of the $\RM$ structure function would be $\beta - 1$ (provided the structure functions are converged). After studying the effect of Gaussian random magnetic fields and lognormal thermal electron densities (constructed independently of each other) on the $\RM$ power spectrum and structure function. In \App{sec:buni}, we explore the effect of uniform random magnetic fields and lognormal thermal electron density and in \App{sec:simdyn} we study the properties of $\RM$ from MHD simulations, where the thermal electron density and magnetic fields can be correlated.

The main summary of the numerical section of the paper is as follows. Assuming a constant thermal electron density, for Gaussian random magnetic fields with a power law spectrum with slope $\alpha$, i.e., $\pk(b) \sim k^{-\alpha}$, the slope of second-order structure function of $\RM$ is $-\alpha - 1$. However, the second-order structure function computed using more than two points per stencil is required for accuracy and convergence (\Sec{sec:simgrf}), depending on the underlying spectral slope of the quantity for which the structure function is computed. The $\RM$ PDFs follow a Gaussian distribution. Then for Gaussian random magnetic fields with $\pk(b) \sim k^{-\alpha}$ and lognormal thermal electron density with $\pk(\ne) \sim k^{-\gamma}$, the slope of the $\RM$ power spectrum can be computed \rev{(assuming no effect of any window or sampling function)} using the empirically derived equation, \Eq{eq:rmspecfit} (\Sec{sec:simlnrf}). In the next section, we use these results and observations to study the properties of small-scale magnetic fields in the SMC and LMC.

\section{$\RM$ structure function and properties of small-scale magnetic fields in the SMC and LMC} \label{sec:obs}
Here we apply the ideas and results from \Sec{sec:sim} to study the properties of the small-scale magnetic fields in the Magellanic clouds. $\RM$ observations for the SMC (79~sources) are taken from \citet{LivingstonEA2021b} and for the LMC (250~sources) from \citet{MaoEA2012} and Livingston et al.~(2021, in prep.) \footnote{\rev{We neglect the large-scale stratification and any radial dependence in the analysis. However, since we are mostly interested in small-scale magnetic field properties, we expect those to not make a significant difference in our results.}}. These $\RM$ observations are due to polarised point sources behind (or within) the SMC and LMC. To estimate the thermal electron density contribution to these $\RM$s, we use the H$\alpha$ observations from \citet{GaustadEA2001}. 

\begin{figure*}
    \includegraphics[width=\columnwidth]{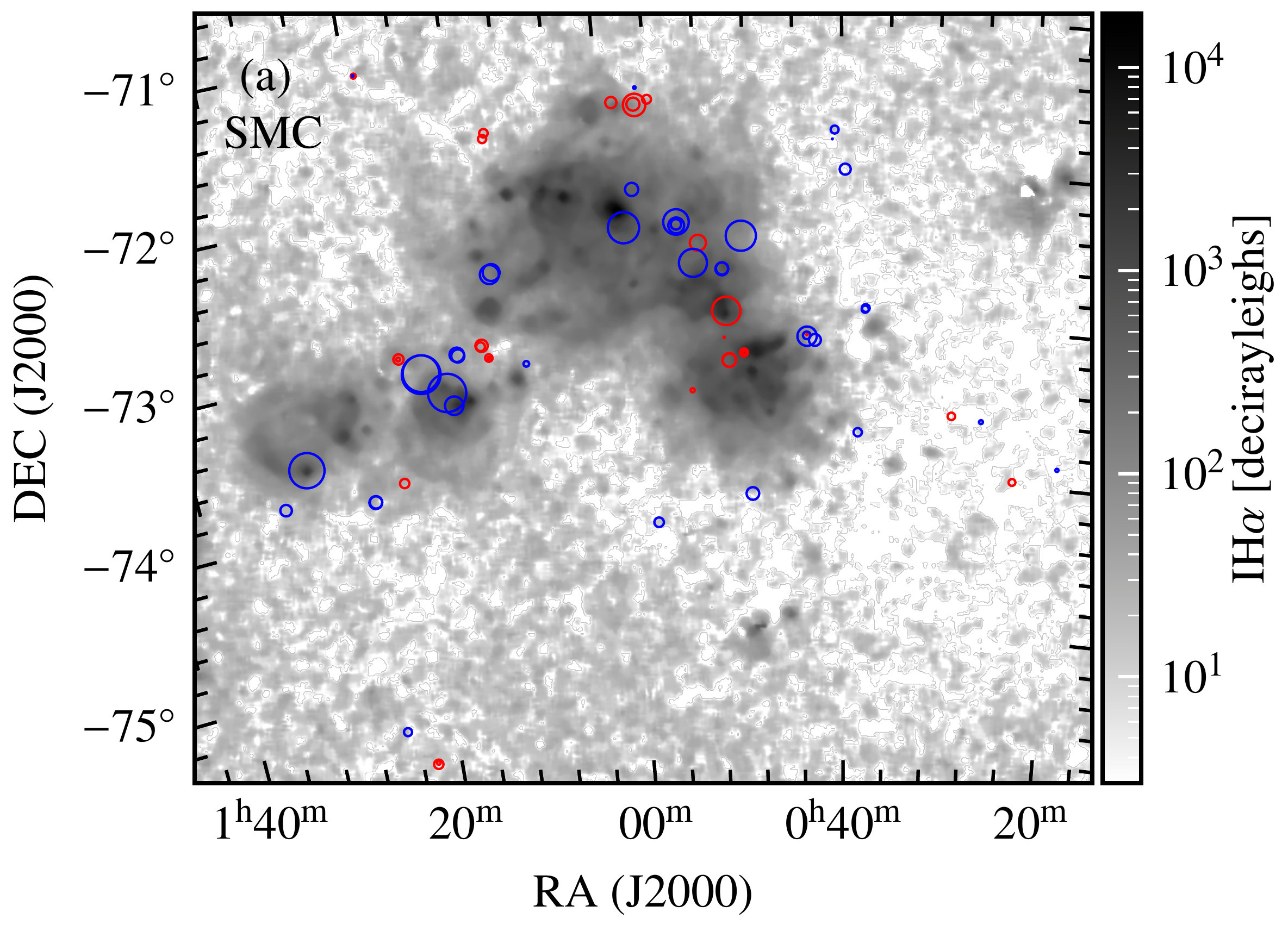} \hspace{0.5cm}
    \includegraphics[width=\columnwidth]{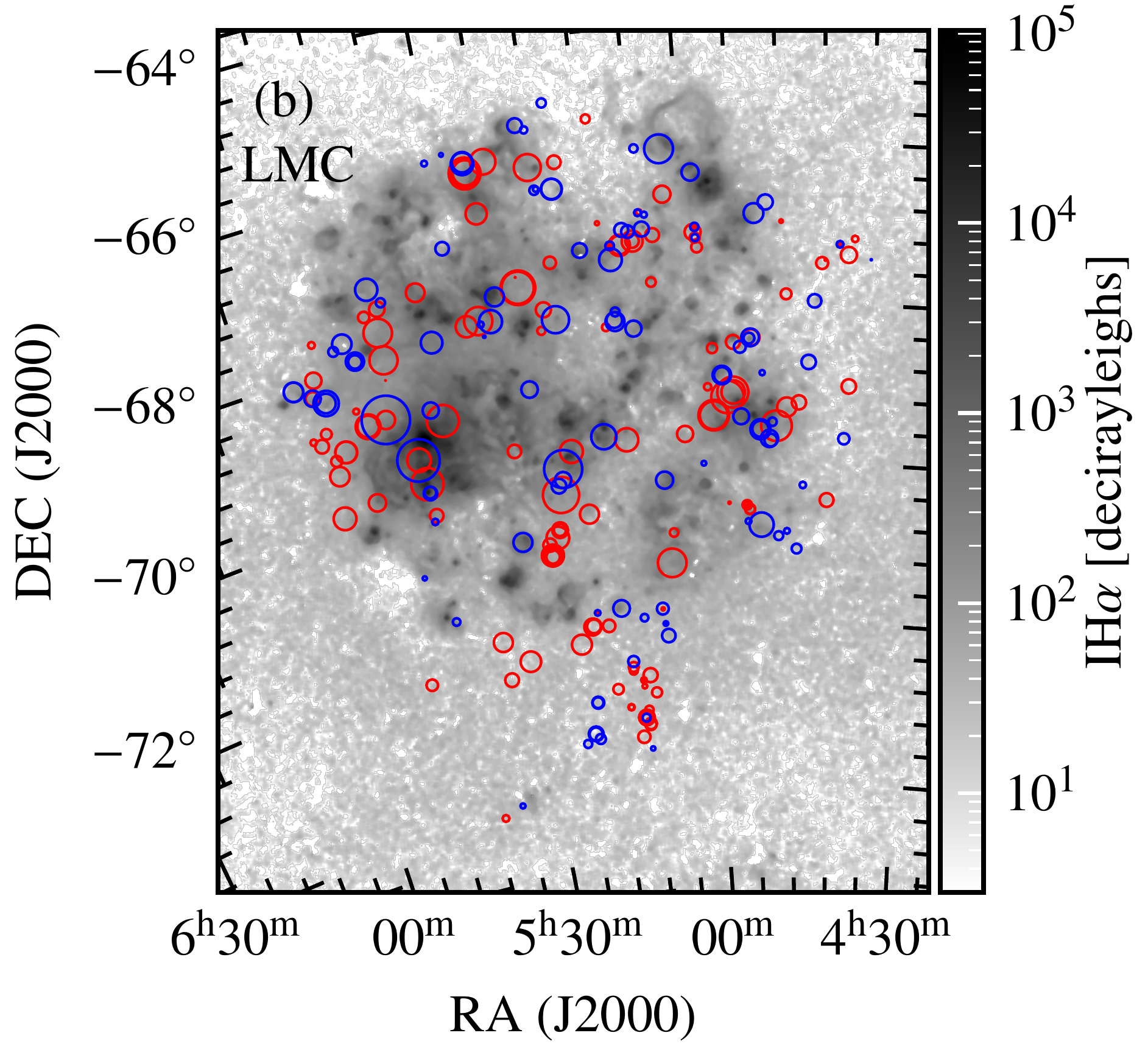}
    \caption{$\RM$ observations from background polarised sources after removing the Milky Way contribution (red circles show positive $\RM$, blue circles show negative $\RM$, and the size of the circle is proportional to the magnitude of $\RM$) for the SMC ((a), 65~sources) and the LMC ((b), 246~sources). The background colour shows $\IHa$ (in decirayleighs) after removing the contribution from the Milky Way and dust in the medium for both cases. Pairs of sources have non-uniform separation, and thus, it would be very difficult to compute an $\RM$ power spectrum for them, so we resort to $\RM$ structure functions to study the small-scale magnetic field properties in the SMC and LMC.}
    \label{fig:smclmcrmiha}
\end{figure*}

\begin{figure*}
    \includegraphics[width=\columnwidth]{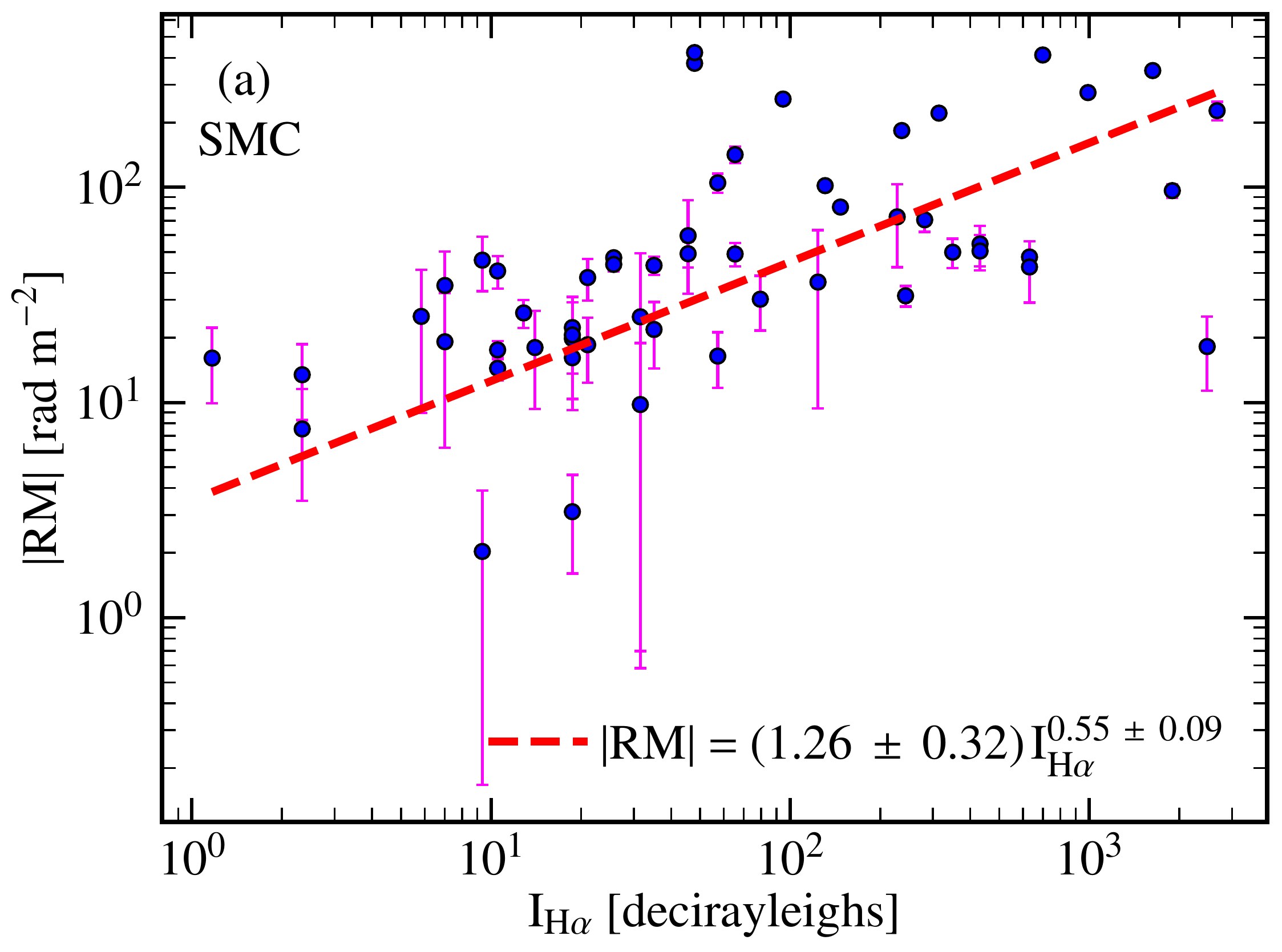} \hspace{0.5cm}
    \includegraphics[width=\columnwidth]{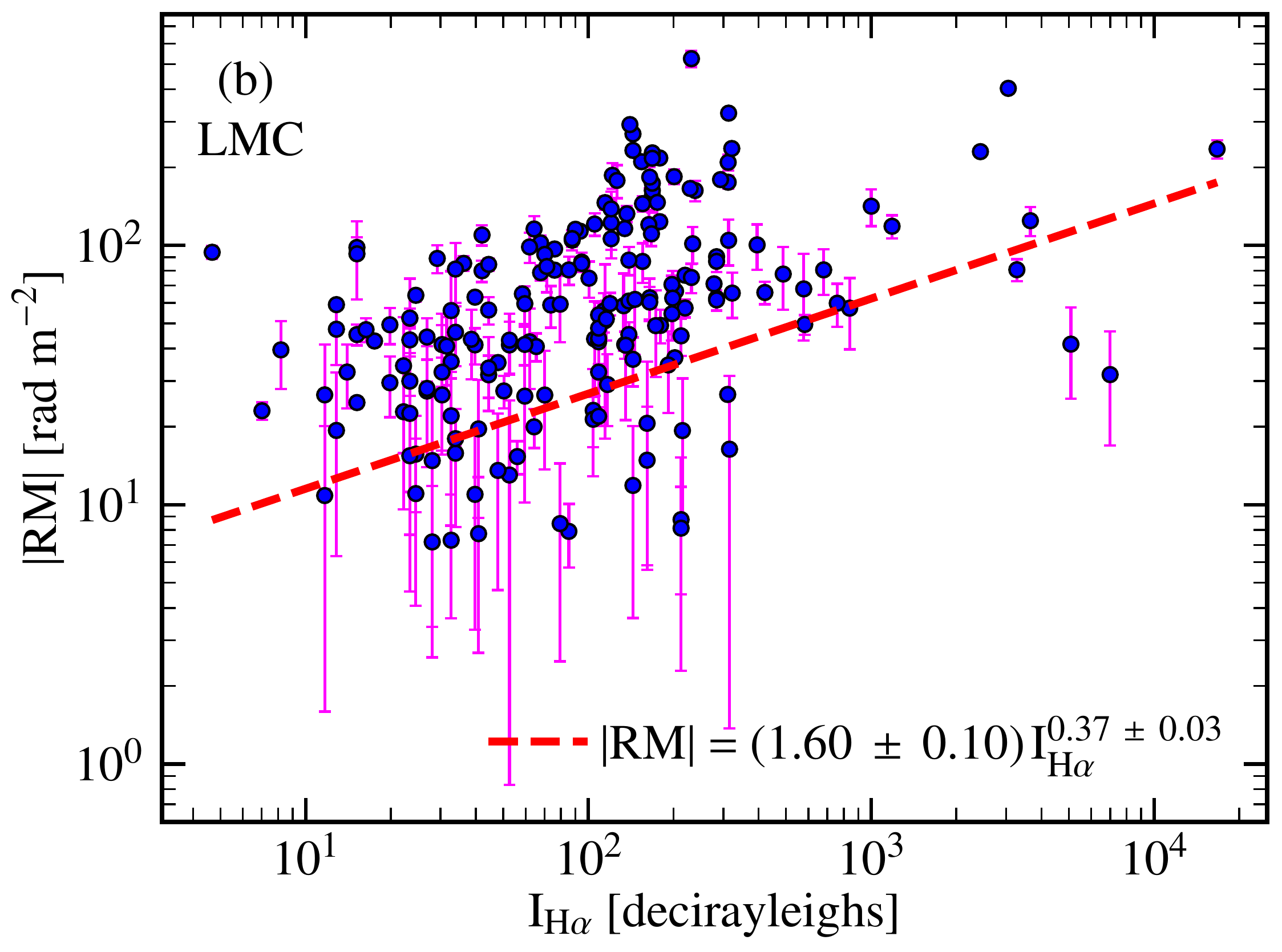}
    \caption{$|\RM|$ vs.~$\IHa$ (blue points with magenta lines showing the fluctuations in $\RM$) and the best-fit relationship \rev{(obtained using the least-squares fitting procedure but also considering the errors in the fitting)} between them (red, dashed line) for the SMC (a) and the LMC (b). The slope of the relationship for the LMC deviates significantly from the theoretically expected slope of $0.5$. This might be because of the clumpy medium and/or large-scale magnetic field variations along the path length. The slope is smaller for the LMC ($\approx 0.37$) than the SMC ($\approx 0.55$), and this is probably because the large-scale field (as reported in the literature) is stronger in the LMC \citep[$\sim 1~\muG$,][]{MaoEA2012} as compared to the SMC \citep[$\sim 0.2~\muG$,][]{MaoEA2008}.}
    \label{fig:smclmcrmihacorr}
\end{figure*}

We remove the Milky Way $\RM$ contribution from the observed $\RM$s \rev{by using analytical models, eq.~5 in \citet{MaoEA2008} for the SMC, and eq.~3 in \citet{MaoEA2012} for the LMC, at their location}. In \App{sec:rmmw}, we discuss three models to account for the Milky Way $\RM$ contribution and our reasoning to choose these analytical models. Besides the thermal electrons in the medium, the H$\alpha$ map also has contributions from the Milky Way and \revb{the H$\alpha$ emission is also attenuated by dust in the Milky Way, SMC, and LMC}. These contributions have been removed using the prescription in \citet{SmartEA2019} (see their Sec.~3.4, where they estimate these contributions for the SMC and we assume the same prescription for the LMC). Furthermore, for both clouds, we chose only those $\RM$ for which H$\alpha$ data is available and the error in $\RM$ is less than the $\RM$ value. This reduces the number of point sources to 65 for the SMC and 246 for the LMC. With this low number of sources, it would not be possible to study the PDF of $\RM$, but we aim to explore small-scale magnetic field properties via the $\RM$ structure function. 

\Fig{fig:smclmcrmiha} shows the $\RM$ (Milky Way corrected) as circles (red for positive $\RM$s, blue for negative $\RM$s, and the size of a circle is proportional to $|\RM|$) and the intensity of H$\alpha$, $\IHa$, in the background for the SMC (\Fig{fig:smclmcrmiha}a) and LMC (\Fig{fig:smclmcrmiha}b). In both cases, it is clear that the sources are located at random positions with non-uniform distances between pairs of sources, and thus, it will be extremely difficult to compute the $\RM$ power spectrum. We, therefore, resort to $\RM$ structure functions to probe the properties of small-scale magnetic fields in the SMC and LMC. Before that, we first need to characterise and account for the contribution of the thermal electron density to the $\RM$ structure function.

In \Fig{fig:smclmcrmihacorr}, we study the dependence of $|\RM|$ (with the Milky Way contribution removed) on $\IHa$ (contributions from the Milky Way and dust in the medium also removed) for the SMC (\Fig{fig:smclmcrmihacorr}a) and LMC (\Fig{fig:smclmcrmihacorr}b). Ideally, we would expect that since $\RM \propto \ne$ and $\IHa \propto \ne^{2}$, the slope of $|\RM|$ vs.~$\IHa$ line would be roughly $0.5$ \citep[this slope is approximately $0.43$ when computed using pulsars in the Milky Way, see Fig. 11~(e) in][]{SetaF2021}.  We find that the slope is approximately $0.55$ for the SMC and $0.37$ for the LMC. The deviation of this slope from the theoretical expectation of $0.5$, especially for the LMC, is probably because of the following two reasons. First, the filling factor of $\ne$ affects $\IHa$ more than $\RM$ because of the quadratic dependence, and thus, a clumpy medium along the line of sight would lead to a deviation from the expected slope. Second and more importantly, the large-scale magnetic field in these clouds could vary along the path length (and even change sign), and this would lead to cancellations \revb{(this also depends on the number and location of those reversals)}, which in turn would lower the $\RM$ values and give a slope $<0.5$ for the $|\RM|$ vs.~$\IHa$ line. The second point is further corroborated by the fact that the large-scale field in the LMC \citep[$\sim 1~\muG$, see][]{MaoEA2012} is stronger (enhanced cancellation) than that in the SMC \citep[$\sim 0.2~\muG$, see][]{MaoEA2008}, and we find that the slope is smaller for the LMC ($0.37$) than the SMC ($0.55$). We use this relation to account for the contribution of $\ne$ while computing the $\RM$ structure function.

\begin{figure*}
    \includegraphics[width=\columnwidth]{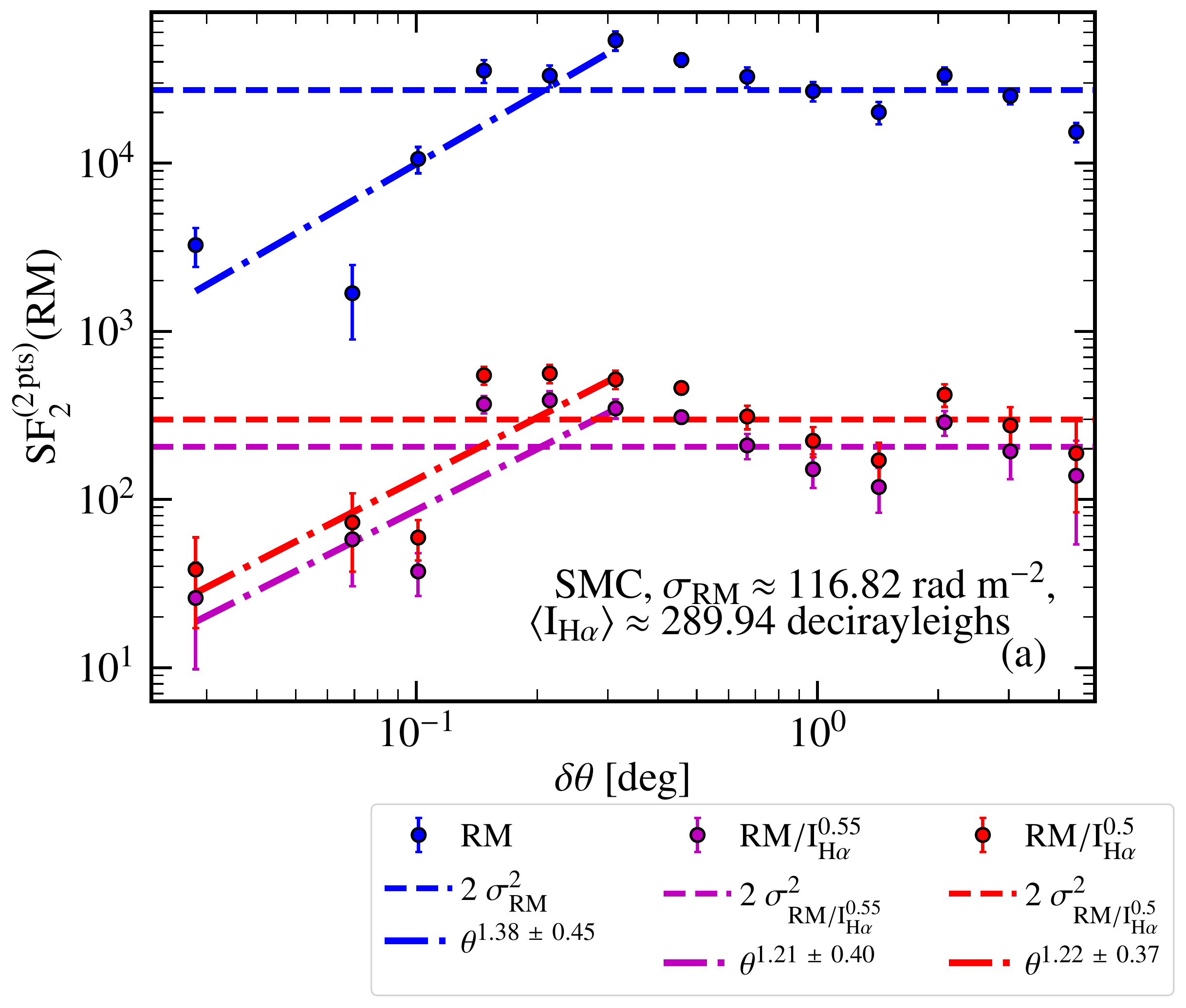} \hspace{0.5cm}
    \includegraphics[width=\columnwidth]{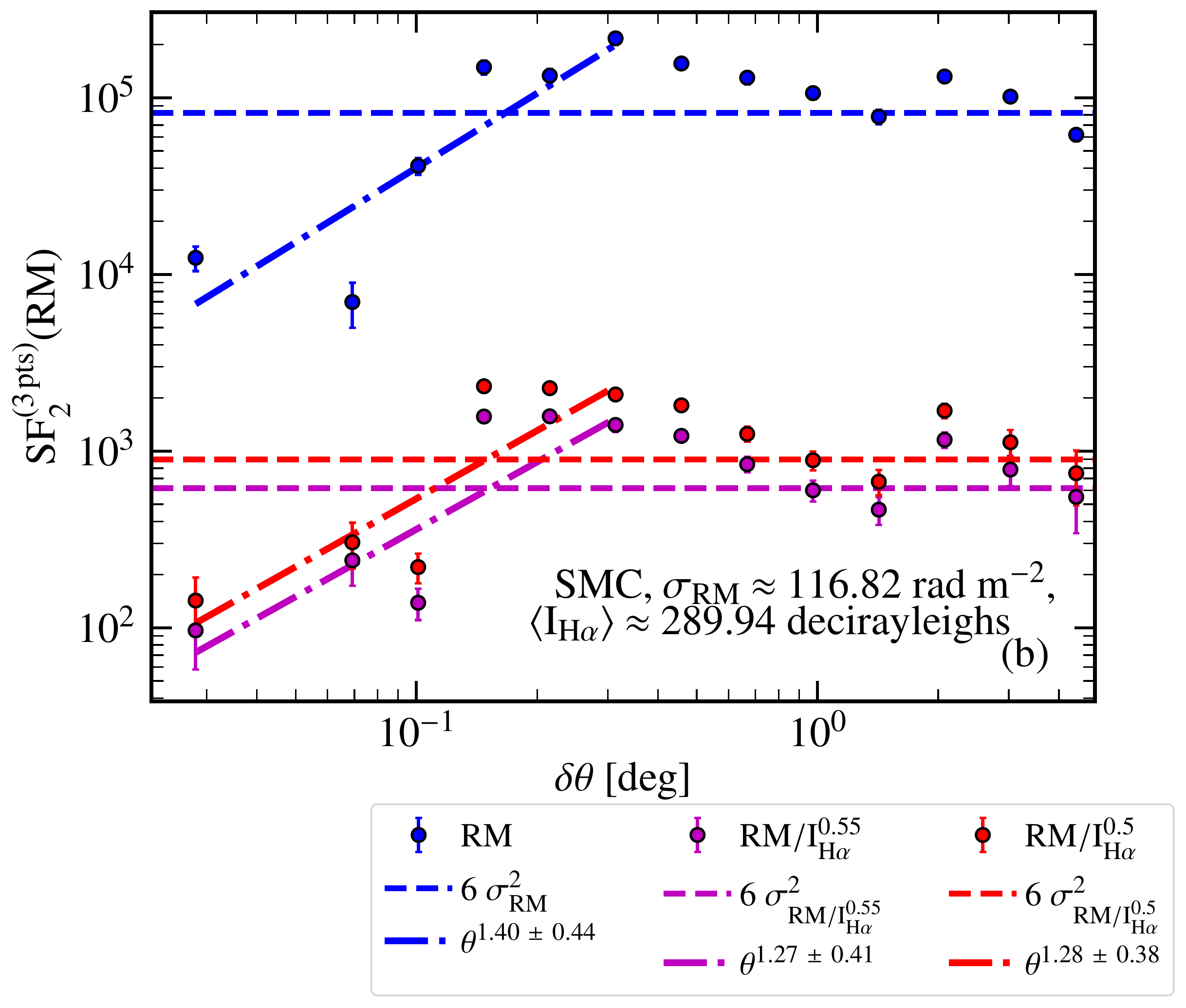}
    \caption{Second-order structure function of $\RM$ (blue), $\RM/\IHa^{0.55}$ (using the relationship from \Fig{fig:smclmcrmihacorr}a, magenta), and $\RM/\IHa^{0.5}$ (red) computed with the two-point stencil ($\sftwo$, (a)) and the three-point stencil ($\sfthree$, (b)) for the SMC. \rev{The error in the second-order structure function of $\RM$ (both stencils) is computed as the one-sigma variation of values in each of the angular separation bin.} We also provide the value of $\sigma_{\RM}$ and $\langle \IHa \rangle$. The dashed lines show the corresponding asymptotic values, $2 \sigma^{2}$ for $\sftwo$ and $6 \sigma^{2}$ for $\sfthree$, at high values of angular separation, $\delta \theta$. The slope of the fitted line (dashed-dotted line, \rev{obtained using the least-squares fitting procedure, which also includes error bars}) is given in the legends. These slopes are similar for both $\sftwo$ and $\sfthree$, indicating convergence.}
    \label{fig:smcrmsf}
\end{figure*}

\begin{figure*}
    \includegraphics[width=\columnwidth]{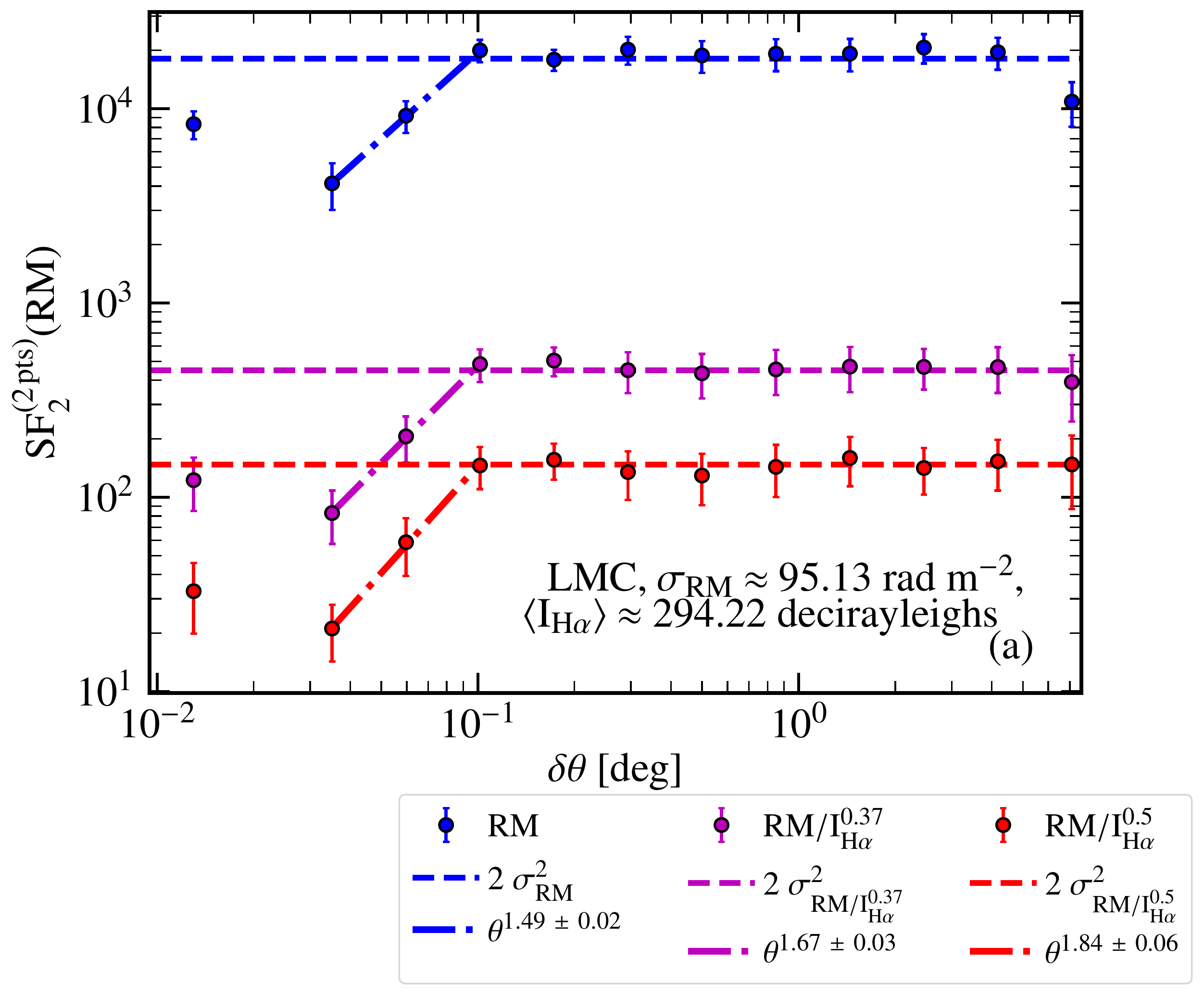} \hspace{0.5cm}
    \includegraphics[width=\columnwidth]{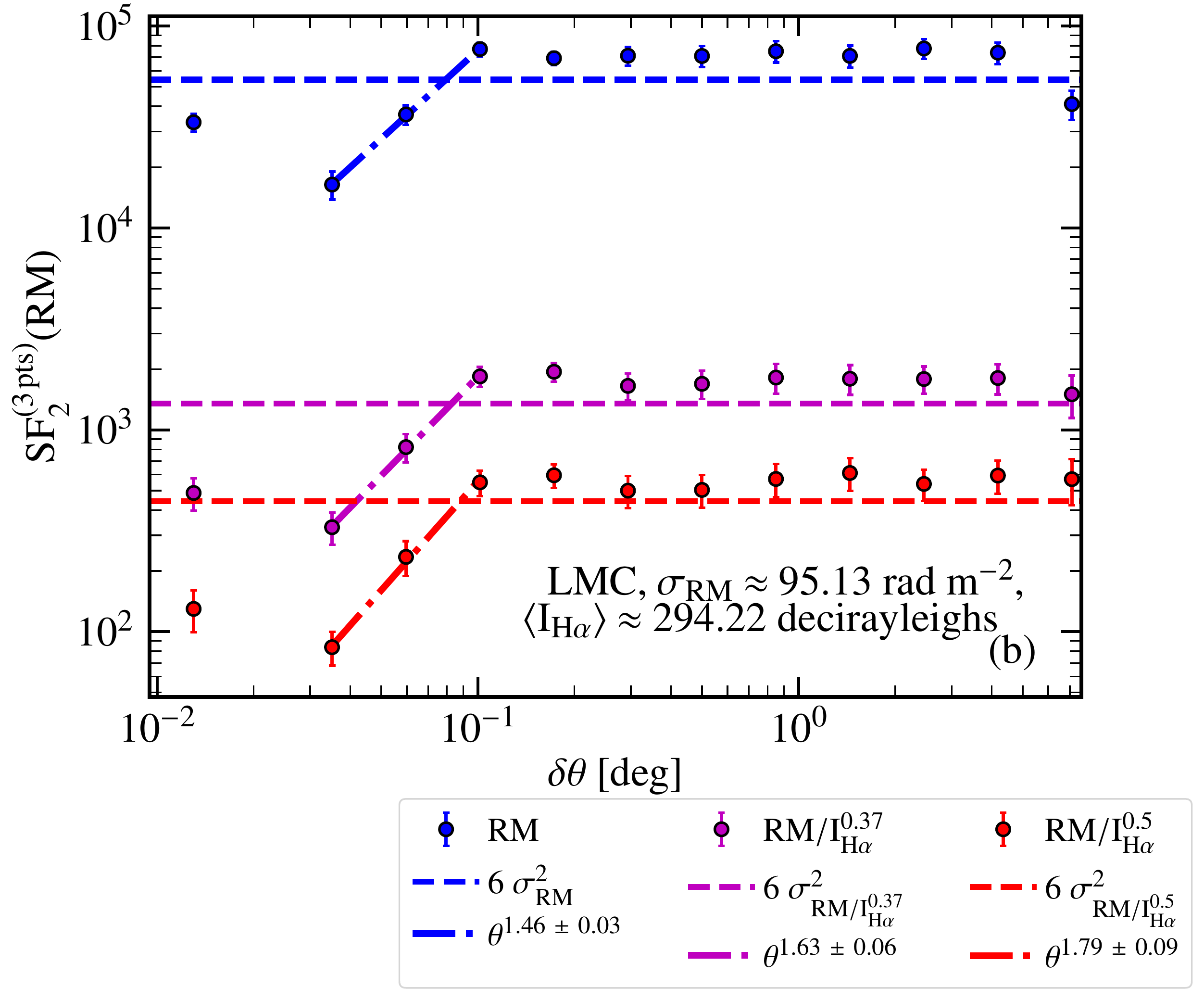}
    \caption{Same as \Fig{fig:smcrmsf}, but for the LMC, and thus includes the structure functions for $\RM/\IHa^{0.37}$ (using the relationship from \Fig{fig:smclmcrmihacorr}b). The slopes are similar for both $\sftwo$ and $\sfthree$, indicating convergence.}
    \label{fig:lmcrmsf}
\end{figure*}

We compute the second-order structure functions with the two-point ($\sftwo$) and the three-point stencil ($\sfthree$), as a function of the angular separation, $\delta \theta$, of $\RM$, $\RM$ compensated by $\IHa$ as per the relationship from \Fig{fig:smclmcrmihacorr}, and $\RM/\IHa^{0.5}$ (theoretically expected slope) for the SMC and LMC. These are shown in \Fig{fig:smcrmsf} and \Fig{fig:lmcrmsf}, respectively. For computing these structure functions, we make sure that there are at least $10$ values for each $\delta \theta$. The fluctuations in the structure functions are estimated by propagating errors in the observed $\RM$ and the Milky Way $\RM$ model. For both clouds and all three structure functions for each cloud, at high values of $\delta \theta$, $\sftwo$ is approximately equal to $2~\sigma_{\RM}^{2}$ and $\sfthree$ is approximately equal to $6~\sigma_{\RM}^{2}$, where the $\sigma_{\RM}$ denotes the standard deviation of $\RM$ in the sample. We also fit the second-order structure function from roughly a few times the minimum $\delta \theta$ to where the structure function starts to saturate (see also the $\RM$ structure functions from the simulations, for example in \Fig{fig:grfrmsf}, for which we fit from a few times the spatial resolution, $L/512$, till the structure function starts to saturate). The details of the fit for each case are given in the legend of \Fig{fig:smcrmsf} and \Fig{fig:lmcrmsf}. The slopes for $\sftwo$ and $\sfthree$ are roughly similar, indicating convergence. This is a benefit of computing $\RM$ structure function using multiple-point stencils. We now use the properties of these structure functions to estimate the following properties of the small-scale ISM in the SMC and LMC: the magnetic correlation length (\Sec{sec:obscor}), the field strength (\Sec{sec:obsstr}), and the slope of the power spectrum of the magnetic field and thermal electron density (\Sec{sec:obsslo}).

\subsection{Correlation length of the small-scale magnetic fields} \label{sec:obscor}
The correlation length of small-scale random magnetic fields is roughly equivalent to the turnover or break scale of the $\RM$ structure function (see \App{sec:kmin}). The approximate turnover separation is $0.15 \deg$ for the SMC (\Fig{fig:smcrmsf}a, b) and $0.1 \deg$ for the LMC (\Fig{fig:lmcrmsf}a, b). This separation remains the same for both $\sftwo$ and $\sfthree$. Assuming the distance to the SMC and LMC at $\approx 61~\kpc$ and $\approx 50~\kpc$, respectively, these separations imply a scale of around $160 \pm 21~\pc$ for the SMC and $87 \pm 17~\pc$ for the LMC. Thus, these are roughly the correlation lengths of the magnetic fields, $\ell(b)$. The scale for the LMC is comparable to that obtained in \citet{Gaenslar2005}, which reports $\approx 90\,\pc$, and that for the SMC is also comparable to that obtained in \citet{BurkhartEA2010}, which reports $\approx 160\,\pc$, albeit using a different method. In the next subsection, we use these scales to estimate the magnetic field strength from $\sigma_{\RM}$ and $\langle \IHa \rangle$.

\subsection{The small-scale magnetic field strengths} \label{sec:obsstr}
For a Gaussian random magnetic field with strength $\brms~[\muG]$ and a mean thermal electron density $\langle \ne \rangle~[\cm^{-3}]$, the standard deviation of $\RM$ \citep[Eq. B9 in][]{MaoEA2008},
\begin{align} \label{eq:sigrm}
\sigma_{\RM} = 0.81 \langle \ne \rangle ({L_{\rm pl}} \ell(b))^{1/2} \left(f [B^{2}_{0 \parallel}(1-f) + \brms^{2}/3]\right)^{1/2},
\end{align}
where $\ell(b)$ is the correlation length of the magnetic field in $\pc$, ${L_{\rm pl}}$ is the path length in $\kpc$, $B_{0 \parallel}$ is the parallel (to the line of sight) component of the large-scale field in $\muG$, and $f$ is the filling factor. The correlation length of small-scale magnetic fields, $\ell(b)$, in the SMC and LMC is estimated from the $\RM$ structure function in \Sec{sec:obscor} and we use $\langle \IHa \rangle$ to estimate the product $\langle \ne \rangle~{L_{\rm pl}}^{1/2}$. Using Eq.~8, 12, and 16 of \citet{MaoEA2008},
\begin{align} \label{eq:nemean}
\langle \ne \rangle~{L_{\rm pl}}^{1/2}  = \left(\frac{10\,f\,\langle \IHa \rangle ~ (\langle T_{\rm e} \rangle / 10^{4} {\rm K})^{0.5}}{0.39[0.92 - 0.34 \ln(\langle T_{\rm e} \rangle / 10^{4} {\rm K})]}\right)^{1/2},
\end{align}
where $f$ is the filling factor, $\langle \ne \rangle$ is in $\cm^{-3}$, ${L_{\rm pl}}$ is in $\pc$, $\langle \IHa \rangle$ is in $\rm decirayleighs$, and  $\langle T_{\rm e} \rangle$ is the electron temperature in $\K$. We take $f \approx 0.4$ and $T_{\rm e} \approx 1.4 \times 10^{4}\,\K$ from \citet{MaoEA2008}. Furthermore, we take $B_{0 \parallel}$ values from the literature, $\approx 0.3\,\muG$ for the SMC \citep{LivingstonEA2021} and $\approx 1\,\muG$ for the LMC \citep{Gaenslar2005}. We can estimate the average $\brms$ over the entire cloud using \Eq{eq:sigrm} and \Eq{eq:nemean}, and this method does not require knowing the path length, ${L_{\rm pl}}$, explicitly. \rev{In both clouds, $\sigma_{\RM}$ ($\approx 117~\rad~\m^{-2}$ for the SMC and $\approx 95~\rad~\m^{-2}$ for the LMC) is significantly greater than the level of fluctuations intrinsic to $\RM$ sources \citep[$< 10~\rad~\m^{-2}$, see][]{Schnitzeler2010,ShahS2021} and thus the intrinsic fluctuations due to background sources can be neglected \revb{(see \App{sec:rmint} for further discussion on the effects of intrinsic fluctuations on the properties of second-order $\RM$ structure functions)}.} In the SMC, for $\sigma_{\RM} \approx 117~\rad~\m^{-2}$, $\langle \IHa \rangle \approx 290~\rm decirayleighs$, and $\ell(b) \approx 160~\pc$, using \Eq{eq:sigrm} and \Eq{eq:nemean}, we obtain $\brms \approx 14 \pm 2~\muG$. This is higher than the value ($\approx 2~\muG$) obtained by \citet{MaoEA2008} and that by \citet{LivingstonEA2021} ($\approx 8~\muG$). Similarly, for the LMC with $\sigma_{\RM} \approx 95~\rad~\m^{-2}$, $\langle \IHa \rangle \approx 294~\rm decirayleighs$, and $\ell(b) \approx 87~\pc$, we obtain $\brms \approx 15 \pm 3~\muG$. This estimate is significantly higher than the value ($\approx 4~\muG$) in \citet{Gaenslar2005}. These differences might be due to differences in the methods of obtaining these strengths, differences in the Milky Way model used, and also because of systematics in different terms. After studying the average strength of the fluctuating component of the magnetic field, we now estimate the slope of the magnetic field and thermal electron density power spectrum.

\subsection{Slope of the small-scale magnetic field and thermal electron density power spectra} \label{sec:obsslo}
To estimate the slope of the magnetic field power spectrum ($\alpha$), we use the second-order structure function of $\RM$ compensated by $\IHa$ according to the relationships derived in \Fig{fig:smclmcrmihacorr} (magenta points and dashed, dotted lines in \Fig{fig:smcrmsf} and \Fig{fig:lmcrmsf}). The slopes of the second-order $\RM$ structure functions are very similar for the two- and three-point stencils and this shows convergence (we use the structure function computed with the three-point stencil for further calculations because it is probably more accurate). \rev{Assuming Gaussian random magnetic fields}, for the SMC, ${\rm SF}^{\rm(3\,pts)}_2 (\RM/\IHa^{0.55}) \sim r^{1.27 \pm 0.41}$ implies $\alpha \approx 1.3 \pm 0.4$ in $\pk(b) \sim k^{-\alpha}$ (as demonstrated in \Sec{sec:simgrf}). Similarly, for the LMC, ${\rm SF}^{\rm(3\,pts)}_2 (\RM/\IHa^{0.37}) \sim r^{1.63 \pm 0.06}$ implies $\alpha \approx 1.6 \pm 0.1$. We can use these values of $\alpha$ and estimated $\beta$ (the slope of the $\RM$ power spectrum, $\pk(\RM) \sim k^{-\beta}$) to compute the slope of the thermal electron density power spectrum, $\gamma$ in $\pk(\gamma) \sim k^{-\gamma}$, using \Eq{eq:rmspecfit} \rev{(which assumes a lognormal thermal electron density distribution and no correlation between thermal electron density and magnetic fields)}. Since ${\rm SF}_2 (\RM) \sim r^{\beta - 1}$, using ${\rm SF}^{\rm(3\,pts)}_2 (\RM)$ in \Fig{fig:smcrmsf} and \Fig{fig:lmcrmsf}, we obtain $\beta \approx 2.4$ and $2.5$ for the SMC and LMC, respectively. Now, given $\alpha$ and $\beta$, we obtain $\gamma$ from \Eq{eq:rmspecfit}. This gives two solutions for $\gamma$ as the equation is quadratic and we choose the negative or smaller of the two solutions\footnote{\revb{This is because we expect the slope of the power spectrum of the small-scale magnetic field and thermal electron density to be negative, i.e., their strength decreases with decreasing length scale. Given the turbulent nature of the ISM, this is a reasonable assumption due to the energy cascade (energy injected at larger scales, cascades to smaller scales, and ultimately dissipates).}}. We find that $\gamma \approx 0.002~(\pm 0.2)$ for the SMC and $\approx -0.02~(\pm 0.09)$ for the LMC. The uncertainty in the estimated slope of the thermal electron density power spectrum is high but the thermal electron density power spectrum has a slope $\approx 0$, implying very small (or no) variations over these scales. Thus, the $\RM$ power spectrum is primarily determined by the magnetic field power spectrum and the contribution from the thermal electron density power spectrum is relatively small or negligible.

A summary of all the estimated parameters is given in \Tab{tab:smclmc}. We find that the correlation length of the magnetic field is higher for the SMC by a factor roughly equal to two but the field strength and the slope of the magnetic field power spectrum are comparable for both the SMC and LMC. 

\begin{table*} 
\caption{Estimated properties of small-scale magnetic fields and thermal electron densities in the SMC and LMC. The columns are as follows: 1.~cloud name, 2.~correlation length of the small-scale magnetic field, $\ell(b)$, 3.~average small-scale magnetic field strength, $\brms$, 4.~slope of the magnetic field power spectrum, $\alpha$ in $\pk(b) \sim k^{-\alpha}$, and 5.~slope of the thermal electron density power spectrum, $\gamma$ in $\pk(\ne) \sim k^{-\gamma}$.}
\label{tab:smclmc}
\begin{tabular}{ccccc} 
\hline
 Cloud & $\ell(b)~[\pc]$ & $\brms~[\muG]$  & $\alpha$ & $\gamma$ \\
\hline
SMC & $160 \pm 21$ & $14 \pm 2$ & $1.3 \pm 0.4$ & $0.002 \pm 0.2$  \\ 
LMC & $87 \pm 17$ & $15 \pm 3$ & $1.6 \pm 0.1$ & $-0.02 \pm 0.09$ \\
\hline
\end{tabular}
\end{table*}

\section{Conclusions} \label{sec:con}
Using numerical simulations, we first comprehensively study the dependence of the $\RM$ structure function on the properties of small-scale random magnetic fields and thermal electron densities. We then use those results to study the properties of small-scale magnetic fields in the SMC and LMC. In particular, we discuss the effects of thermal electron density on the statistical properties of $\RM$. For numerical simulations, we use Gaussian random magnetic fields and lognormal thermal electron densities (in \App{sec:simdyn}, we also explore magnetic fields and electron densities from MHD simulations, where the magnetic field-thermal electron density correlation is varied). Our results and conclusions are summarised below.

\begin{itemize}
\item For Gaussian random magnetic fields with a power-law spectrum of slope $\alpha$, i.e., $\pk(b) \sim k^{-\alpha}$, the slope of the $\RM$ power spectrum is $-\alpha - 1$ (assuming a constant thermal electron density). The corresponding slope of the second-order $\RM$ structure function is $\alpha$ (\Sec{sec:simgrf}). This has to be confirmed by computing second-order structure functions with different numbers of points per stencil (\Sec{sec:met}). We compute the structure function with two-, three-, four-, and five-point stencils, and show that agreement with theoretical expectations is best for the five-point stencil. Thus, to accurately capture variations in the $\RM$ structure function, its computation with a high-point stencil (points greater than the two-point stencil) may be required. \rev{This is particularly crucial for observational data, where we usually have only one snapshot and a limited range in the probed scales.} Furthermore, we show that the length scale around which the structure function approaches an asymptotic value (which depends on the number of points per stencil) is approximately equal to the correlation length of the random magnetic field (\App{sec:kmin}). The PDF of $\RM$ also follows a Gaussian distribution.

\item With the Gaussian random magnetic fields, we consider lognormal thermal electron density PDFs, with a power-law spectrum of slope $\gamma$, i.e., $\pk(\ne) \sim k^{-\gamma}$. The slope of the $\RM$ power spectrum, $\beta$, i.e., $\pk(\RM) \sim k^{-\beta}$, depends on both $\alpha$ and $\gamma$. We derive an empirical relationship, \Eq{eq:rmspecfit}, to determine $\beta$, given $\alpha$ and $\gamma$ (\Sec{sec:simlnrf}). Once $\beta$ is known, the second-order structure function of $\RM$, has a slope approximately equal to $\beta - 1$. The $\RM$ PDFs, when a lognormal thermal electron density is included, are non-Gaussian. 


\item We use $\RM$ and H$\alpha$ observations for the SMC and LMC to determine the fluctuating, small-scale magnetic field properties in the SMC and LMC (\Sec{sec:obs}). From these observations, we remove the contribution of the Milky Way from $\RM$ and $\IHa$, and also the contribution of dust in the SMC and LMC from $\IHa$. First, we find the dependence of $|\RM|$ on $\IHa$, $|\RM| \propto \IHa^{0.55}$ for the SMC and $|\RM| \propto \IHa^{0.37}$ for the LMC. Any deviations from the theoretical expectation, $|\RM| \propto \IHa^{0.5}$, are probably because of the non-homogeneity of the interstellar medium and/or because of variations in the large-scale magnetic field along the path length.  

The number of $\RM$ sources for computing $\RM$ statistics is 65 for the SMC and 246 for the LMC. We study the properties of the $\RM$ power spectrum via the $\RM$ structure function (since the pair of sources have non-uniform separation). For the SMC, the low number of points means that there is a higher uncertainty in some of the results (especially the slope of the magnetic power spectrum; \revb{see \App{sec:nsamp} for further discussion on the effect of the number of sources on the properties of second-order $\RM$ structure functions}) and might be biased towards a particular region. Future observations from the Polarisation Sky Survey of the Universe's Magnetism (POSSUM) survey \citep{GaenslerEA2010} and eventually the Square Kilometre Array (SKA) \citep{HealdEA2020} would provide data with higher sensitivity for a significantly larger number of sources, which would make it possible to study the $\RM$ spectra \rev{(probe of magnetic scales)} and PDFs \rev{(probe of magnetic field structure, i.e., Gaussian vs. non-Gaussian)} in much greater detail. Using the data at hand for the Magellanic clouds, \rev{it is difficult to study $\RM$ PDFs and thus we cannot explore the magnetic field structure.} However, we compute the second-order structure function of $\RM$, $\RM$ compensated by $\IHa$ according to the derived relation, and the theoretical relation. We show that the slope of the $\RM$ structure functions for the two- and three-point stencils are roughly similar, for both the SMC (\Fig{fig:smcrmsf}) and LMC (\Fig{fig:lmcrmsf}). This shows the convergence of the $\RM$ structure function. Based on the $\RM$ structure functions, we derive the following three properties of the magnetic fields in the SMC and LMC \rev{(assuming Gaussian random magnetic fields, lognormal thermal electron density, and no correlation between them)}: the correlation length ($160 \pm 21~\pc$ for the SMC and $87 \pm 17~\pc$ for the LMC), the magnetic field strength ($14 \pm 2~\muG$ for the SMC and $15 \pm 3~\muG$ for the LMC), and the slope of the magnetic power spectrum ($-1.3 \pm 0.4$ for the SMC and $-1.6 \pm 0.1$ for the LMC). Furthermore, we find that the thermal electron density is practically constant over the magnetic field correlation scales and the $\RM$ power spectrum is primarily decided by the magnetic field power spectrum.

\end{itemize} 

\section*{Acknowledgements}
C.~F.~acknowledges funding provided by the Australian Research Council (Future Fellowship FT180100495), and the Australia-Germany Joint Research Cooperation Scheme (UA-DAAD). We further acknowledge high-performance computing resources provided by the Leibniz Rechenzentrum and the Gauss Centre for Supercomputing (grants~pr32lo, pn73fi and GCS Large-scale project~22542), and the Australian National Computational Infrastructure (grant~ek9) in the framework of the National Computational Merit Allocation Scheme and the ANU Merit Allocation Scheme.

\section*{Data Availability}
The $\RM$ observations for the SMC are taken from \citet{LivingstonEA2021b} and that for the LMC are from \citet{MaoEA2012} and Livingston et al.~(2021, in prep.). The H$\alpha$ map for the Magellanic clouds is taken from \citet{GaustadEA2001}. The Milky Way $\RM$ models are taken from \citet{OppermannEA2015} and \citet{HutschenreuterEA2020}. The MHD simulation data and the data derived from the observations is available upon reasonable request to the corresponding author, Amit Seta (\href{mailto:amit.seta@anu.adu.au}{amit.seta@anu.adu.au}).


\bibliographystyle{mnras}
\bibliography{rmsf}



\appendix

\section{Turnover or break scale of the $\RM$ structure function} \label{sec:kmin}
\begin{figure*}
    \includegraphics[width=\columnwidth]{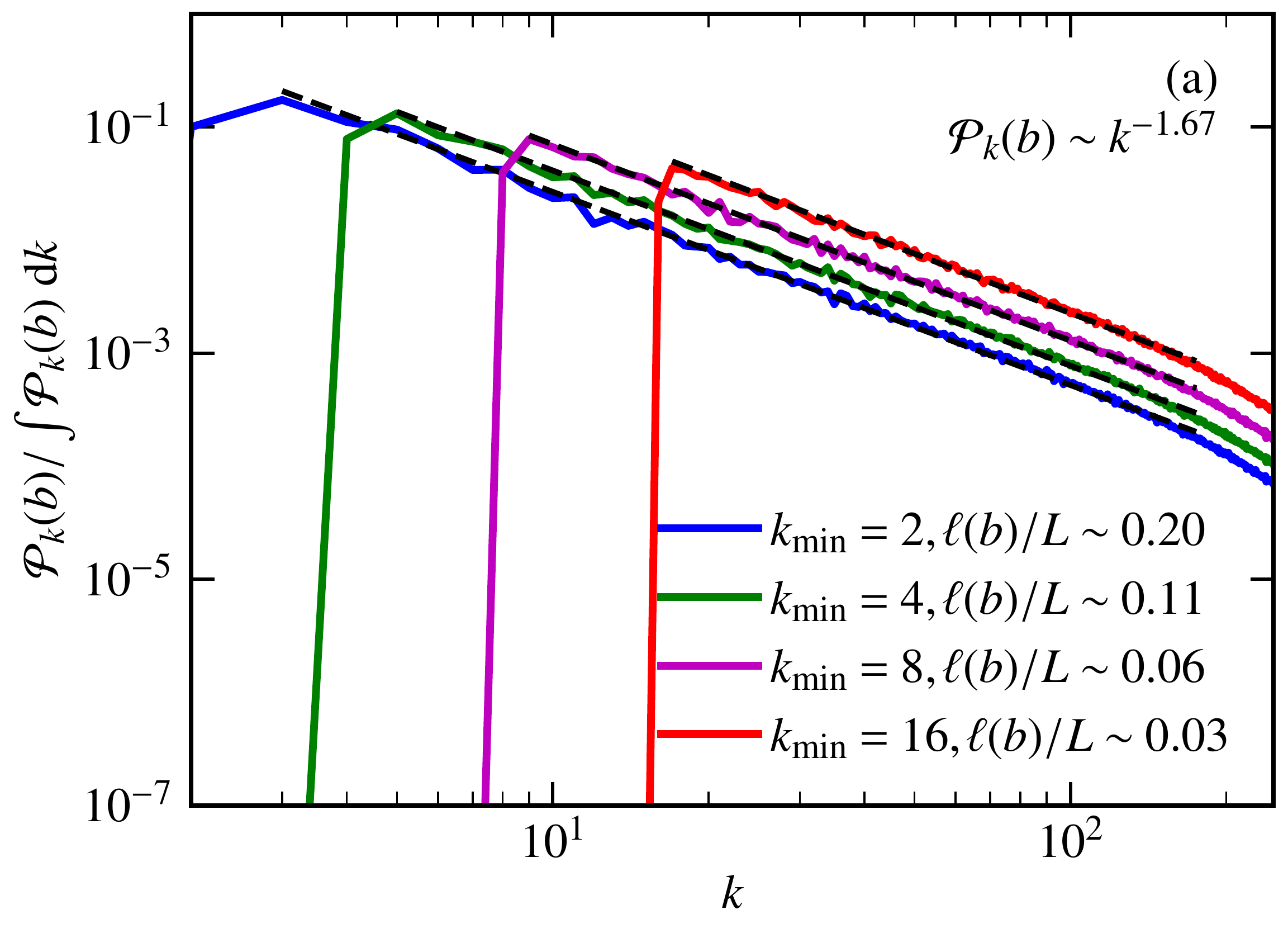} \hspace{0.5cm}
    \includegraphics[width=\columnwidth]{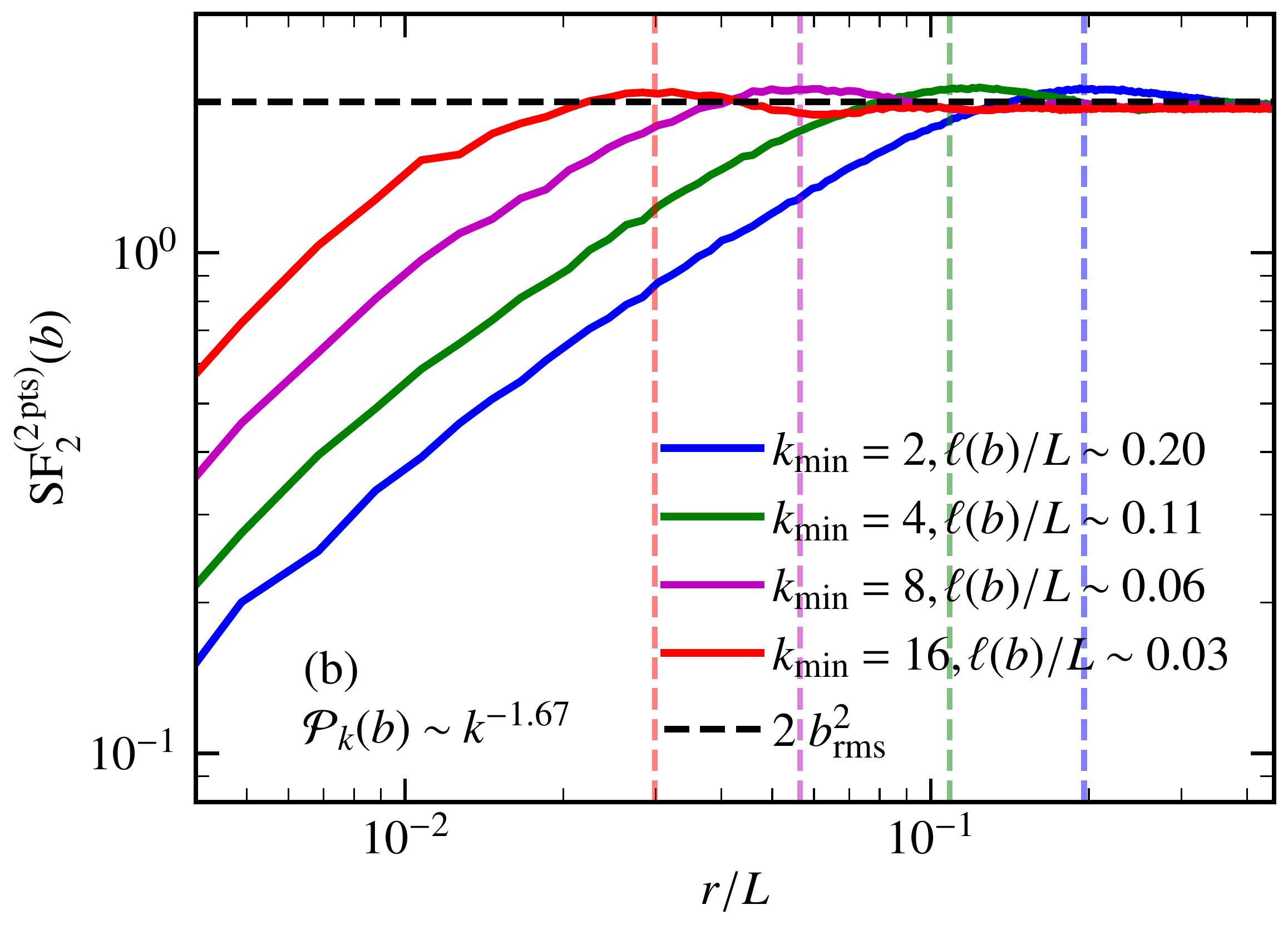}
    \includegraphics[width=\columnwidth]{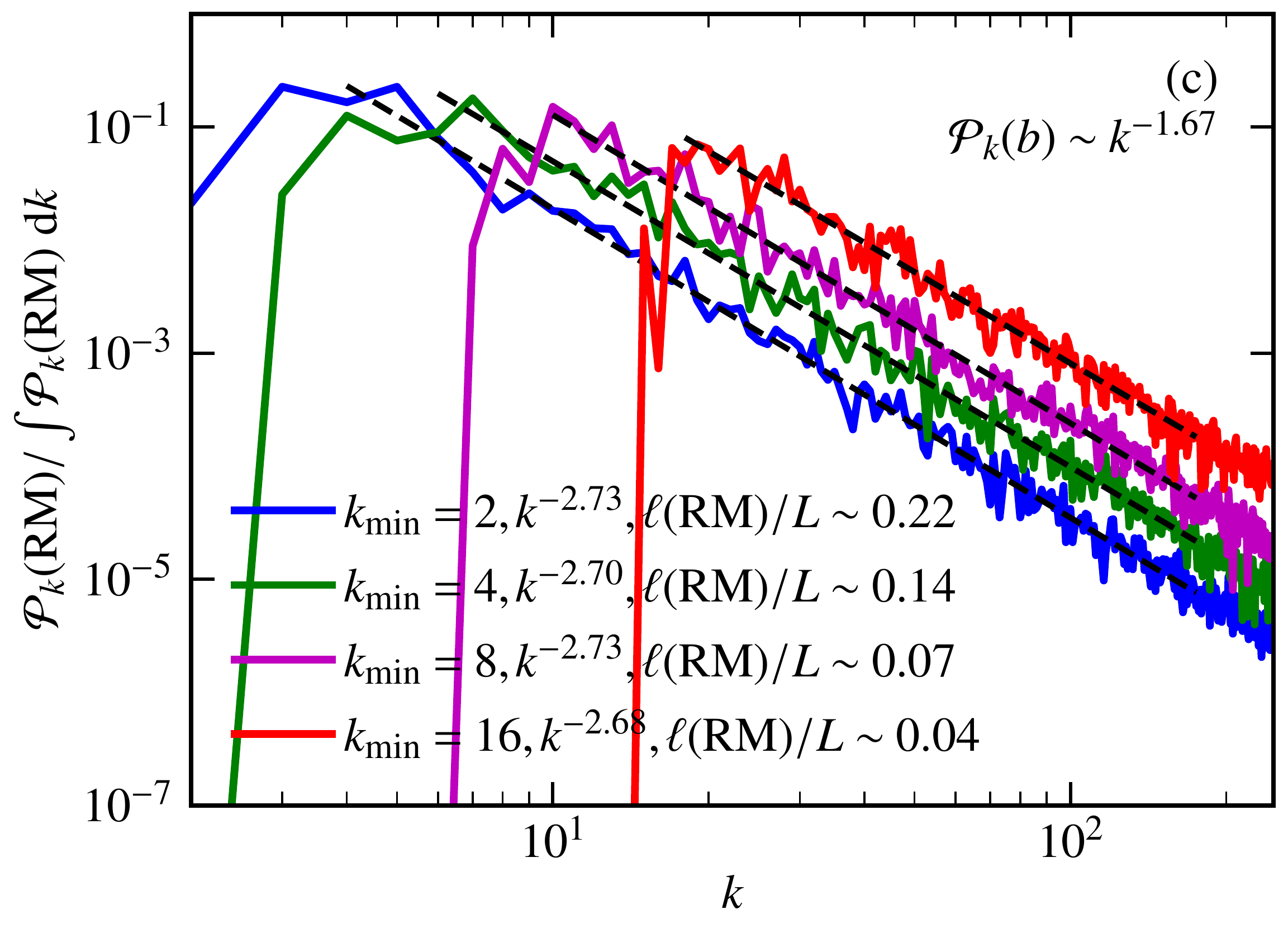} \hspace{0.5cm}
    \includegraphics[width=\columnwidth]{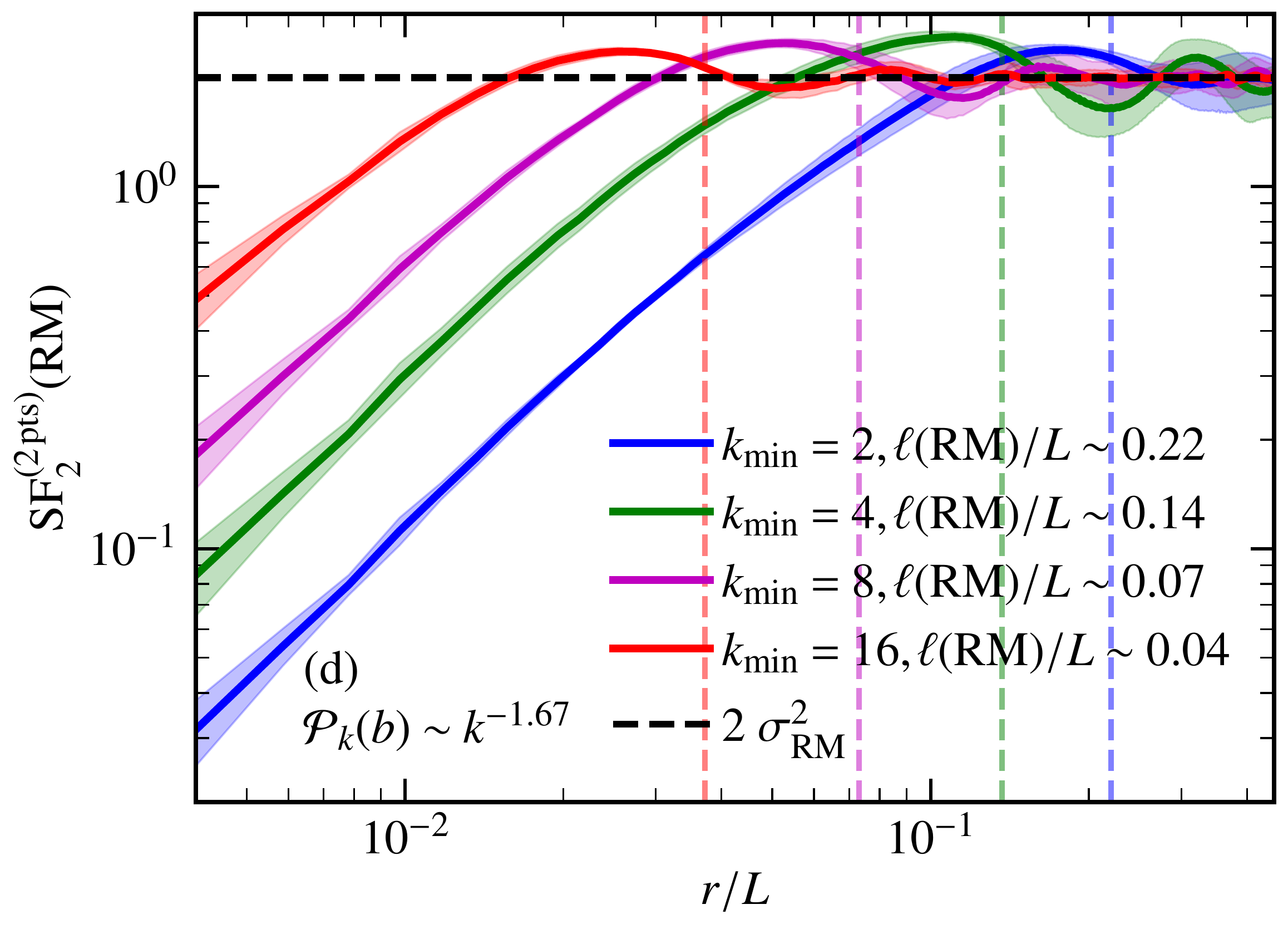}
    \caption{One-dimensional power spectrum ((a), (c)) and second-order structure function computed with two points ((b), (d)) for magnetic field ((a), (b)) and $\RM$ ((c), (d)) for a Gaussian random magnetic field with power spectrum, $\pk(b) \sim k^{-1.67}$, and different $k_{\rm min}$ (largest scale of the magnetic field). The corresponding correlation length of magnetic fields, $\ell(b)$, and $\RM$, $\ell(\RM)$, is given in the legends. The slope of $\RM$ power spectrum is $\approx 2.7$, as expected from \Sec{sec:simgrf}. The turnover scale of $\RM$ structure function is approximately equal to $\ell(\RM)$, which in turn is roughly equal to $\ell(b)$. This shows that the turnover scale of $\RM$ structure function probes the correlation length of the small-scale random magnetic fields.}
    \label{fig:grfkmin}
\end{figure*}

Here, we study how the turnover or break scale of $\RM$ structure function depends on the properties of the small-scale random magnetic fields. For a Gaussian random magnetic field with a power law magnetic power spectrum of slope $-1.67$, we vary the minimum wavenumber, $k_{\rm min}$ (which corresponds to the largest scale in the random field), keeping the size of the numerical domain, $L$, and the number of grid points, $512^{3}$, same. The power spectrum, $\pk(b)$ for $k_{\rm min} = 2, 4, 8,$ and $16$ is shown in \Fig{fig:grfkmin}a. We calculate the correlation length of the small-scale magnetic field, $\ell(b)$, using \Eq{eq:lbrm} with $\pk(b)$ and the corresponding values are given in the legend of \Fig{fig:grfkmin}a. In \Fig{fig:grfkmin}b, we show the second-order structure function computed with two points for the magnetic field and it shows that the scale at which the structure function turns over is approximately equal to $\ell(b)$. \Fig{fig:grfkmin}c and \Fig{fig:grfkmin}d shows the $\RM$ power spectrum and second-order structure function computed with two points for different $k_{\rm min}$. The corresponding correlation length of $\RM$ computed using \Eq{eq:lbrm} is given in the legend. The $\RM$ structure function roughly turns over at a scale that is approximately equal to the $\ell(\RM)$, which in turn is roughly equal to $\ell(b)$. Thus, for a Gaussian random magnetic field, the turnover scale of $\RM$ structure function is approximately the correlation length of the magnetic field. \revb{If the scale of the region probed is smaller than the correlation length of the magnetic field, the $\RM$ structure function does not turn over and does not saturate at the expected value.}

\section{Rotation measure structure function for uniform random magnetic fields and lognormal thermal electron density.} \label{sec:buni}

\begin{figure*}
    \includegraphics[width=\columnwidth]{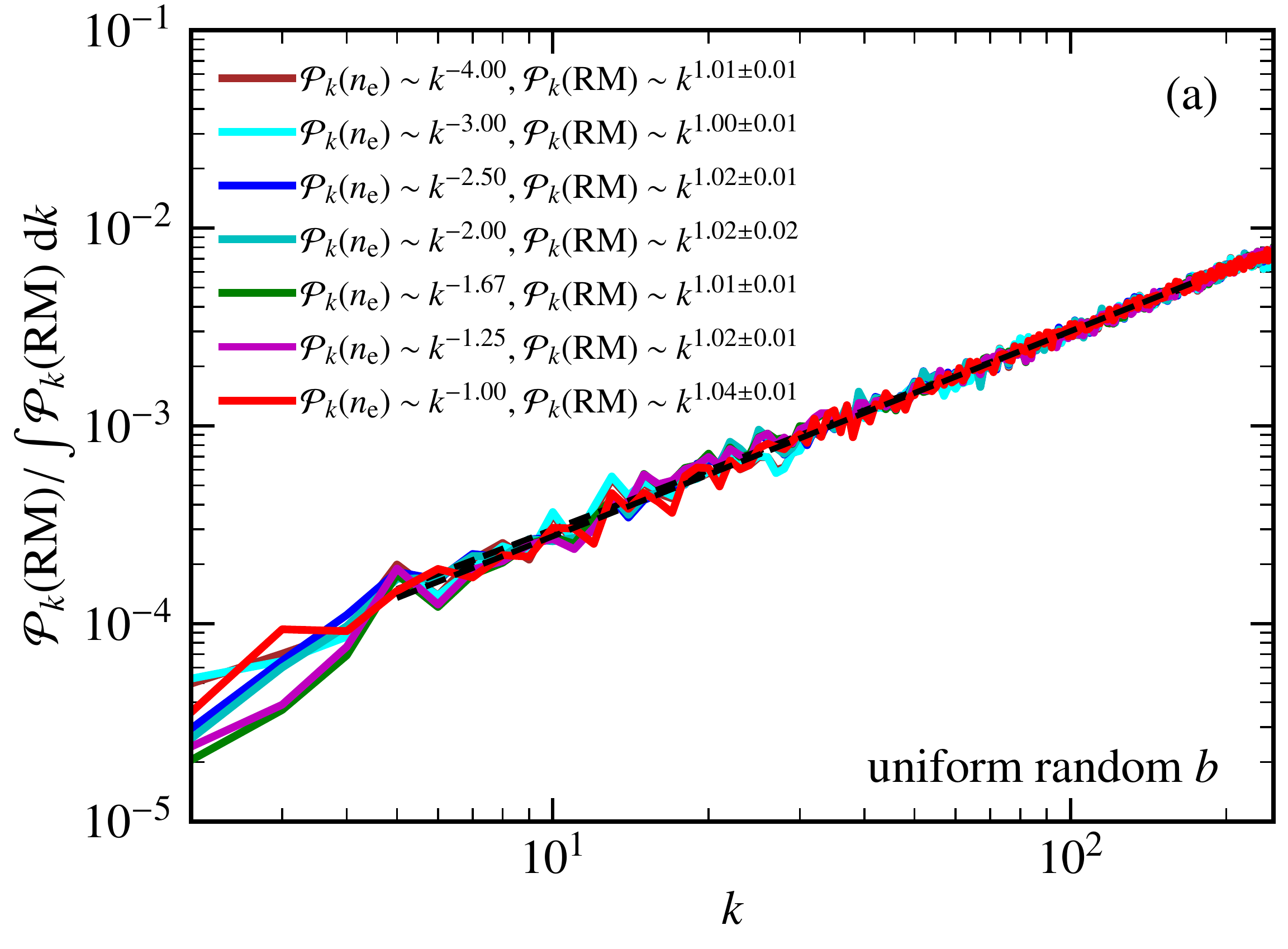} \hspace{0.5cm}
    \includegraphics[width=1.008\columnwidth]{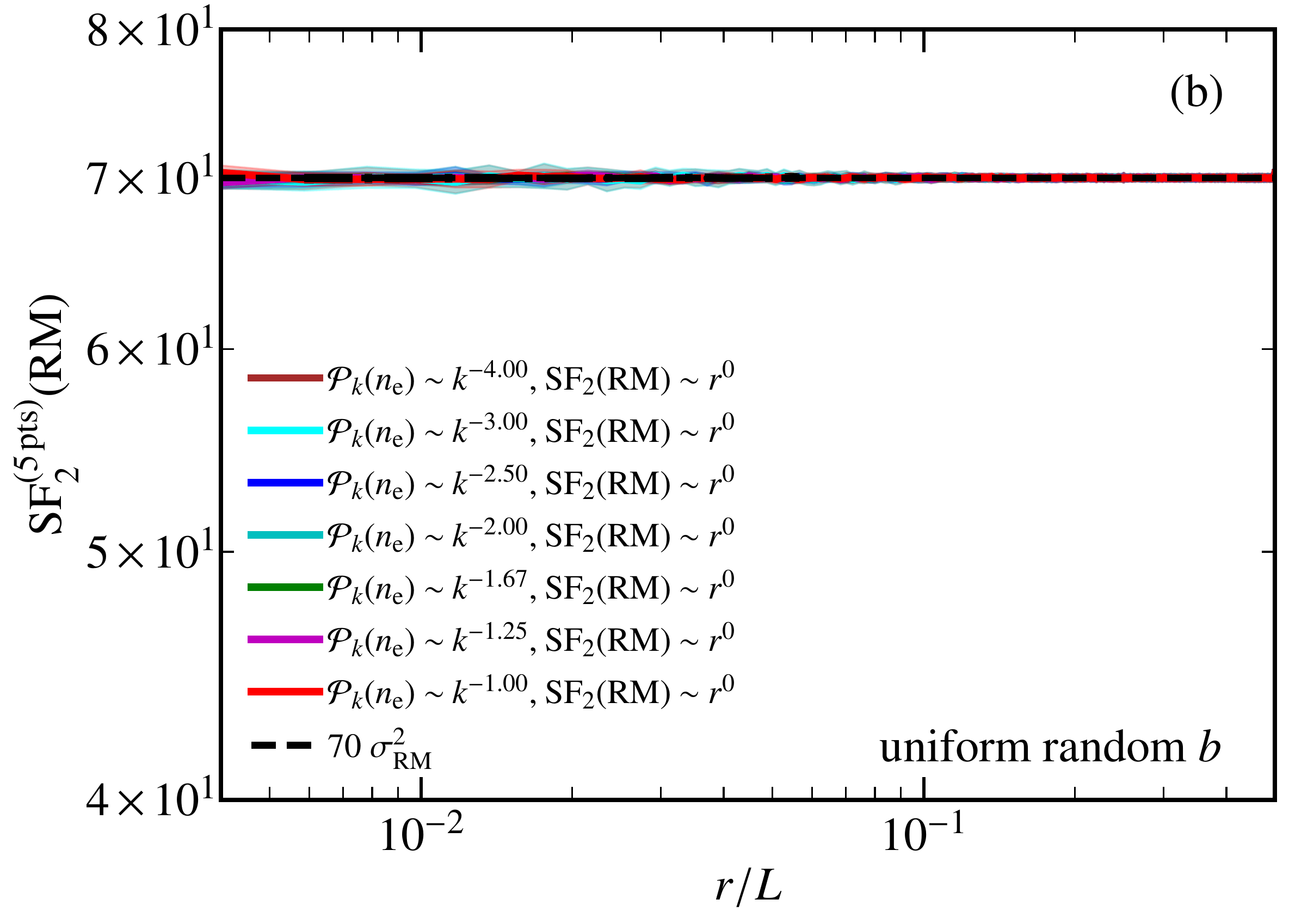}
 \caption{\rev{One-dimensional power spectrum (panel~a) and second-order structure function (panel~b), $\sffive$, computed with five-point stencils, for a random magnetic field, $b$, drawn from a uniform distribution (no particular power spectrum), and lognormal thermal electron densities, $\ne$, with a power-law spectrum and different values of the slope, $\gamma$, in the range $[1, 4]$. For all $\gamma$, the $\RM$ power spectrum follows a power law with slope $1$ and the $\RM$ structure function is flat at $70~\sigma^{2}_{\RM}$. This further shows that the scales of $\RM$ fluctuations are primarily controlled by the magnetic field spectra.}}
 \label{fig:buni}
\end{figure*}

\rev{Here, we explore $\RM$ structure functions for lognormal thermal electron densities with a power-law spectrum ($\pk(\ne) \sim k^{-\gamma}$ with $\gamma=[1, 4]$, normalised to the same $\langle \ne \rangle$ for all $\gamma$s) and for a random magnetic field drawn from a uniform distribution with no particular power spectrum (normalised to the same $\brms$ for all $\gamma$s). \Fig{fig:buni} shows the corresponding $\RM$ power spectra and second-order structure functions ($\sffive$, calculated with the highest number of points for the stencil, see \Sec{sec:simgrf}) for different values of $\gamma$. For all $\gamma$, the $\RM$ power spectrum follows a $k^{1}$ power law and the $\RM$ structure function is flat (at $70~\sigma^{2}_{\RM}$; see \Sec{sec:met}). Thus, we do not get any $\RM$ fluctuations for a uniform random magnetic field, even on varying the slope of the thermal electron density power spectrum (unlike \Sec{sec:simlnrf}, where we get $\RM$ fluctuations even for a constant $\ne$). This shows that the magnetic field spectrum and distribution are important for studying $\RM$ fluctuation scales (also see \App{sec:kmin}).
}

\section{Correlated magnetic fields and thermal electron density from compressible MHD simulations with zero mean field and weak random seed field} \label{sec:simdyn}
In this Appendix, we explore $\RM$ constructed from thermal electron densities and magnetic fields, which are numerical solutions of the MHD equations, not independent of each other, and can also be correlated. We use numerically driven MHD turbulence simulations, where we solve the continuity equation, induction equation, and Navier-Stokes equation with a prescribed forcing on a three-dimensional triply-periodic uniform grid consisting of $512^{3}$ grid points. We also consider explicit viscous and resistive terms. The flow is driven on the scale, $L/2$, where $L$ is the side length of the cubic numerical domain. The simulation is initialised with a very weak random seed field (with mean zero) and undergoes small-scale dynamo action. Thus, the magnetic field first grows exponentially and then saturates once it becomes strong enough to react back on the turbulent flow. The full details of these numerical simulations are given in Sec.~2 and Sec.~3 of \citet{SetaF2021}. The main parameter that controls the correlation between the thermal electron density and magnetic fields is the Mach number, $\Mach$, of the turbulent flow and the correlation increases with $\Mach$. In this paper, we use the thermal electron density and magnetic fields for four Mach numbers ($\Mach=0.1, 2, 5,$ and $10$) and all other parameters (such as viscosity, resistivity, solenoidal nature of the turbulent forcing, driving scale of turbulence, and length of the domain) remain the same across all Mach numbers. These simulations are labelled as $\Mach0.1, \Mach2.0, \Mach5.0,$ and $\Mach10.0$. Moreover, for each $\Mach$, we take $10$ statistically independent snapshots in time, once the field has saturated, for statistical averaging. 

\begin{figure*}
    \includegraphics[width=2\columnwidth]{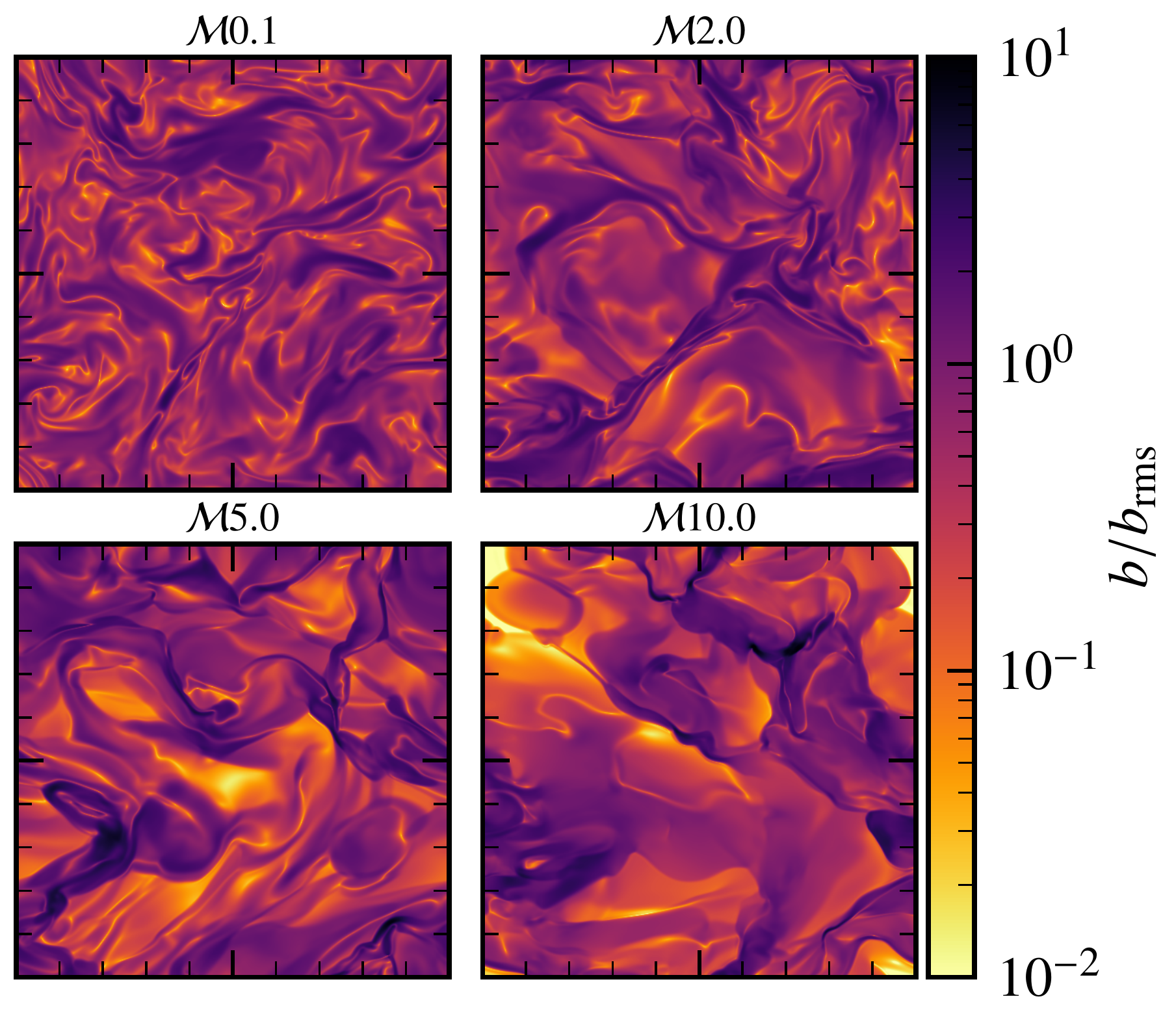}
    \caption{Two-dimensional slices of normalised magnetic field strengths, $b/\brms$, obtained from the MHD simulations at various different Mach numbers, $\Mach=0.1, 2, 5, 10$. Magnetic fields have complex structures and spatially vary over many orders of magnitude in comparison to the Gaussian random magnetic fields explored in \Sec{sec:simgrf} (c.f.~\Fig{fig:grfb2d}).}
    \label{fig:dynb2d}
\end{figure*}

\begin{figure*}
    \includegraphics[width=2\columnwidth]{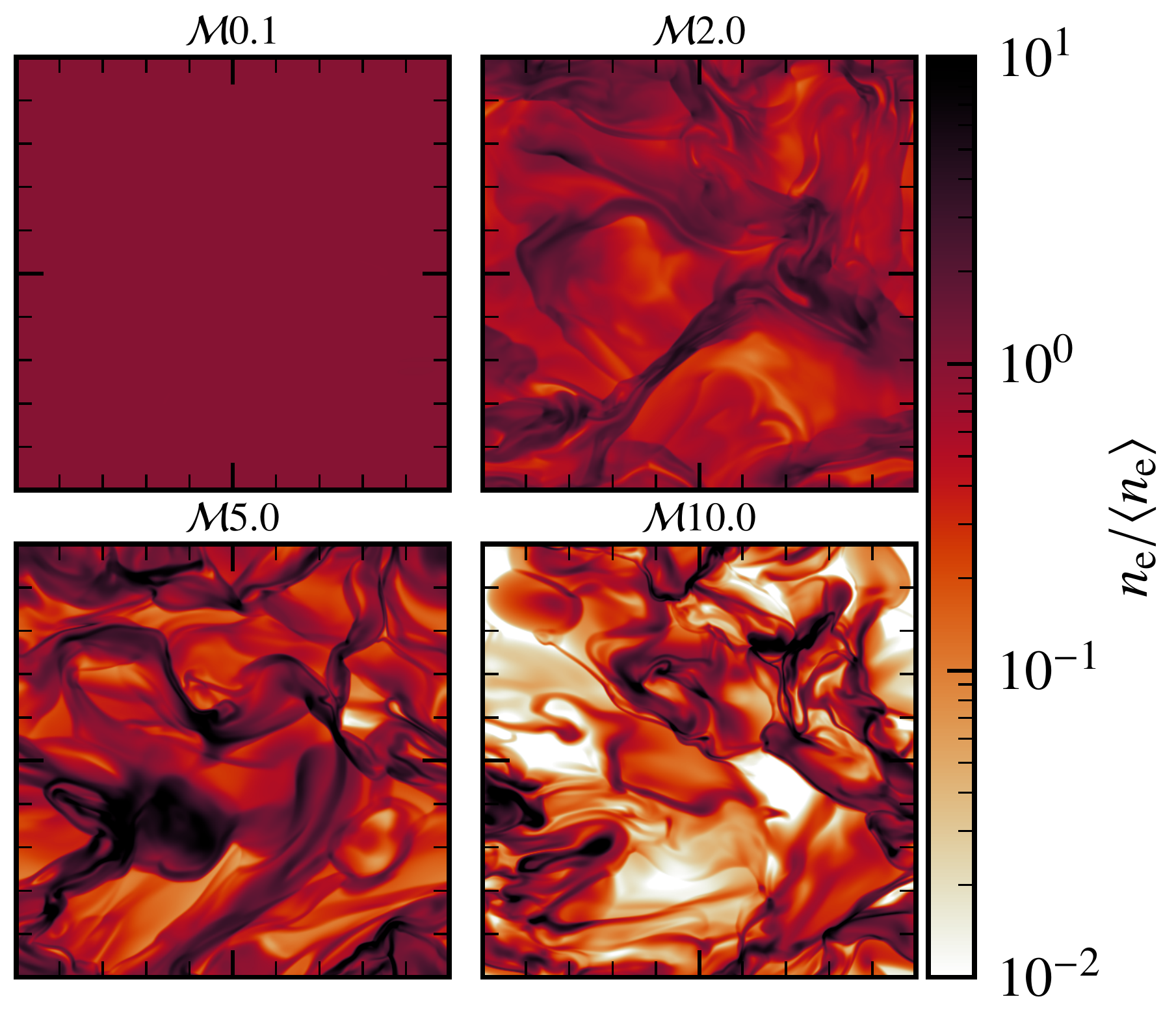}
    \caption{Two-dimensional slices of normalised thermal electron densities, $\ne/\langle \ne \rangle$, obtained from the MHD simulations at various different Mach numbers, $\Mach=0.1, 2, 5, 10$. For $\Mach0.1$, the thermal electron density is practically constant. As the Mach number increases, the thermal electron densities spatially vary over a large range of scales and the structures show correlation with the magnetic structures in \Fig{fig:dynb2d}.}
    \label{fig:dynne2d}
\end{figure*}

\begin{figure*}
    \includegraphics[width=2\columnwidth]{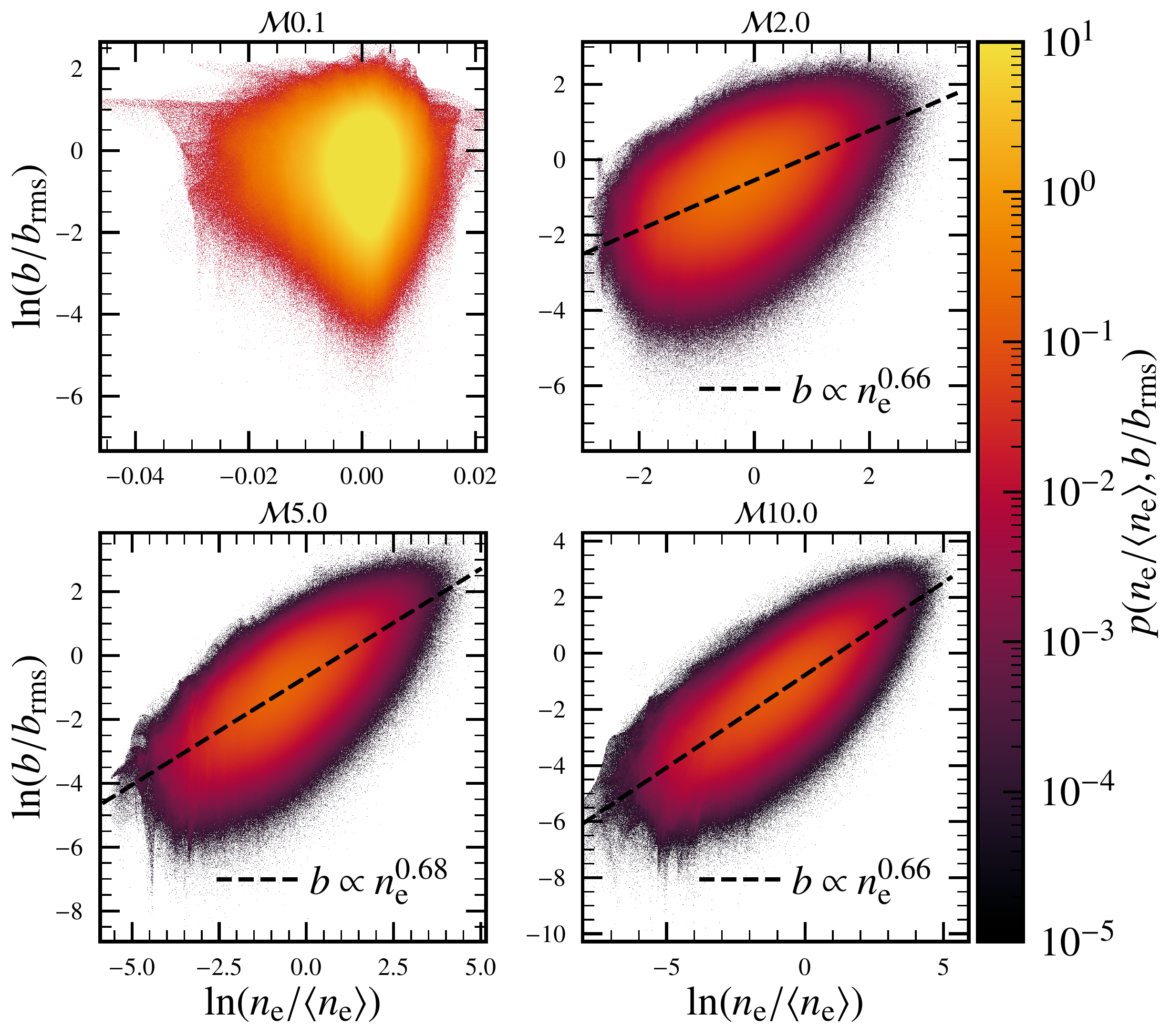}
    \caption{Two-dimensional PDF of normalised thermal electron density and magnetic fields for various Mach numbers. No significant correlation is seen for the $\Mach0.1$ case, but as the Mach number increases, the correlation between the two quantities is enhanced. For $\Mach=2, 5, 10$, the best-fit line (dashed, black line; \rev{obtained using the least-squares fitting procedure}) approximately follows the flux freezing constraint, $b \propto \ne^{2/3}$. However, as the Mach number increases, the points fit the constraint more tightly, which shows that the correlation increases with $\Mach$.}
    \label{fig:dynbnecorr}
\end{figure*}

\begin{figure*}
    \includegraphics[width=\columnwidth]{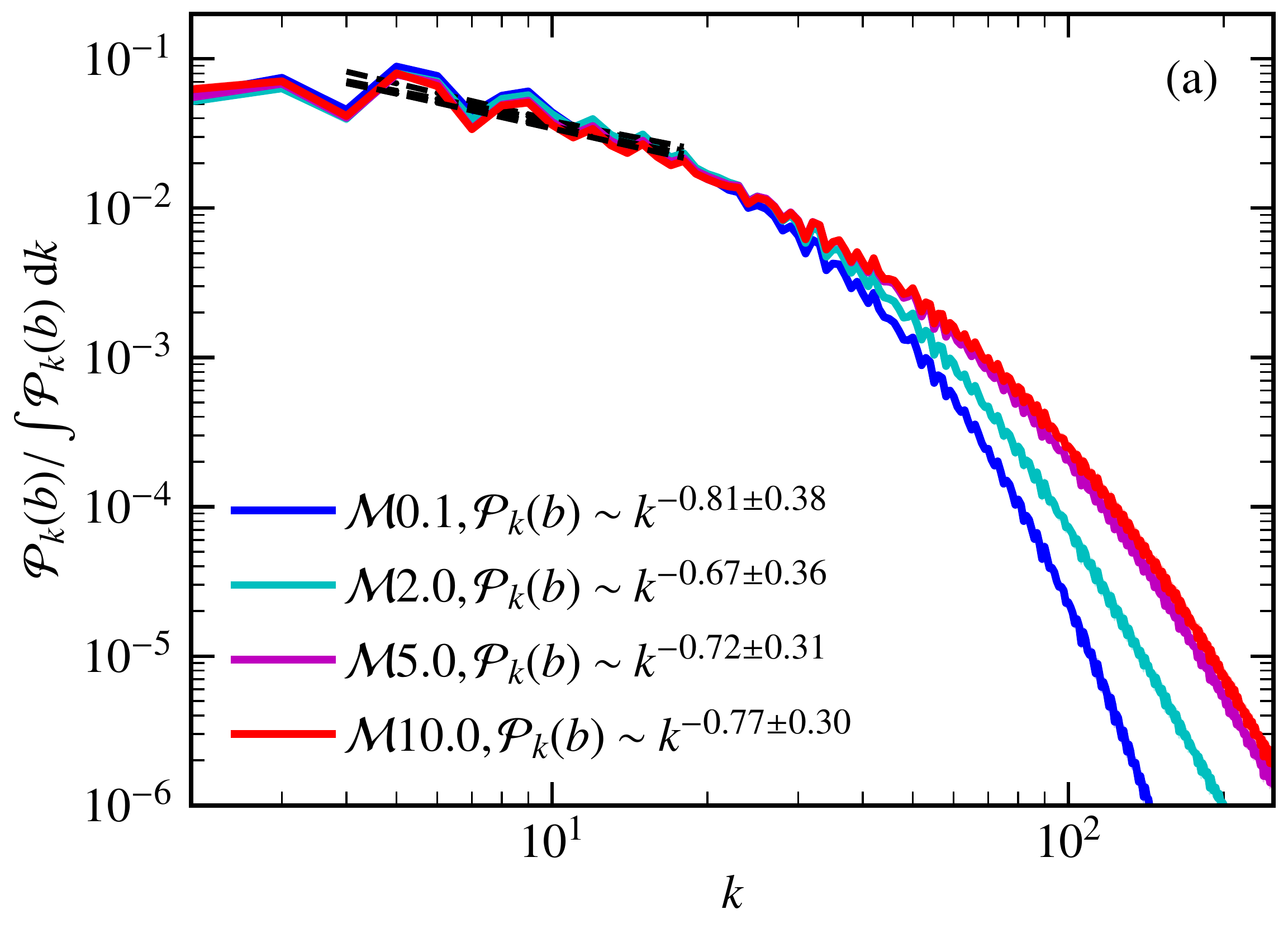} \hspace{0.5cm}
    \includegraphics[width=\columnwidth]{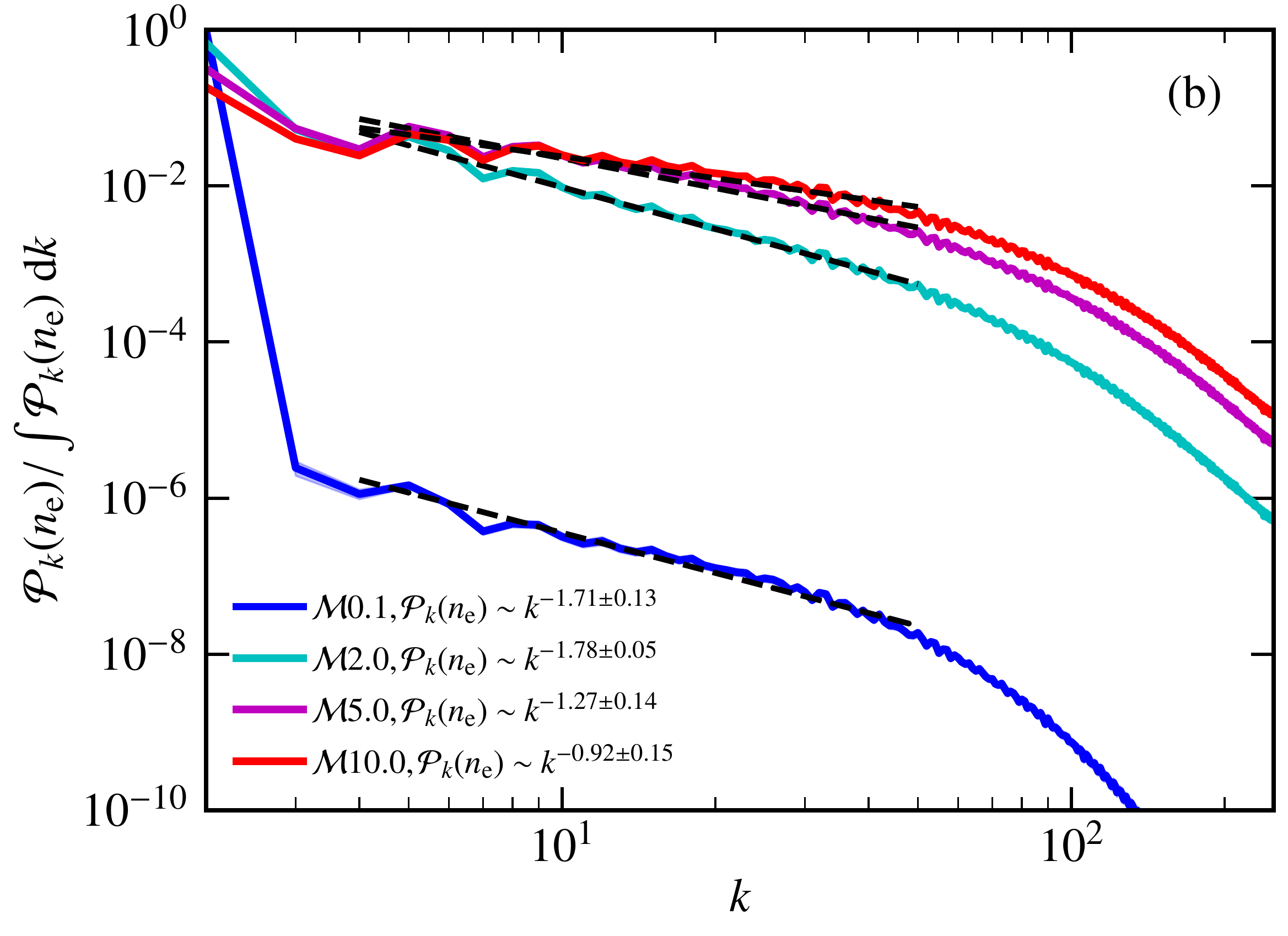}
    \caption{One-dimensional shell-averaged power spectrum of magnetic fields (a) and thermal electron densities (b) for all Mach numbers (averaged over $10$ statistically independent time snapshots). The spectral slope of the magnetic field is largely independent of the Mach number on larger scales ($k \lesssim 20$), but the slope of the thermal electron density spectrum becomes shallower with increasing Mach number.}
    \label{fig:dynspec}
\end{figure*}

\Fig{fig:dynb2d} and \Fig{fig:dynne2d} shows two-dimensional slices of normalised magnetic field strengths, $b/\brms$, and thermal electron densities, $\ne/\langle \ne \rangle$, centered on the middle of the three-dimensional numerical domain for different Mach numbers. The magnetic field is random but shows complex structures. It also spatially varies over a much larger range of magnetic field strengths than the Gaussian random magnetic fields in \Fig{fig:grfb2d}. The thermal electron density is practically constant for $\Mach=0.1$ (subsonic case), but varies spatially as the Mach number increases. Even visually, the structures in thermal electron densities and magnetic fields seem correlated for $\Mach=2, 5,$ and $10$. The correlation and enhancement in correlation with $\Mach$ is shown by two-dimensional PDFs of $\ne$ and $b$ in \Fig{fig:dynbnecorr}. For $\Mach=0.1$, the thermal electron density does not change much but the magnetic field varies significantly (see \Fig{fig:dynb2d}), so both distributions are largely uncorrelated. However, as the Mach number increases, the correlation between $\ne$ and $b$ is enhanced. The best-fit line for $\Mach=2, 5,$ and $10$ (\Fig{fig:dynbnecorr}) approximately agrees with the relationship obtained from the flux-freezing constraint, $b \propto \ne^{2/3}$, and more number of points tightly follow the relation for higher Mach numbers (compare the distribution of points for the $\Mach2.0$ and $\Mach10.0$ cases in \Fig{fig:dynbnecorr}). 

In \Fig{fig:dynspec}, we show power spectra of the magnetic field and the thermal electron density for different Mach numbers. Unlike our numerical experiments in \Sec{sec:simgrf} and \Sec{sec:simlnrf}, the inertial range in these MHD simulations is quite small. The magnetic field spectra at larger scales have roughly the same slope across all Mach numbers. By contrast, the slope of the thermal electron density power spectrum becomes shallower as the Mach number increases. This shows that power at a smaller scale for the thermal electron density increases as the Mach number increases. This result agrees with previous studies of compressible hydrodynamic turbulence \citep{KimR2005} and MHD turbulence with a mean field \citep{FederrathK2013}.

\begin{figure}
    \includegraphics[width=\columnwidth]{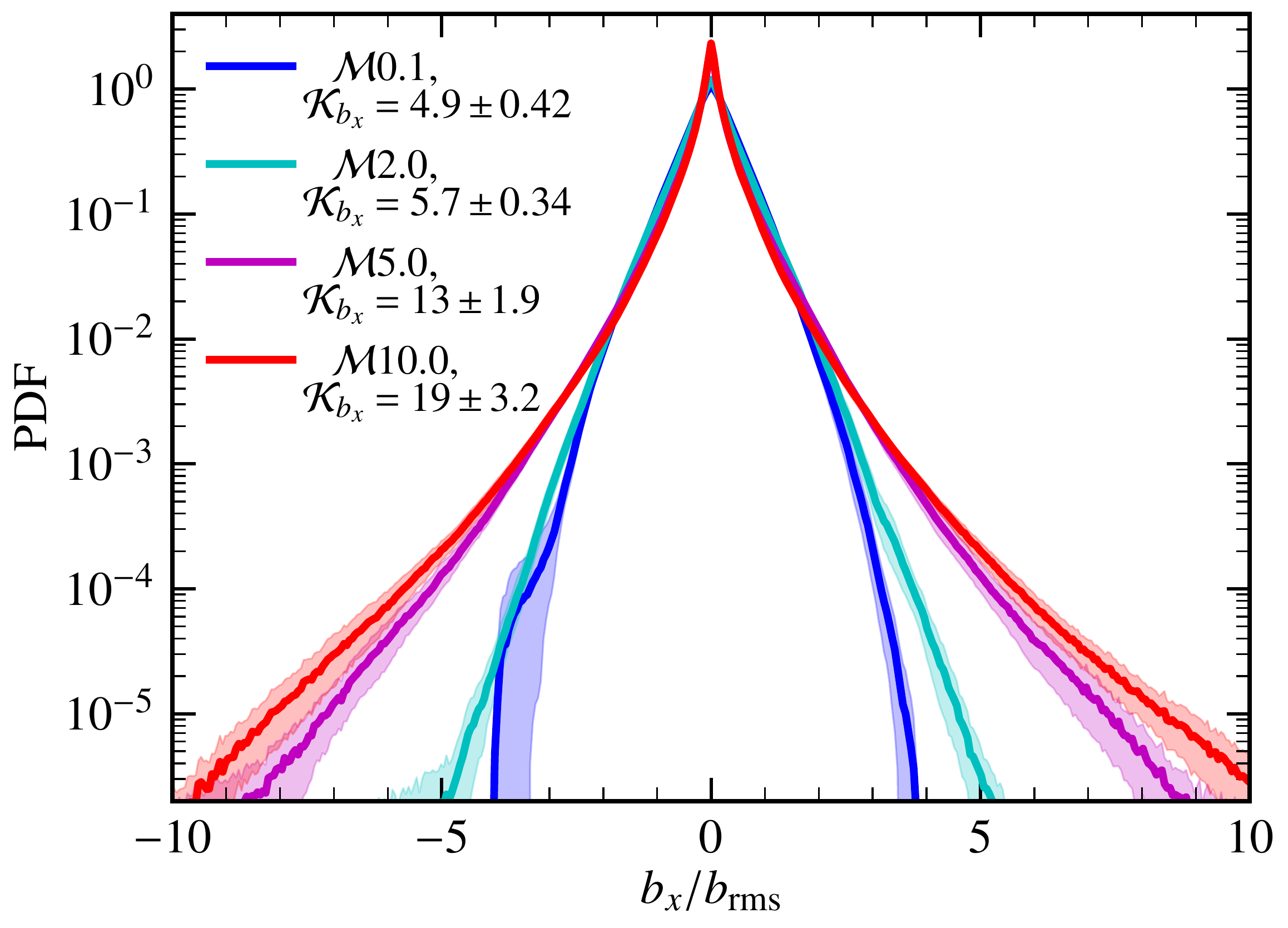}
    \caption{PDF of a single component of the magnetic field at $\Mach=0.1, 2, 5,$ and $10$. The lines and shaded regions show the mean and standard deviation obtained after averaging over $10$ snapshots. The distributions are non-Gaussian and the computed kurtosis (given in the legend) increases with the Mach number. This shows that the magnetic field distribution becomes more non-Gaussian as the Mach number increases. }
    \label{fig:dynbpdf}
\end{figure}

\begin{figure*}
    \includegraphics[width=2\columnwidth]{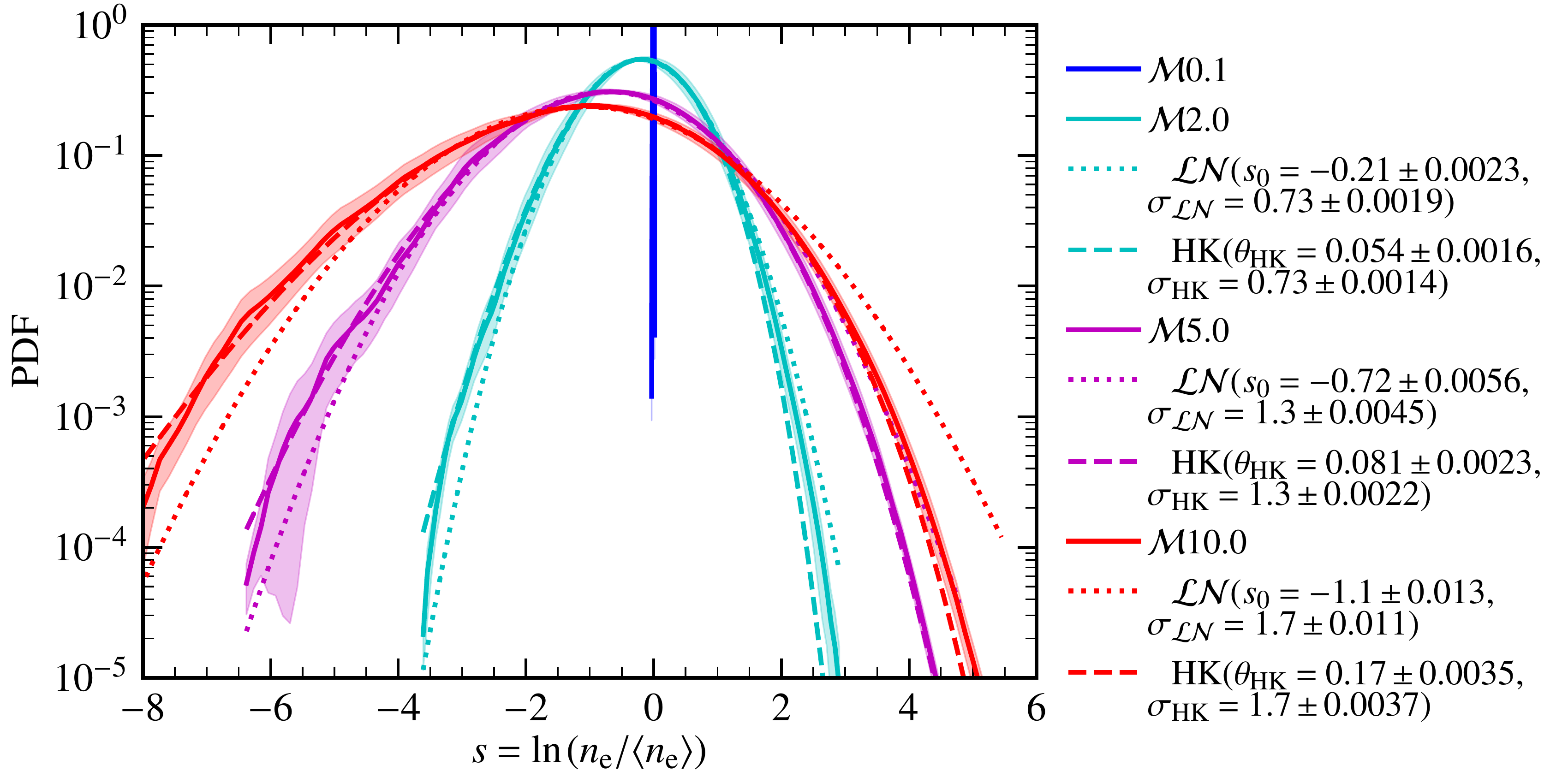}
    \caption{PDF of $s = \ln(\ne/\langle \ne \rangle$ for $\Mach=0.1, 2, 5, 10$ (solid lines show the mean and the shaded region show the one-sigma fluctuations, when averaged over $10$ statistically independent time snapshots). The thermal electron density is practically a delta function for the $\Mach0.1$ case. For $\Mach=2, 5, 10$, $\ne / \langle \ne \rangle$ varies by orders of magnitude. We fit the lognormal (\Eq{eq:lognormal}, dotted line) and Hopkins PDF model (\Eq{eq:Hopkins}, dashed line), and show that the Hopkins PDF model provides a better fit, especially at higher Mach numbers and at higher values of $s$. The parameter, $\theta_{\rm HK}$, is a measure of the spatial intermittency of the thermal electron density and it increases with increasing Mach number.}
    \label{fig:dynnepdf}
\end{figure*}

\begin{table*} 
\caption{Fit parameters for the lognormal and Hopkins fits of the thermal electron density PDF (\Fig{fig:dynnepdf}, except for the $\Mach0.1$ case, where the thermal electron density is practically a delta function) and the computed kurtosis for the magnetic field PDF (\Fig{fig:dynbpdf}). The columns are: 1.~Mach number of the simulation, $\Mach$, 2.~mean in the lognormal fit, $s_{0}$, 3.~variance in the lognormal fit, $\sigma_{\mathcal{LN}}$, 4.~$-\sigma_{\mathcal{LN}}^{2} / 2$, which is roughly equivalent to column~2 due to the mass conservation constraint, 5.~intermittency parameter in the Hopkins PDF model, $\theta_{\rm HK}$, 6.~variance in the Hopkins PDF model, $\sigma_{\rm HK}$, and 7.~kurtosis of the x-component of the magnetic field, $\ku_{b_{x}}$ (the $y$- and $z$-components are very similar to the $x$-component, as the simulations analysed here do not have a mean field). \rev{Further in column~8, we provide the variance of $s$, $\sigma_{s}$, obtained using the analytical expression, \Eq{eq:densvar}. The variance estimated from the fit (column~3) is close to that obtained using the analytical expression (column~8).}}
\label{tab:nebfits}
\begin{tabular}{lcccccccc} 
\hline
 $\mathcal{M}$ & $s_0$ & $\sigma_{\mathcal{LN}}$ & $-\sigma_{\mathcal{LN}}^2/2$ & $\theta_{\rm HK}$ & $\sigma_{\rm HK}$ & $\mathcal{K}_{b_x}$ & $\sigma_{s}$ \\ 
 \hline
$ 0.1$& $--$& $--$& $--$& $--$& $--$& $4.9 \pm 0.42$ & $0.03$ \\ 
$ 2.0$& $-0.21 \pm 0.0023$& $0.73 \pm 0.0019$ & $-0.27$ & $0.054 \pm 0.0016$& $0.73 \pm 0.0014$& $5.7 \pm 0.34$ & $0.6$\\
$ 5.0$& $-0.72 \pm 0.0056$ & $1.3 \pm 0.0045$ & $-0.84$ &  $0.081 \pm 0.0023$& $1.3 \pm 0.0022$& $13 \pm 1.9$ & $1.2$ \\
$10.0$& $-1.1 \pm 0.013$ & $1.7 \pm 0.011$ & $-1.4$ & $0.17 \pm 0.0035$& $1.7 \pm 0.0037$& $19 \pm 3.2$ & $1.6$ \\
\hline
\end{tabular}
\end{table*}

 \Fig{fig:dynbpdf} shows the PDF of a single component of the magnetic fields, $b_{x}/\brms$, at different Mach numbers. The distribution is far from Gaussian (even for the $\Mach0.1$ case), and this is also confirmed by the computed kurtosis (using \Eq{eq:kurt} but for $b_{x}/\brms$), which is $>3$ (kurtosis of a Gaussian distribution is equal to 3) for all Mach number cases. The kurtosis increases with the Mach number and this shows that the non-Gaussianity in the magnetic field increases with increasing $\Mach$. Thus, the magnetic field PDFs in these MHD simulations are highly non-Gaussian; a characteristic of small-scale dynamo-generated magnetic fields \citep[see][]{SetaEA2020}.
 
 In \Fig{fig:dynnepdf}, we show the $\ne / \langle \ne \rangle$ PDFs, and fit them with a lognormal \citep{Vazquez1994,PassotV1998,FederrathEA2008} and Hopkins \citep{Hopkins2013,FederrathB2015} model (except for the $\Mach0.1$ case, where $\ne$ does not vary much). For the lognormal distribution, we fit $s = \ln(\ne/\langle \ne \rangle)$ with the following distribution,
\begin{align} \label{eq:lognormal}
\mathcal{LN}(s) = \frac{1}{\left(2 \pi \sigma_{\mathcal{LN}}^{2}\right)^{1/2}} \exp\left(-\frac{(s-s_{0})^{2}}{2 \sigma_{\mathcal{LN}}^{2}}\right),
\end{align}
where $s_{0}$ and $\sigma_{\mathcal{LN}}$ are the mean and variance of the distribution, respectively, to be determined from the fit. Also, due to mass conservation, $s_{0} = - {\sigma_{\mathcal{LN}}}^{2}/2$ \citep{Vazquez1994}. For the Hopkins model, which includes the effects of intermittency, we fit $s$ with the following distribution, 
\begin{align} \label{eq:Hopkins}
{\rm HK}(s) = I_{1} \left(2 ~(\lambda \omega(s))^{-1/2}\right) \exp[-(\lambda + \omega(s))] \left(\frac{\lambda}{\theta_{\rm HK}^{2} \omega(s)}\right)^{-1}, \\ \nonumber
\lambda \equiv \sigma_{\rm HK}^{2} / (2~ \theta_{\rm HK}^{2}), \quad \omega(s) \equiv \lambda/(1 + \theta_{\rm HK}) - s/\theta_{\rm HK}, \, (\omega \ge 0),
\end{align}
where $I_{1}$ is the modified Bessel function of the first kind, $\sigma_{\rm HK}$ is the variance, and $\theta_{\rm HK}$ is the intermittency parameter. For $\theta_{\rm HK} \rightarrow 0$, the Hopkins PDF model is equivalent to the lognormal distribution, and the higher the $\theta_{\rm HK}$, the larger the deviation from the lognormal distribution \citep[i.e., stronger intermittency; see also][]{KritsukEA2007,FederrathEA2010,KonstandinEA2012,FederrathEA2021}. Both $\sigma_{\rm HK}$ and $\theta_{\rm HK}$ are to be determined from the fit.

In \Fig{fig:dynnepdf}, we show the PDF of $s$ of the simulations together with the fitted distributions for $\Mach=2, 5, $ and $10$. The best-fit parameters for both fit models are given in \Tab{tab:nebfits}. For the lognormal distribution, as expected, $s_{0} \approx - {\sigma_{\mathcal{LN}}}^{2}/2$. For the Hopkins PDF model, $\theta_{\rm HK}$ increases as the Mach number increases. The Hopkins PDF model provides a better fit to $\ne$ than the lognormal distribution. This is especially true for higher values of $\ne / \langle \ne \rangle$, where the lognormal PDF overestimates the data. Most of the previous simulations studied this in the absence of a magnetic field, or in the presence of a weak mean magnetic field \citep{Hopkins2013} (although Beattie et al., 2021, in prep., is currently exploring the intermittency in the case of a strong mean field). We show that the statistics of the density field are similar to the cases studied in \citet{Hopkins2013}, but here in the absence of a mean field.

\rev{Besides the $\ne$ PDF, the density variance can also be computed analytically as \citep{PadoanN2011,PriceEA2011,MolinaEA2012,KonstandinEA2012,FederrathKlessen2012,FederrathB2015},
\begin{align} \label{eq:densvar}
\sigma_{s}  =  \sqrt{\ln\left(1 + \frac{\plasmabeta}{\plasmabeta +1} b^2 \Mach^{2}\right)},
\end{align}
where $\plasmabeta$ is the plasma beta \citep[values computed in Table~1 of][]{SetaF2021}, $b$ is the driving parameter \citep[$b\sim1/3$ for our purely solenoidally driven turbulence; see][]{FederrathEA2008}, and $\Mach$ is the Mach number. We provide $\sigma_{s}$ computed using \Eq{eq:densvar} in \Tab{tab:nebfits}, which is close to that obtained from fitting the $\ne$ distributions ($\sigma_{\mathcal{LN}}$).}

\begin{figure*}
    \includegraphics[width=2\columnwidth]{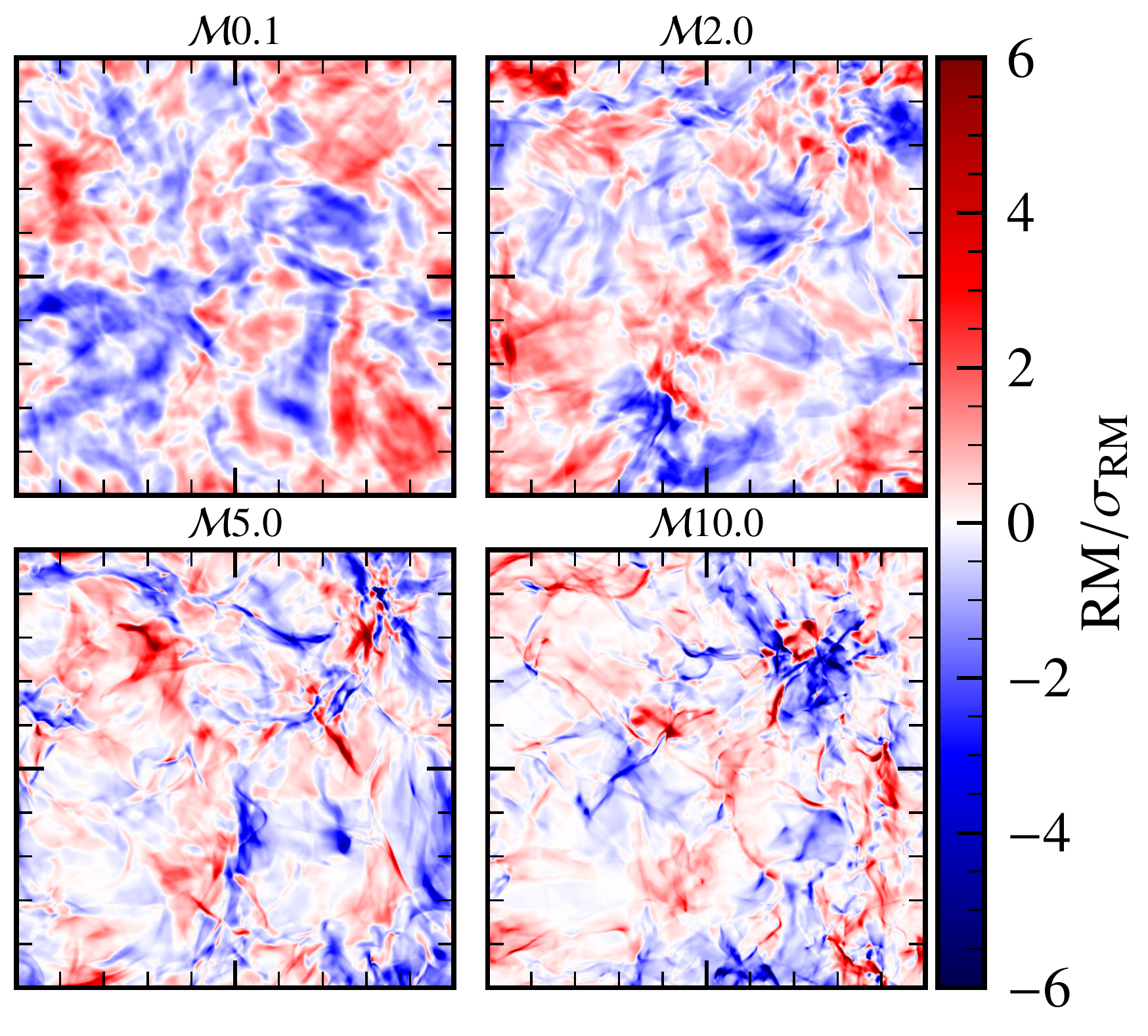}
    \caption{$\RM$ maps obtained using the thermal electron density and magnetic fields at four Mach numbers, $\Mach=0.1, 2, 5,$ and $10$. The $\RM$ structures look filamentary, especially for higher Mach numbers and at higher values of $|\RM|$.}
    \label{fig:dynrm}
\end{figure*}

\begin{figure*}
    \includegraphics[width=\columnwidth]{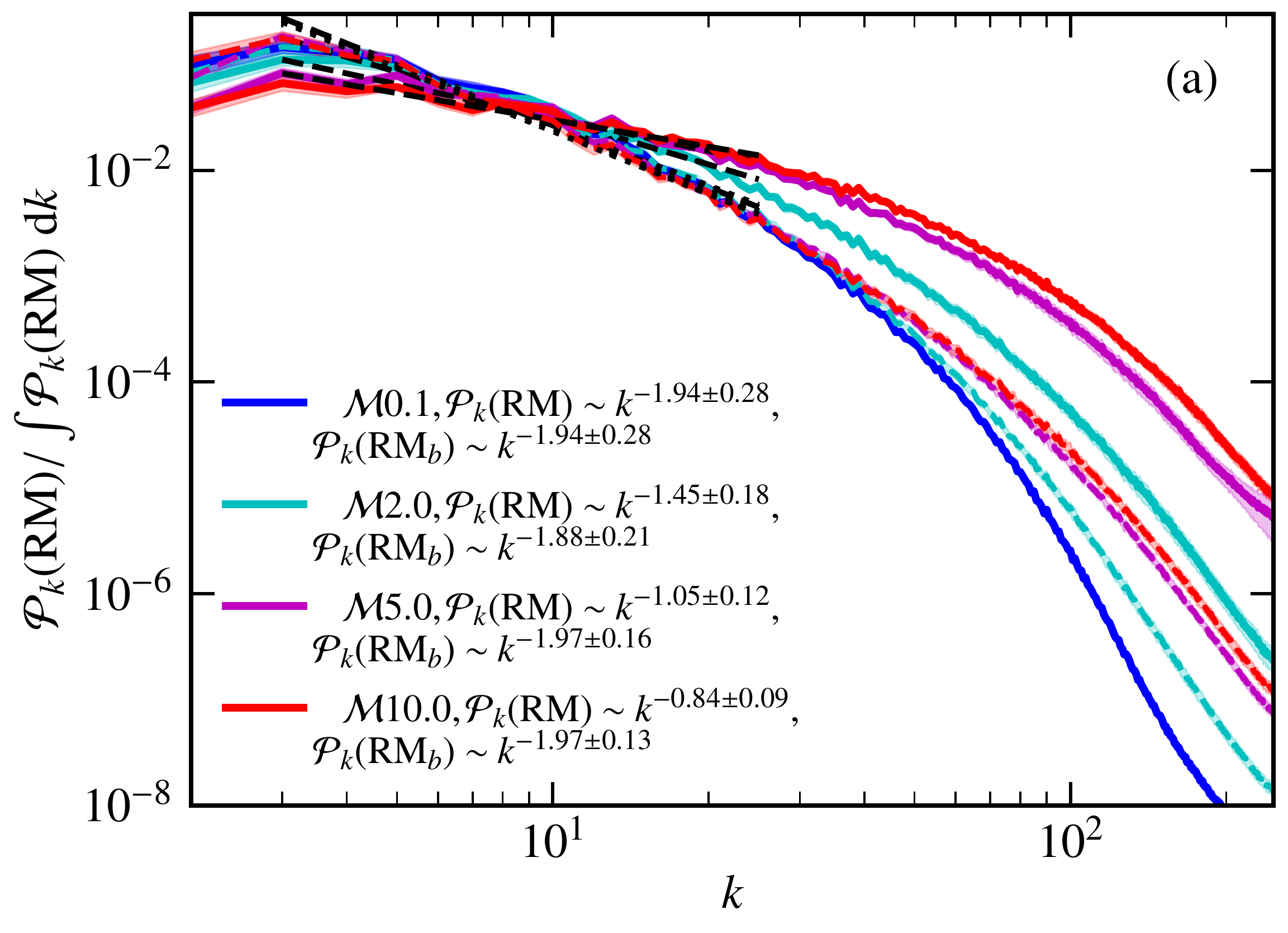} \hspace{0.5cm}
    \includegraphics[width=\columnwidth]{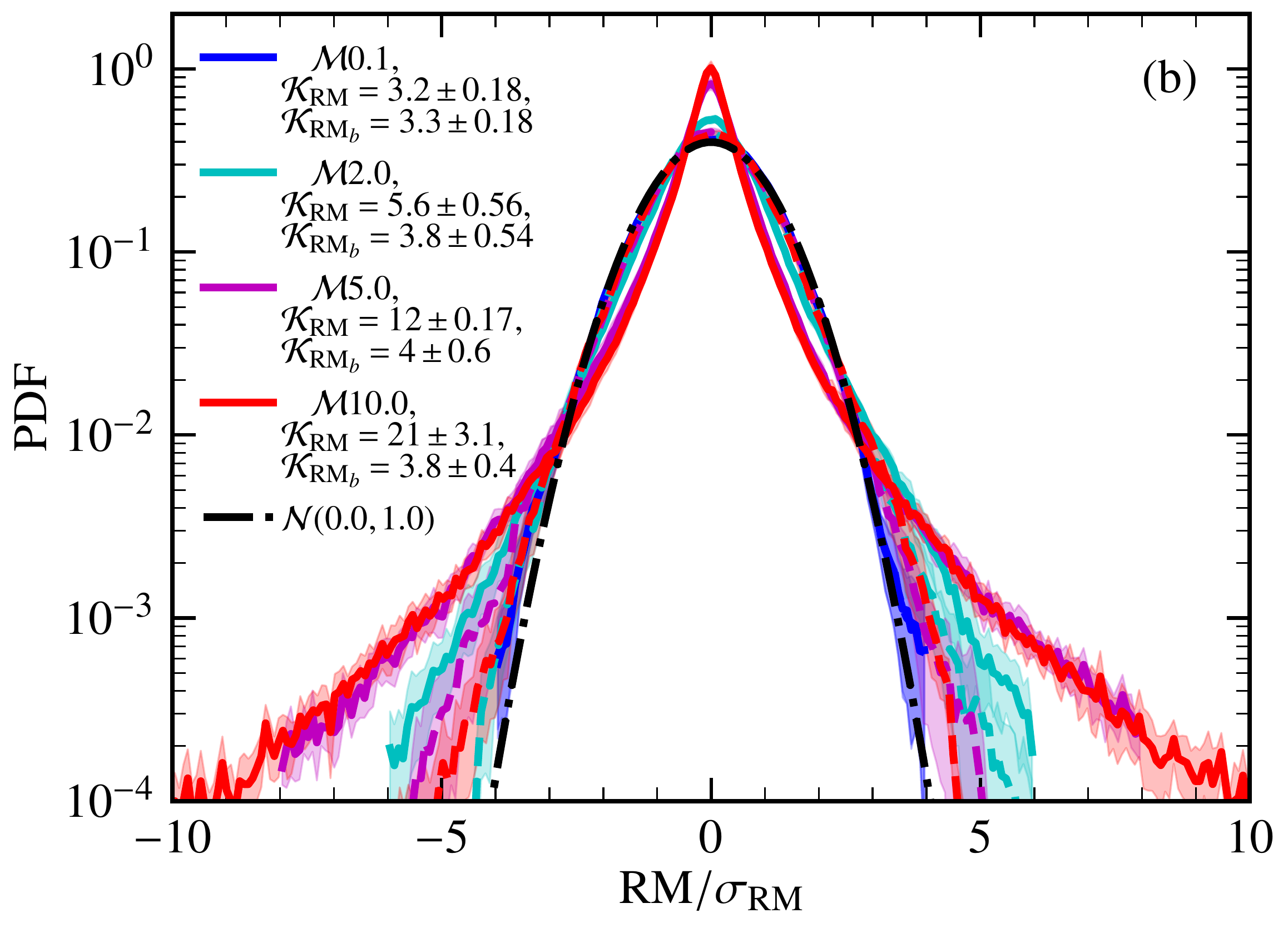}
    \caption{$\RM$ spectra (a) and PDFs (b) of $\RM$ (computed using \Eq{eq:rm}, solid lines) and $\RM_{b}$ (computed using \Eq{eq:rmb}, without considering $\ne$, dashed lines) for MHD simulations with all four Mach numbers. For all Mach numbers, the slope of the spectra of $\RM_{b}$ (at larger scales) is roughly the same as that of $\RM$ and the $\RM_{b}$ PDF roughly follows a Gaussian distribution. On the other hand, for $\RM$, the slope of the magnetic spectrum becomes shallower as the Mach number increases and also the non-Gaussianity of the $\RM$ PDF is enhanced.}
    \label{fig:dynrmspecpdf}
\end{figure*}

After studying the spectra and PDFs of $\ne$ and $b$ from the MHD simulations, we now compute $\RM$ from them using \Eq{eq:rm}. \Fig{fig:dynrm} shows the $\RM$ maps from all four Mach numbers. The $\RM$ structures are filamentary (especially at a higher value of $\RM/\sigma_{\RM}$) and this is due to elongated structures in the magnetic fields and thermal electron densities \rev{along the line of sight} (specifically at high $\Mach$). To study the effect of $\ne$ and its correlation with the magnetic fields on the $\RM$ spectra and the PDFs, for all Mach numbers, we also compute 
\begin{align} \label{eq:rmb}
\frac{\RM_{b}}{\rad~\m^{-2}} = 0.812 \int_{L_{\rm pl}/\pc} \frac{1}{\cm^{-3}} \,  \frac{b_{\parallel}}{\muG} \, \dd \left(\frac{l}{\pc}\right), 
\end{align}
where we assume $\ne/\langle \ne \rangle=\mathrm{const}$, and all other quantities are same as in \Eq{eq:rm}. Now, we compare the spectra and PDFs of $\RM$ and $\RM_{b}$.

\Fig{fig:dynrmspecpdf} shows the spectra and PDFs of $\RM$ and $\RM_{b}$ for $\Mach=0.1, 2, 5, 10$. The spectrum for the $\Mach0.1$ case is roughly the same for both $\RM$ and $\RM_{b}$. The power spectrum of $\RM_{b}$ does not change much with the Mach number. This is because the magnetic field power spectra, at larger scales, have a very similar slope for all Mach numbers (\Fig{fig:dynspec}a). For $\Mach=2, 5, 10$, the $\RM_{b}$ power spectrum is different from the $\RM$ power spectrum, the slope for $\RM$ is shallower than $\RM_{b}$. Also, the $\RM$ power spectrum becomes shallower as the Mach number increases. This is because of the contribution of $\ne$ to $\RM$ (see \Fig{fig:dynspec}b for the power spectrum of $\ne$). \Fig{fig:dynrmspecpdf}b shows the PDF of $\RM$ and $\RM_{b}$ for all Mach numbers. The PDF of $\RM_{b}$ follows a near Gaussian distribution  ($\ku_{\RM_{b}} \approx 3$) for all Mach numbers, whereas the PDF of $\RM$ is non-Gaussian and the level of non-Gaussianity increases with the Mach number ($\ku_{\RM}$ increases with $\Mach$). Thus, even for a non-Gaussian magnetic field (see \Fig{fig:dynbpdf}), the PDF of $\RM$ is approximately a Gaussian distribution if the thermal electron density is assumed to be a constant. The non-Gaussianity in the PDF of $\RM$ is primarily due to $\ne$ and correlated $\ne-b$ structures \citep[also, see Fig. 3~(a) and Table 2 in][]{SetaF2021}. Thus, the thermal electron density and its correlation with magnetic fields affect both the spectra (and by extension the structure function) and PDF of $\RM$.

\section{Models for the Milky Way $\RM$ contribution} \label{sec:rmmw}
\begin{figure*}
    \includegraphics[width=\columnwidth]{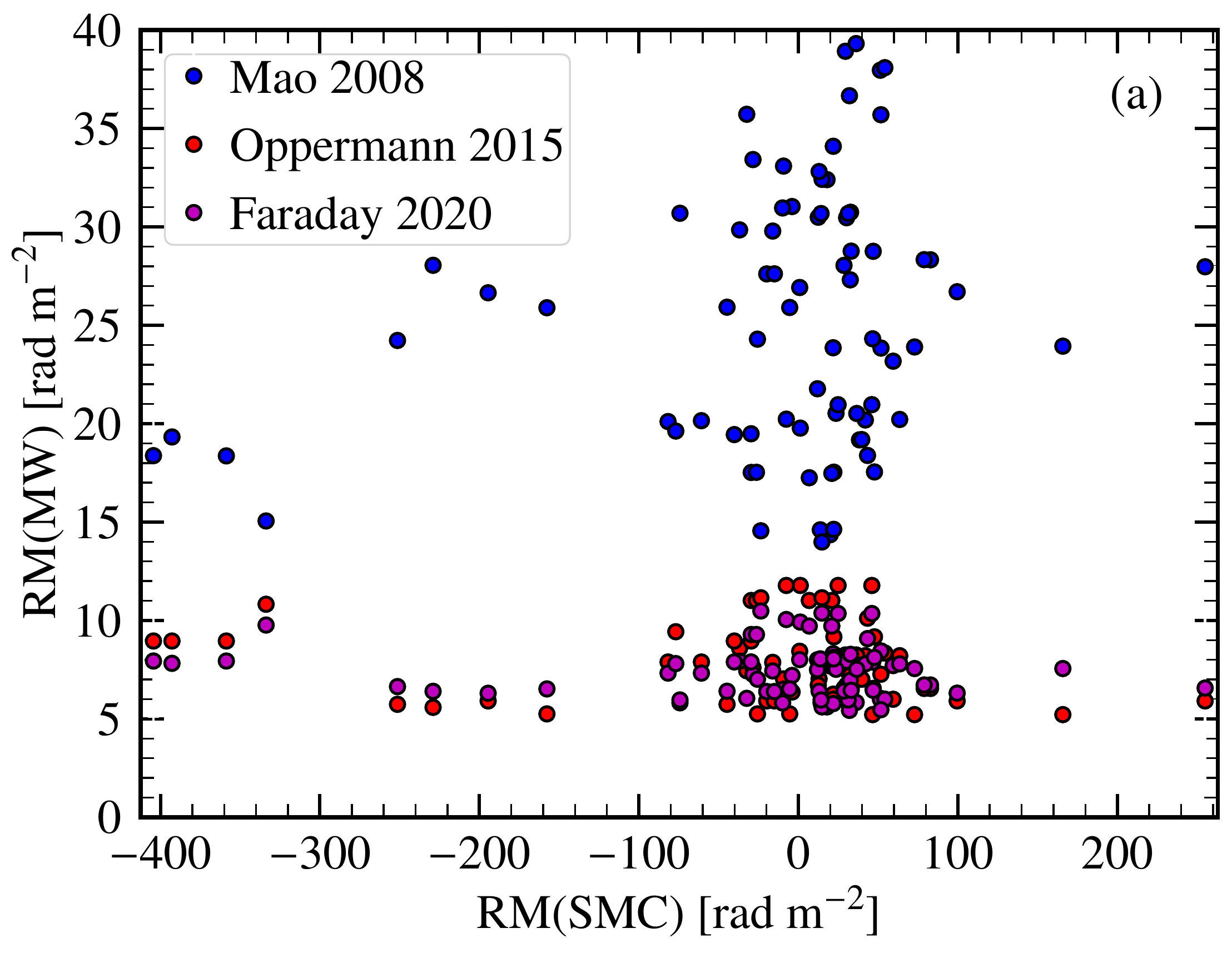} \hspace{0.5cm}
    \includegraphics[width=1.008\columnwidth]{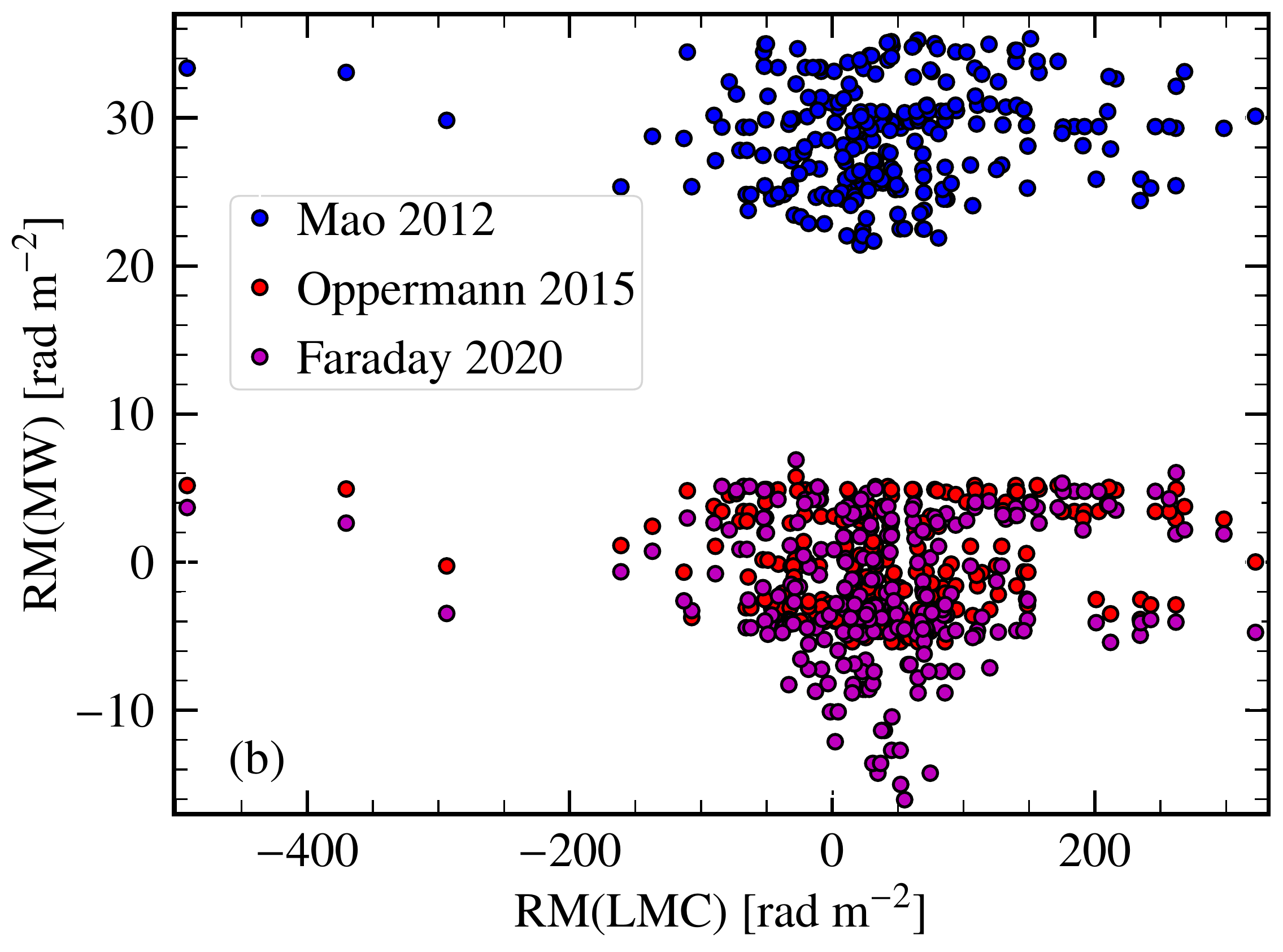}
    \caption{Comparison between the three different Milky Way $\RM$ models, analytical model (blue points, using \citet{MaoEA2008} for the SMC and \citet{MaoEA2012} for the LMC), Oppermann 2015 (red points, \citet{OppermannEA2015}), and Faraday 2020 (magenta points, \citet{HutschenreuterEA2020}), for the $\RM$s observed from the SMC ((a), $\RM_{\rm SMC}$) and LMC ((b), $\RM_{\rm LMC}$). The $\RM_{\rm MW}$ predicted by the analytical model is always higher than the other two, which gives comparable numbers in that region of the sky. These differences in models might affect the estimated properties of small-scale magnetic fields.}
    \label{fig:rmmw}
\end{figure*}

We remove the $\RM$ contribution due to the Milky Way from the observed $\RM$s using a known model. Here, we compare three different models for the Milky Way $\RM$ at sky positions where $\RM$ sources for the SMC and LMC are located. These models are following: analytical model from the literature (Eq. 5 in \citet{MaoEA2008} for the SMC and Eq. 3 in \citet{MaoEA2012} for the LMC), a Bayesian model by \citet{OppermannEA2015}, and a more recent Bayesian model by \citet{HutschenreuterEA2020} (this model is refereed as `Faraday 2020'). \Fig{fig:rmmw} shows the comparison between these three models and \Tab{tab:rmmw} summarises the mean and standard deviation of $\RM (\rm MW)$ for each model. We find that the analytical models always give the highest $\RM (\rm MW)$ and also have a larger variation (the mean and standard deviation both are higher, see \Tab{tab:rmmw}). The other two models give comparable results for the SMC and slightly different numbers for the LMC. Thus, these models can introduce differences in computed quantities and by extension to the estimates of small-scale magnetic field properties. For our work, we chose analytical models as the data from \cite{MaoEA2012} for the LMC, which we use, is also used to generate the Bayesian Milky Way $\RM$ models and thus might show some correlation between the $\RM$ of the LMC and $\RM (\rm MW)$ \citep[see Sec. 4.2.2 in][]{HutschenreuterEA2020}.

\begin{table*} 
\caption{Table showing the mean ($\mu$) and standard deviation ($\sigma$) for $\RM(\rm MW)$ computed at locations of $\RM$s for the SMC and LMC for each of the three different models: analytical (\citet{MaoEA2008} for the SMC and  \citet{MaoEA2012} for the LMC), Oppermann 2015 \citep{OppermannEA2015}, and Faraday 2020 \citep{HutschenreuterEA2020}. All numerical values are in $\rad~\m^{-2}$.}
\label{tab:rmmw}
\begin{tabular}{ccccccc} 
\hline 
\multirow{2}{*}{Cloud} &
 \multicolumn{2}{|c|}{Analytical} &\multicolumn{2}{|c|}{Oppermann 2015} & \multicolumn{2}{|c|}{Faraday 2020} \\
 & $\mu_{\RM(\rm MW)}$ & $\sigma_{\RM(\rm MW)}$ & $\mu_{\RM(\rm MW)}$ & $\sigma_{\RM(\rm MW)}$ & $\mu_{\RM(\rm MW)}$ & $\sigma_{\RM(\rm MW)}$ \\
 
\hline
SMC & 25.1 & 6.6 & 7.7 & 1.9 & 7.5 & 1.3 \\
LMC & 28.8 & 3.5 & -0.043 & 3.4 & -2.0 & 4.7 \\
\hline
\end{tabular}
\end{table*}

\section{Effect of fluctuations in the intrinsic $\RM$ distribution of sources} \label{sec:rmint}
\begin{figure*}
    \includegraphics[width=\columnwidth]{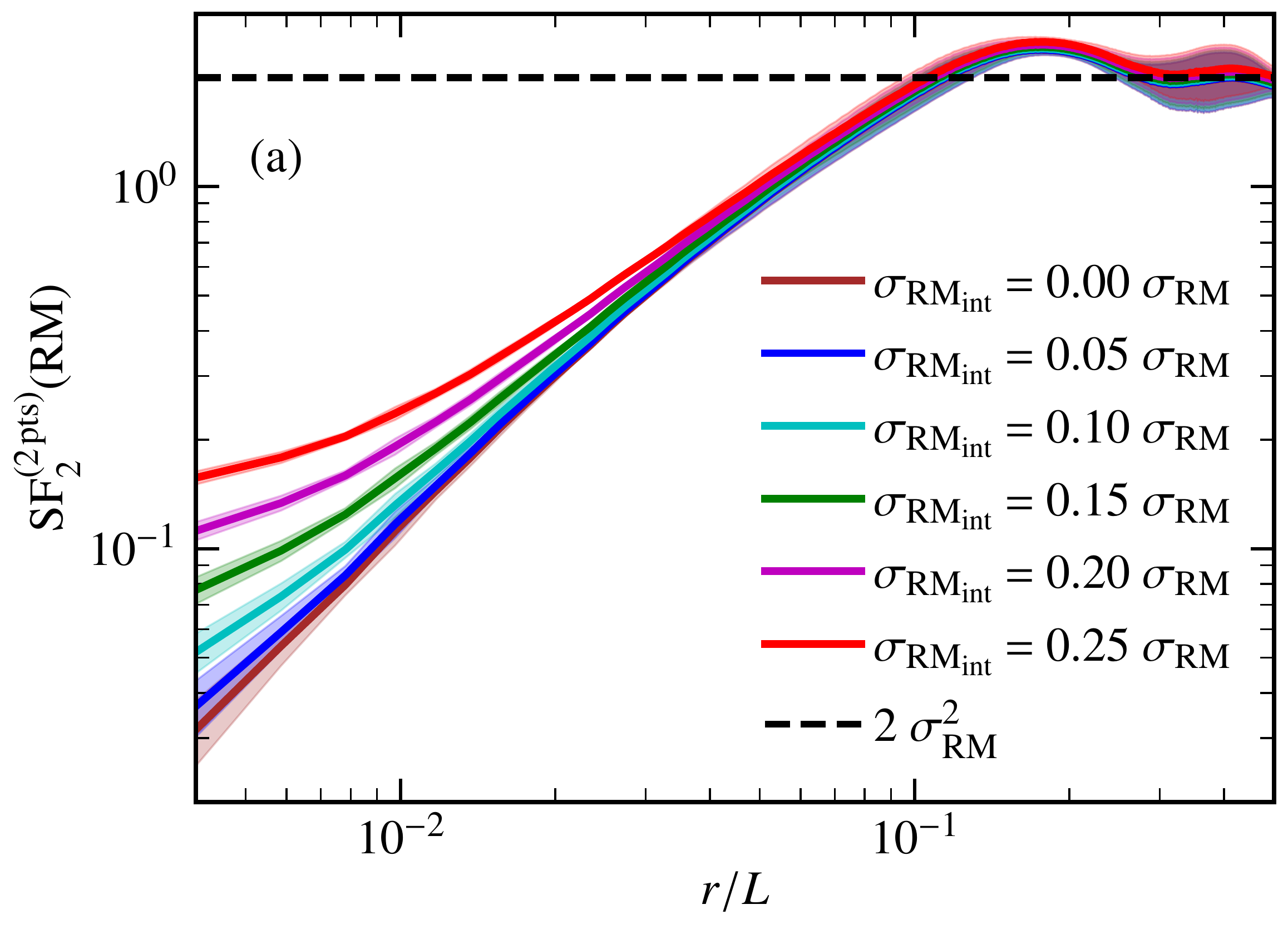} \hspace{0.5cm}
    \includegraphics[width=\columnwidth]{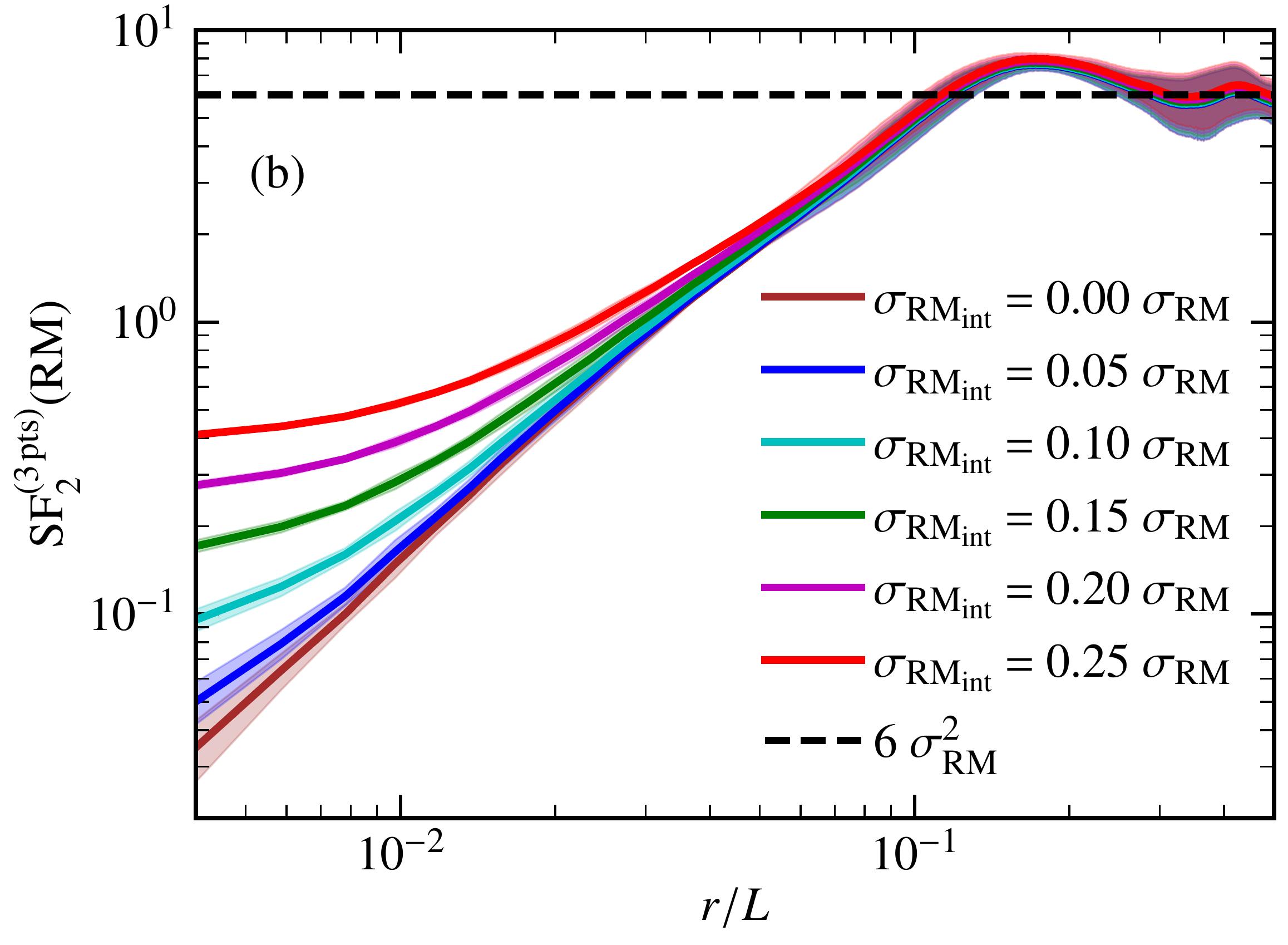} 
    \includegraphics[width=\columnwidth]{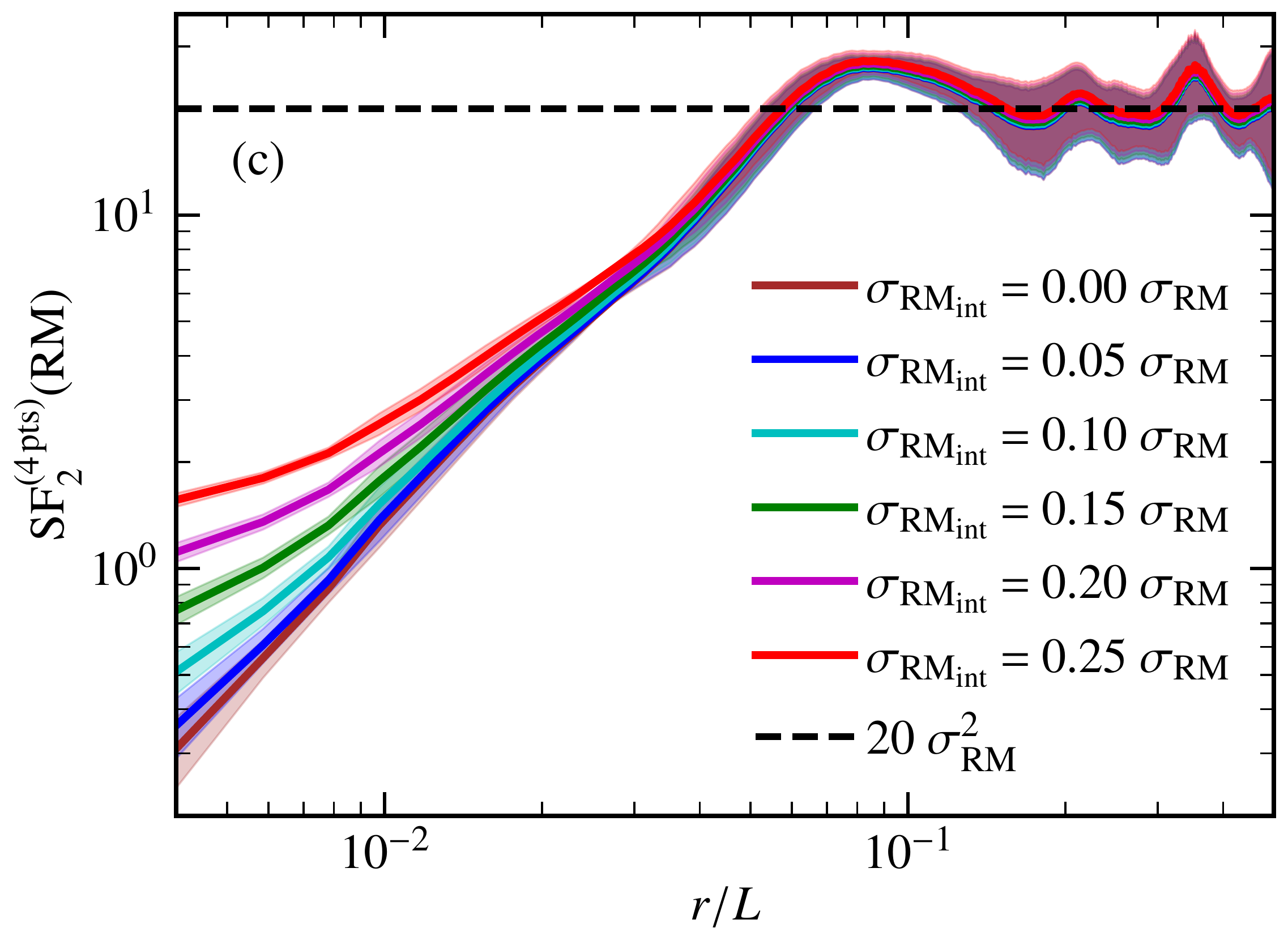} \hspace{0.5cm}
    \includegraphics[width=\columnwidth]{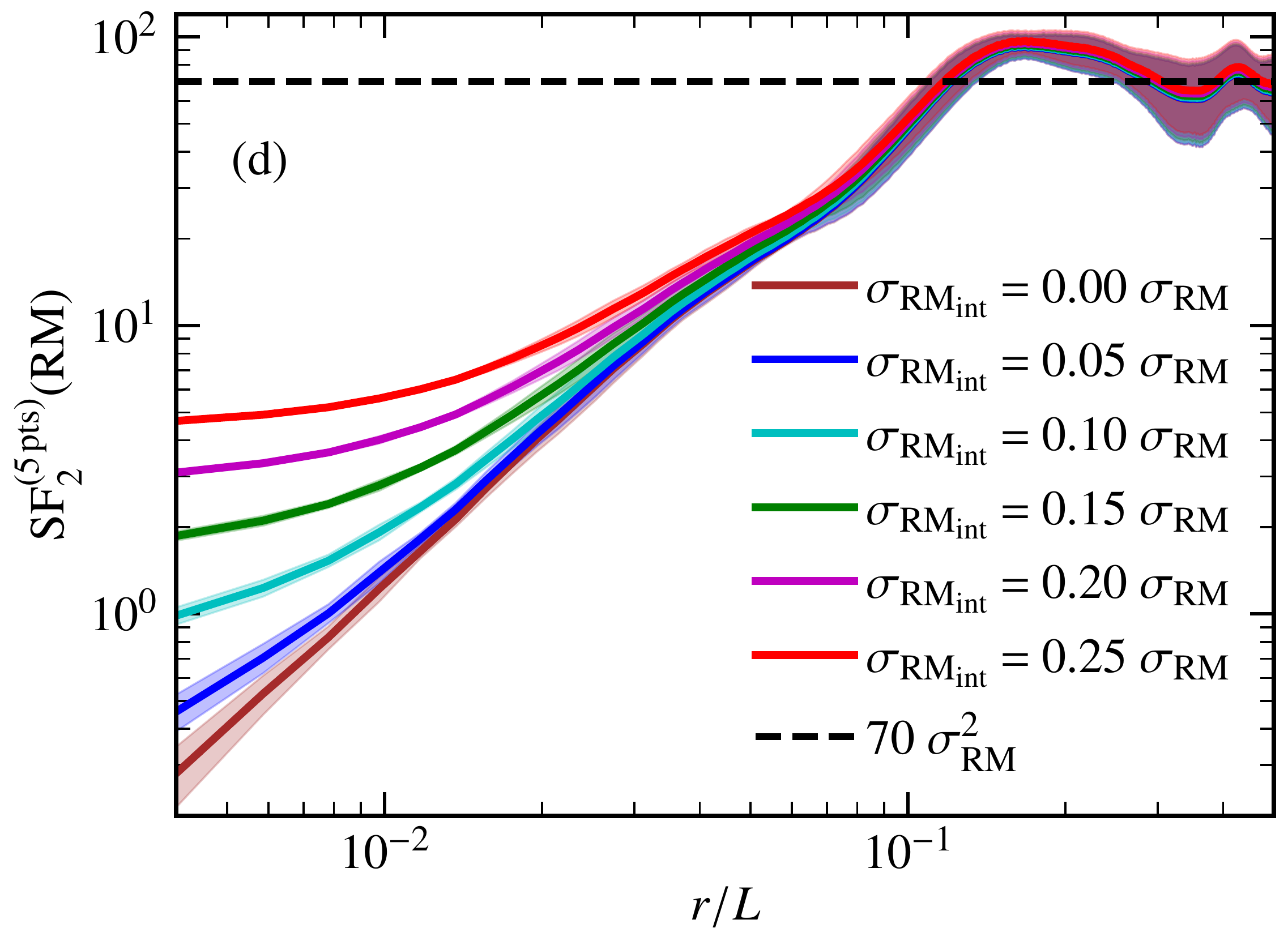} 
    \caption{\revb{Second-order $\RM$ structure function computed using stencils with two (a), three (b), four (c), and five (d) points for a Gaussian random magnetic field with power spectrum, $\pk(b) \sim k^{-1.67}$ and different levels of intrinsic $\RM$ fluctuations (solid lines, corresponding values shown in the legend), $\sigma_{\RM_{\rm int}}$ in units of $\sigma_{\RM}$ (which is due to the intervening medium). The dashed, black line shows the expected values of the second-order $\RM$ structure function at large length scales (depends on the number of points in the stencil, see \Sec{sec:met}). For all the cases (i.e., different number of points in the stencil), the following properties of the second-order $\RM$ structure function: the slope at larger scales, the turnover scale, and the expected values at large length scales are unaffected by the intrinsic $\RM$ fluctuations. Only the slope of the second-order $\RM$ structure function at smaller scales is affected (the structure function at smaller scales becomes flatter) as $\sigma_{\RM_{\rm int}}$ increases.}}
\label{fig:rmint}
\end{figure*}

\revb{In principle, the background $\RM$ distribution can also have some dispersion, say $\sigma_{\RM_{\rm int}}$, which is intrinsic to the source distribution. This $\sigma_{\RM_{\rm int}}$ is usually small compared to the $\RM$ due to the intervening medium, $\sigma_{\RM}$. For the case of the SMC and LMC, $\sigma_{\RM_{\rm int}}$ is $\lesssim 10\%$ of the $\sigma_{\RM}$ (see \Sec{sec:obsstr}). So, practically, its effect on the analysis is small and can be neglected. However, for completeness, here we show how varying $\sigma_{\RM_{\rm int}}$ affects the second-order $\RM$ structure function.}

\revb{For a Gaussian random magnetic field with power spectrum, $\pk(b) \sim k^{-1.67}$, in \Fig{fig:rmint}, we show the second-order $\RM$ structure function computed using stencils with two, three, four, and five points and varying levels of intrinsic $\RM$ fluctuations (in units of $\sigma_{\RM_{\rm int}}$, ranging from $0$ to $25 \%$). For higher $\sigma_{\RM_{\rm int}} / \sigma_{\RM}$, at smaller scales, the second-order $\RM$ structure function gets flatter and this affects the slope of the $\RM$ structure function. However, the slope, turnover scale, and the expected value at large length scales are not significantly affected. These conclusions hold true for any number of points in the stencil.}

\section{Effect of varying number of sources} \label{sec:nsamp}
\begin{figure*}
    \includegraphics[width=\columnwidth]{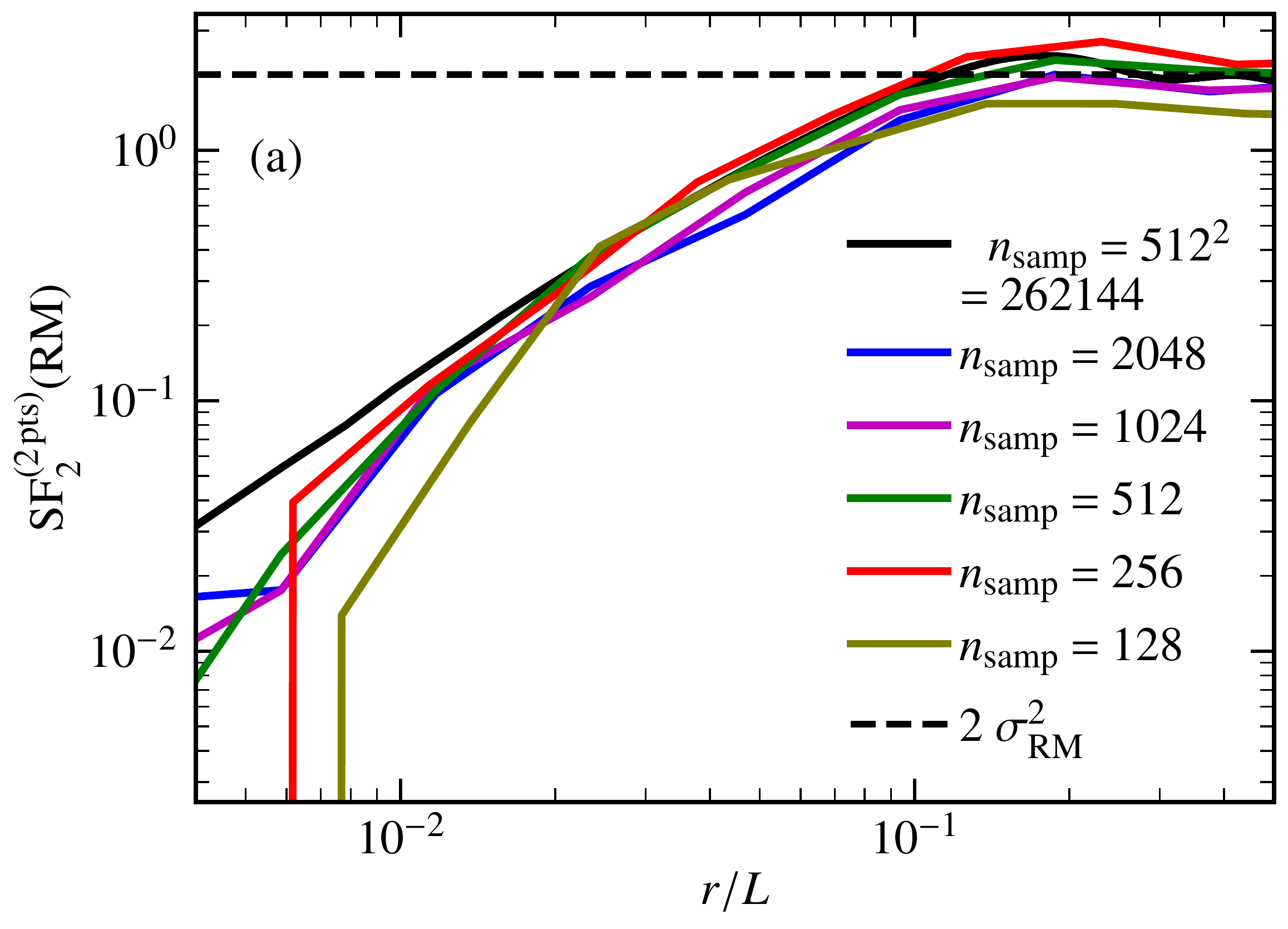} \hspace{0.5cm}
    \includegraphics[width=\columnwidth]{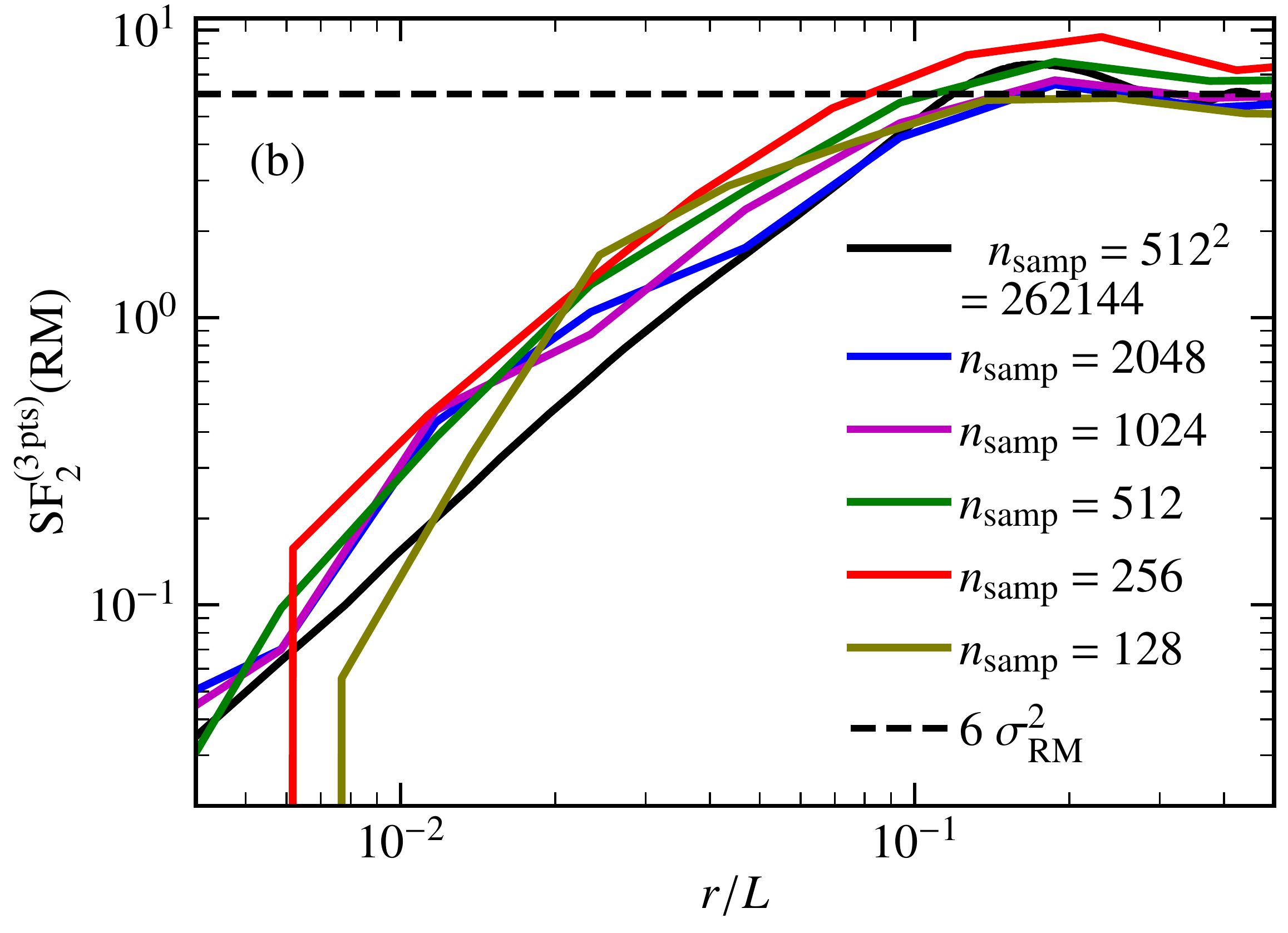}
    \caption{\revb{Second-order $\RM$ structure function computed using stencils with two (a) and three (b) points for a Gaussian random magnetic field with power spectrum, $\pk(b) \sim k^{-1.67}$ and varying number of $\RM$ sources in the sample, $n_{\rm samp}$ (motivated by \Fig{fig:smcrmsf} and \Fig{fig:lmcrmsf}). The dashed, black line shows the expected values of the second-order $\RM$ structure function at large length scales. The solid, black line shows the entire sample ($512^{2}$) and coloured lines show the different number of samples chosen randomly from the entire sample. The slope at smaller scales is severely affected by the number of samples. However, with a few hundred points, the slope of the second-order $\RM$ structure function at larger scales, the turnover scale, and the expected values at large length scales remain roughly the same.}}
    \label{fig:nsamp}
\end{figure*}

\revb{Usually, $\RM$ observations have a much smaller sample size in comparison to numerical experiments. Here, we test how varying the number of samples affects the second-order $\RM$ structure function. In \Fig{fig:nsamp}, we compare the second-order $\RM$ structure function obtained for the whole sample (using $512^{2}$ points, shown in solid, black line) and that computed for a random sample with varying sample size, $n_{\rm samp}$ (coloured lines). For a lower number of sources, the second-order $\RM$ structure function is severely affected at smaller scales. However, with a sample size of a few hundred sources, the slope on larger scales, the turnover scale, and the expected value at the large scales remain largely unaffected.
}


\bsp
\label{lastpage}
\end{document}